\RequirePackage{fix-cm} 
\documentclass[a4paper, twoside, reqno, dvips, 12pt]{amsart}
\usepackage{fixltx2e}   

\usepackage{etex}

\usepackage[latin1]{inputenc}
\usepackage[T1]{fontenc}

\usepackage{esint}
\usepackage{dsfont}
\usepackage{xspace}
\usepackage{amsgen}
\usepackage{amsthm}
\usepackage{amssymb}
\usepackage{amsmath}
\usepackage{wasysym}
\usepackage{upgreek}
\usepackage{amsfonts}
\usepackage{stmaryrd}
\usepackage{scalerel}
\usepackage{mathtools}
\usepackage{stackengine}

\usepackage{relsize}
\usepackage[mathcal, mathscr]{euscript}

\usepackage{mathrsfs}
\DeclareMathAlphabet{\mathscrbf}{OMS}{mdugm}{b}{n}

\usepackage{a4wide}

\usepackage[dvipsnames, table]{xcolor}
\definecolor{bckg}{RGB}{20.8, 20.8, 20.8}
\definecolor{oneblue}{rgb}{0.0, 0.0, 0.85}
\definecolor{Lightblue}{RGB}{214, 214, 214}
\definecolor{bluepigment}{rgb}{0.2, 0.2, 0.6}
\definecolor{charcoal}{rgb}{0.21, 0.27, 0.31}
\definecolor{denimblue}{rgb}{0.08, 0.38, 0.74}
\definecolor{Lightgray}{rgb}{0.89, 0.89, 0.89}
\definecolor{darkgrey}{rgb}{0.273, 0.281, 0.30}
\definecolor{darkelectricblue}{rgb}{0.33, 0.41, 0.47}

\usepackage[sort&compress, comma, square, numbers]{natbib}

\usepackage{psfrag}
\usepackage{graphicx}
\usepackage{adjustbox}
\usepackage[tight]{subfigure}
\usepackage{morefloats}
\usepackage{indentfirst}

\usepackage[usenames, dvipsnames, pdf]{pstricks}
\usepackage{epsfig}
\usepackage{pst-grad} 
\usepackage{pst-plot} 

\usepackage{acronym}
\usepackage{microtype}
\usepackage[labelsep=period,%
            labelfont={bf,sf,color=bluepigment},%
            justification=raggedright]{caption}

\usepackage[perpage, symbol]{footmisc}

\usepackage[colorlinks,
           urlcolor=oneblue,
           linkcolor=denimblue,
           citecolor=NavyBlue,
           bookmarksopen=false,
           pdfpagemode=UseNone,
           pagebackref]{hyperref}

\usepackage[explicit]{titlesec}

\titleformat{\section}[block]
  {\color{NavyBlue}\Large\sffamily\bfseries}
  {}
  {0.0em}
  {\colorbox{bckg!5}{\strut\parbox{\dimexpr\linewidth-2\fboxsep\relax}{\thesection. #1}}}
  [\vspace*{0.33em}]

\titleformat{name=\section,numberless}[block]
  {\color{NavyBlue}\Large\sffamily\bfseries}
  {}
  {0.0em}
  {\colorbox{bckg!5}{\strut\parbox{\dimexpr\linewidth-2\fboxsep\relax}{#1}}}
  [\vspace*{0.33em}]

\titleformat{\subsection}
  {\color{NavyBlue}\large\sffamily\bfseries}
  {}
  {0.0em}
  {\colorbox{bckg!5}{\parbox{\dimexpr\linewidth-2\fboxsep\relax}{\thesubsection. #1}}}
  [\vspace*{0.33em}]

\titleformat{name=\subsection,numberless}
  {\color{NavyBlue}\Large\sffamily\bfseries}
  {}
  {0em}
  {\colorbox{bckg!5}{\parbox{\dimexpr\linewidth-2\fboxsep\relax}{#1}}}
  [\vspace*{0.33em}]

\titleformat{\subsubsection}
  {\color{bluepigment}\sffamily\normalsize\bfseries}
  {\thesubsubsection}
  {0.5em}
  {#1}
  [\vspace*{0.33em}]

\titleformat{\paragraph}[runin]
  {\color{bluepigment}\sffamily\small\bfseries}
  {}
  {0em}
  {#1}

\titlespacing{\section}{0.0em}{1.5em plus 2pt minus 2pt}%
{1.0em plus 2pt minus 2pt}[0em]
\titlespacing{\subsection}{0.5em}{1.5em plus 2pt minus 2pt}%
{1.0em}[0em]
\titlespacing{\subsubsection}{0.5em}{1.5em plus 2pt minus 2pt}%
{1.0em plus 2pt minus 2pt}[0em]

\usepackage{titletoc}

\contentsmargin{0.5em}
\newlength{\tocsep} 
\titlecontents{section}[\tocsep]
  {\addvspace{10pt}\bfseries\sffamily}
  {\contentslabel[\thecontentslabel]{\tocsep}}
  {}
  {\ \titlerule*[0.75pc]{.}\ \thecontentspage}
  []
\titlecontents{subsection}[\tocsep]
  {\addvspace{8pt}\sffamily}
  {\contentslabel[\thecontentslabel]{\tocsep}}
  {}
  {\ \titlerule*[0.5pc]{.}\ \thecontentspage}
  []
\titlecontents*{subsubsection}[\tocsep]
  {\addvspace{2pt}\footnotesize\sffamily}
  {}
  {}
  {\ \titlerule*[0.35pc]{.}\ \thecontentspage}
  [\\*]

\makeatletter
\def\@setauthors{%
  \begingroup
  \def\thanks{\protect\thanks@warning}%
  \trivlist
  \centering\footnotesize \@topsep30\p@\relax
  \advance\@topsep by -\baselineskip
  \item\relax
  \author@andify\authors
  \def\\{\protect\linebreak}%
  \textsc{\normalsize\textcolor{darkelectricblue}{\authors}}%
  \ifx\@empty\contribs
  \else
    ,\penalty-3 \space \@setcontribs
    \@closetoccontribs
  \fi
  \endtrivlist
  \endgroup
}
\def\@settitle{\begin{center}%
  \baselineskip14\p@\relax
    \bfseries
    \textsc{\Large\textcolor{charcoal}{\@title}}
  \end{center}%
}
\makeatother

\usepackage{enumitem}
\setlist[description]{%
  topsep=30pt,               
  itemsep=5pt,               
  font={\bfseries\sffamily\color{NavyBlue}}, 
}

\usepackage{fancyhdr}
\usepackage{lastpage}

\newcommand*\Title{\textcolor{bluepigment}{Dispersive shallow water wave modelling. Part II}}
\newcommand*\Authors{\textcolor{bluepigment}{G.~Khakimzyanov, D.~Dutykh, et al.}}
\newcommand*{\plogo}{\textcolor{gray}{{\texttt{arXiv.org} / \textsc{hal}}}} 

\numberwithin{equation}{section}

\newtheorem{lemma}{Lemma}

\newtheorem{remark}{Remark}

\newtheorem{theorem}{Theorem}

\newcommand{\up}[1]{$^{\mathrm{\small\textsf{#1}}}$} 

\newcommand{\p}{\Bbbk}

\newcommand{\m}{{\sf m}\xspace}
\newcommand{\s}{{\sf s}\xspace}

\newcommand{\cm}{{\sf cm}\xspace}
\newcommand{\ub}{\bar{\u}}

\newcommand{\pc}{\check{p}}

\newcommand{\N}{\mathds{N}}
\newcommand{\R}{\mathds{R}}

\newcommand{\Id}{\mathbb{I}}
\newcommand{\A}{\mathscr{A}}

\newcommand{\I}{\mathcal{I}}

\newcommand{\ud}{\mathrm{d}}
\newcommand{\ui}{\mathrm{i}}
\newcommand{\ue}{\mathrm{e}}
\newcommand{\E}{\mathscr{E}}
\newcommand{\J}{\mathscr{U}}
\newcommand{\F}{\mathcal{F}}
\newcommand{\Q}{\mathcal{Q}}

\newcommand{\Cs}{\mathscr{C}}
\newcommand{\Di}{\mathcal{D}}

\newcommand{\Ru}{\mathcal{R}}

\newcommand{\Pp}{\mathscr{P}}
\newcommand{\Rr}{\mathscr{R}}
\renewcommand{\beta}{\upbeta}

\renewcommand{\leq}{\leqslant}
\renewcommand{\geq}{\geqslant}
\newcommand{\eps}{\varepsilon}
\renewcommand{\O}{\mathcal{O}}

\renewcommand{\H}{\mathcal{H}}

\newcommand{\D}{\mathscrbf{D}}
\newcommand{\U}{\boldsymbol{U}}
\renewcommand{\alpha}{\upalpha}

\newcommand{\x}{\boldsymbol{x}}
\newcommand{\vO}{\boldsymbol{0}}

\renewcommand{\u}{\boldsymbol{u}}
\renewcommand{\v}{\boldsymbol{v}}
\newcommand{\const}{\mathrm{const}}

\newcommand{\No}{$\mathrm{N}^\circ$}

\newcommand{\Ac}{\mathring{\mathscr{A}}}
\newcommand{\Fc}{\mathring{\mathcal{F}}}
\newcommand{\Hc}{\mathring{\mathcal{H}}}

\newcommand{\vc}{\mathring{\boldsymbol{v}}}

\newcommand{\h}{\Delta q}
\newcommand{\Pnh}{\power}
\newcommand{\pb}{\varrho}
\newcommand{\dx}{\Delta x}
\newcommand{\om}{\upvarpi}
\renewcommand{\r}{\Upsilon}
\newcommand{\Ff}{\mathbf{F}}
\newcommand{\K}{\mathcal{K}}
\newcommand{\Lb}{\mathbf{L}}
\newcommand{\Rb}{\mathbf{R}}
\newcommand{\Jj}{\mathcal{J}}
\newcommand{\Fl}{\mathscr{F}}
\newcommand{\Gl}{\mathscr{G}}
\newcommand{\hc}{\mathring{h}}
\newcommand{\uc}{\mathring{u}}
\newcommand{\coef}{\upvartheta}
\newcommand{\Maltese}{\bigstar}
\newcommand{\etac}{\mathring{\eta}}
\newcommand{\Lab}{\mathbf{\Lambda}}
\newcommand{\Pnhc}{\mathring{\power}}
\newcommand{\pbc}{\mathring{\varrho}}
\newcommand{\rc}{\mathring{\Upsilon}}
\newcommand{\Pc}{\mathring{\mathbb{P}}}
\newcommand{\Kc}{\mathring{\mathcal{K}}}
\newcommand{\Lbc}{\mathring{\mathbf{L}}}
\newcommand{\Rbc}{\mathring{\mathbf{R}}}
\newcommand{\Flc}{\mathring{\mathscr{F}}}
\newcommand{\Glc}{\mathring{\mathscr{G}}}
\newcommand{\Rrc}{\mathring{\mathscr{R}}}
\newcommand{\Labc}{\mathring{\mathbf{\Lambda}}}
\newcommand{\Labbc}{\bar{\mathring{\mathbf{\Lambda}}}}
\newcommand{\power}{\raisebox{.15\baselineskip}{\Large\ensuremath{\wp}}}

\newcommand\openbigstar[1][0.7]{%
  \scalerel*{%
    \stackinset{c}{-.125pt}{c}{}{\scalebox{#1}{\color{white}{$\bigstar$}}}{%
      $\bigstar$}%
  }{\bigstar}
}

\newcommand{\ie}{\emph{i.e.}\xspace}
\newcommand{\eg}{\emph{e.g.}\xspace}
\newcommand{\etc}{\emph{etc.}\xspace}

\renewcommand{\sim}{\thicksim}
\renewcommand{\div}{\grad\scal}
\newcommand{\sech}{\mathrm{sech}}
\newcommand{\Mat}{\mathrm{Mat}\,}
\newcommand{\scal}{\boldsymbol{\cdot}}
\newcommand{\grad}{\boldsymbol{\nabla}}

\newcommand{\abs}[1]{\lvert\, #1\, \rvert}

\newcommand{\norm}[1]{\lVert\, #1\, \rVert}
\newcommand{\otimesb}{\boldsymbol{\otimes}}

\newcommand{\pd}[2]{\frac{\partial #1}{\partial\/ #2}}
\newcommand{\od}[2]{\frac{\mathrm{d} #1}{\mathrm{d}\/#2}}

\newcommand{\eqdef}{\mathop{\stackrel{\,\mathrm{def}}{:=}\,}}

\newcommand{\half}{{\textstyle{1\over2}}}

\newcommand{\sixth}{{\textstyle{1\over6}}}
\newcommand{\fourth}{{\textstyle{1\over4}}}

\newcommand{\threeeights}{{\textstyle{3\over8}}}

\usepackage{acronym}
\acrodef{bvp}[BVP]{Boundary Value Problem}
\acrodef{NSWE}{Nonlinear Shallow Water Equations}

\begin{document}

\title[\Title]{Dispersive shallow water wave modelling. Part II: Numerical simulation on a globally flat space}

\author[G.~Khakimzyanov]{Gayaz Khakimzyanov}
\address{\textbf{G.~Khakimzyanov:} Institute of Computational Technologies, Siberian Branch of the Russian Academy of Sciences, Novosibirsk 630090, Russia}
\email{Khak@ict.nsc.ru}

\author[D.~Dutykh]{Denys Dutykh$^*$}
\address{\textbf{D.~Dutykh:} LAMA, UMR 5127 CNRS, Universit\'e Savoie Mont Blanc, Campus Scientifique, F-73376 Le Bourget-du-Lac Cedex, France}
\email{Denys.Dutykh@univ-smb.fr}
\urladdr{http://www.denys-dutykh.com/}
\thanks{$^*$ Corresponding author}

\author[O.~Gusev]{Oleg Gusev}
\address{\textbf{O.~Gusev:} Institute of Computational Technologies, Siberian Branch of the Russian Academy of Sciences, Novosibirsk 630090, Russia}
\email{gusev\_oleg\_igor@mail.ru}

\author[N.~Yu.~Shokina]{Nina Yu. Shokina}
\address{\textbf{N.~Yu.~Shokina:} Institute of Computational Technologies, Siberian Branch of the Russian Academy of Sciences, Novosibirsk 630090, Russia}
\email{Nina.Shokina@googlemail.com}
\urladdr{https://www.researchgate.net/profile/Nina\_Shokina/}



\begin{titlepage}
\thispagestyle{empty} 
\noindent
{\Large Gayaz \textsc{Khakimzyanov}}\\
{\it\textcolor{gray}{Institute of Computational Technologies, Novosibirsk, Russia}}
\\[0.02\textheight]
{\Large Denys \textsc{Dutykh}}\\
{\it\textcolor{gray}{CNRS--LAMA, Universit\'e Savoie Mont Blanc, France}}
\\[0.02\textheight]
{\Large Oleg \textsc{Gusev}}\\
{\it\textcolor{gray}{Institute of Computational Technologies, Novosibirsk, Russia}}
\\[0.02\textheight]
{\Large Nina \textsc{Shokina}}\\
{\it\textcolor{gray}{Institute of Computational Technologies, Novosibirsk, Russia}}
\\[0.08\textheight]

\vspace*{1.1cm}

\colorbox{Lightblue}{
  \parbox[t]{1.0\textwidth}{
    \centering\huge\sc
    \vspace*{0.7cm}
    
    \textcolor{bluepigment}{Dispersive shallow water wave modelling. Part II: Numerical simulation on a globally flat space}
    
    \vspace*{0.7cm}
  }
}

\vfill 

\raggedleft     
{\large \plogo} 
\end{titlepage}


\newpage
\thispagestyle{empty} 
\par\vspace*{\fill}   
\begin{flushright} 
{\textcolor{denimblue}{\textsc{Last modified:}} \today}
\end{flushright}


\newpage
\maketitle
\thispagestyle{empty}


\begin{abstract}

In this paper we describe a numerical method to solve numerically the weakly dispersive fully nonlinear \textsc{Serre}--\textsc{Green}--\textsc{Naghdi} (SGN) celebrated model. Namely, our scheme is based on reliable finite volume methods, proven to be very efficient for the hyperbolic part of equations. The particularity of our study is that we develop an adaptive numerical model using moving grids. Moreover, we use a special form of the SGN equations where non-hydrostatic part of pressure is found by solving a nonlinear elliptic equation. Moreover, this form of governing equations allows to determine the natural form of boundary conditions to obtain a well-posed (numerical) problem.


\bigskip\bigskip
\noindent \textbf{\keywordsname:} nonlinear dispersive waves; moving adaptive grids; finite volumes; conservative finite differences \\


\end{abstract}


\newpage
\thispagestyle{empty}
\tableofcontents
\thispagestyle{empty}


\newpage
\section{Introduction}

In 1967 D.~\textsc{Peregrine} derived the first two-dimensional \textsc{Boussinesq}-type system of equations \cite{Peregrine1967}. This model described the propagation of long weakly nonlinear waves over a general non-flat bottom. From this landmark study the modern era of long wave modelling started. On one hand researchers focused on the development of new models and in parallel the numerical algorithms have been developed. We refer to \cite{Brocchini2013} for a recent `reasoned' review of this topic.

The present manuscript is the continuation of our series of papers devoted to the long wave modelling. In the first part of this series we derived the so-called base model \cite{Khakimzyanov2016c}, which encompasses a number of previously known models (but, of course, not all of nonlinear dispersive systems). The governing equations of the base model are
\begin{align}\label{eq:base1}
  \H_t\ +\ \div[\,\H\U\,]\ &=\ 0\,, \\
  \ub_t\ +\ (\ub\scal\grad)\ub\ +\ \frac{\grad\Pp}{\H}\ &=\ \frac{\pc}{\H}\;\grad h\ -\ \frac{1}{\H}\;\Bigl[\,(\H\J)_t\ +\ (\ub\scal\grad)(\H\J)\nonumber \\
  &\ +\ \H(\J\scal\grad)\,\ub\ +\ \H\J\div\ub\,\Bigr]\,, \label{eq:base2}
\end{align}
where $\U\ \eqdef\ \ub\ +\ \J$ is the modified horizontal velocity and $\J\ =\ \J(\H,\,\ub)$ is the closure relation to be specified later. Depending on the choice of this variable various models can be obtained (see \cite[Section~\textsection 2.4]{Khakimzyanov2016c}). Variables $\Pp$ and $\pc$ are related to the fluid pressure. The physical meaning of these variables is reminded below in Section~\ref{sec:model}. In the present paper we propose an adaptive numerical discretization for a particular, but very popular nowadays model which can be obtained from the base model \eqref{eq:base1}, \eqref{eq:base2}. Namely, if we choose $\J\ \equiv\ \vO$ (thus, $\U$ becomes the depth-averaged velocity $\u$) then we obtain equations equivalent to the celebrated \textsc{Serre}--\textsc{Green}--\textsc{Naghdi} (SGN) equations \cite{Serre1956, Serre1953a, Green1974} (rediscovered later independently by many other researchers). This system will be the main topic of our numerical study. Most often, adaptive techniques for dispersive wave equations involve the so-called Adaptive Mesh Refinement (AMR) \cite{Sadaka2012} (see also \cite{Berger2011} for nonlinear shallow water equations). The particularity of our study is that we conserve the total number of grid points and the adaptivity is achieved by judiciously redistributing them in space \cite{Huang2001a, Huang2001}. The ideas of redistributing grid nodes is stemming from the works of \textsc{Bakhvalov} \cite{Bakhvalov1969}, \textsc{Il'in} \cite{Ilin1969} and others \cite{Alalykin1970, Thomas1979}.

The base model \eqref{eq:base1}, \eqref{eq:base2} admits an elegant conservative form \cite{Khakimzyanov2016c}:
\begin{align}\label{eq:base3}
  \H_t\ +\ \div[\,\H\U\,]\ &=\ 0\,, \\
  (\H\,\U)_t\ +\ \div\Bigl[\,\H\ub\otimesb\U\ +\ \Pp(\H,\,\ub)\cdot\Id\ +\ \H\,\J\otimesb\ub\,\Bigr]\ &=\ \pc\,\grad h\,, \label{eq:base4}
\end{align}
where $\Id\ \in\ \Mat_{2\,\times\,2}(\R)$ is the identity matrix and the operator $\otimesb$ denotes the tensorial product. We note that the pressure function $\Pp(\H,\,\ub)$ incorporates the familiar hydrostatic pressure part $\dfrac{g\,\H^{\,2}}{2}$ well-known from the Nonlinear Shallow Water Equations (NSWE) \cite{SV1871, Barthelemy2004}. By setting $\J\ \equiv\ \vO$ we obtain readily from \eqref{eq:base3}, \eqref{eq:base4} the conservative form of the SGN equations (one can notice that the mass conservation equation \eqref{eq:base1} was already in conservative form).

Nonlinear dispersive wave equations represent certain numerical difficulties since they involve mixed derivatives (usually of the horizontal velocity variable, but sometimes of the total water depth as well) in space and time. These derivatives have to be approximated numerically, thus leaving a lot of room for the creativity. Most often the so-called \emph{Method Of Lines} (MOL) is employed \cite{Schiesser1994, Reddy1992, Shampine1994, Kreiss1992}, where the spatial derivatives are discretized first and the resulting system of coupled Ordinary Differential Equations (ODEs) is then approached with more or less standard ODE techniques, see \eg \cite{Hairer2009, Hairer1996}. The MOL separates the choice of time discretization from the procedure of discretization in space, even if the interplay between two schemes might be important. For example, it would be natural to choose the same order of accuracy for both schemes.

Let us review the available spatial discretization techniques employed in recent numerical studies. We focus essentially on fully nonlinear weakly dispersive models, even if some interesting works devoted to \textsc{Boussinesq}-type and unidirectional equations will be mentioned. First of all, dispersive wave equations with the dispersion relation given by a rational function (\emph{\`a la} BBM \cite{Peregrine1966, bona}) usually involve the inversion of an elliptic operator. This gives the first idea of employing the splitting technique between the hyperbolic and elliptic operators. This idea was successfully realized in \eg \cite{Barakhnin1997, Barakhnin1999, Horrillo2006, Bonneton2011}. Historically, perhaps the finite difference techniques were applied first to dispersive (and more general non-hydrostatic) wave equations \cite{Chubarov1987, Madsen1991, Madsen1992, Wei1995a, Casulli1999, Chubarov2000, Cienfuegos2007, Zhao2015}. Then, naturally we arrive to the development of continuous \textsc{Galerkin}/Finite Element type discretizations \cite{Dougalis1985, Bona1986, AntunesDoCarmo1993, Walkley2002, Sorensen2004, DMS1, Mitsotakis2014}. See also a recent review \cite{DMII} and references therein. Pseudo-spectral \textsc{Fourier}-type methods can also be successfully applied to the SGN equations \cite{Dutykh2011a}. See \cite{Fabien2014} for a pedagogical review of pseudo-spectral and radial basis function methods for some shallow water equations. More recently, the finite volume type methods were applied to dispersive equations \cite{LeMetayer2010, Dutykh2011e, ChazelLannes2010, Dutykh2011a, Dutykh2010e, Kazolea2013}. In the present study we also employ a predictor--corrector finite volume type scheme \cite{Shokin2006}, which is described in details below.

The present article is organized as follows. In Section~\ref{sec:model} we present the governing equations in 2D and 1D spatial dimensions. The numerical method is described in Section~\ref{sec:num}. Several numerical illustrations are shown in Section~\ref{sec:res} including the solitary wave/wall or bottom interactions and even a realistic underwater landslide simulation. Finally, in Section~\ref{sec:concl} we outline the main conclusions and perspectives of the present study. In Appendix~\ref{app:press} we provide some details on analytical derivations used in this manuscript.


\section{Mathematical model}
\label{sec:model}

\begin{figure}
  \centering
  \includegraphics[width=0.79\textwidth]{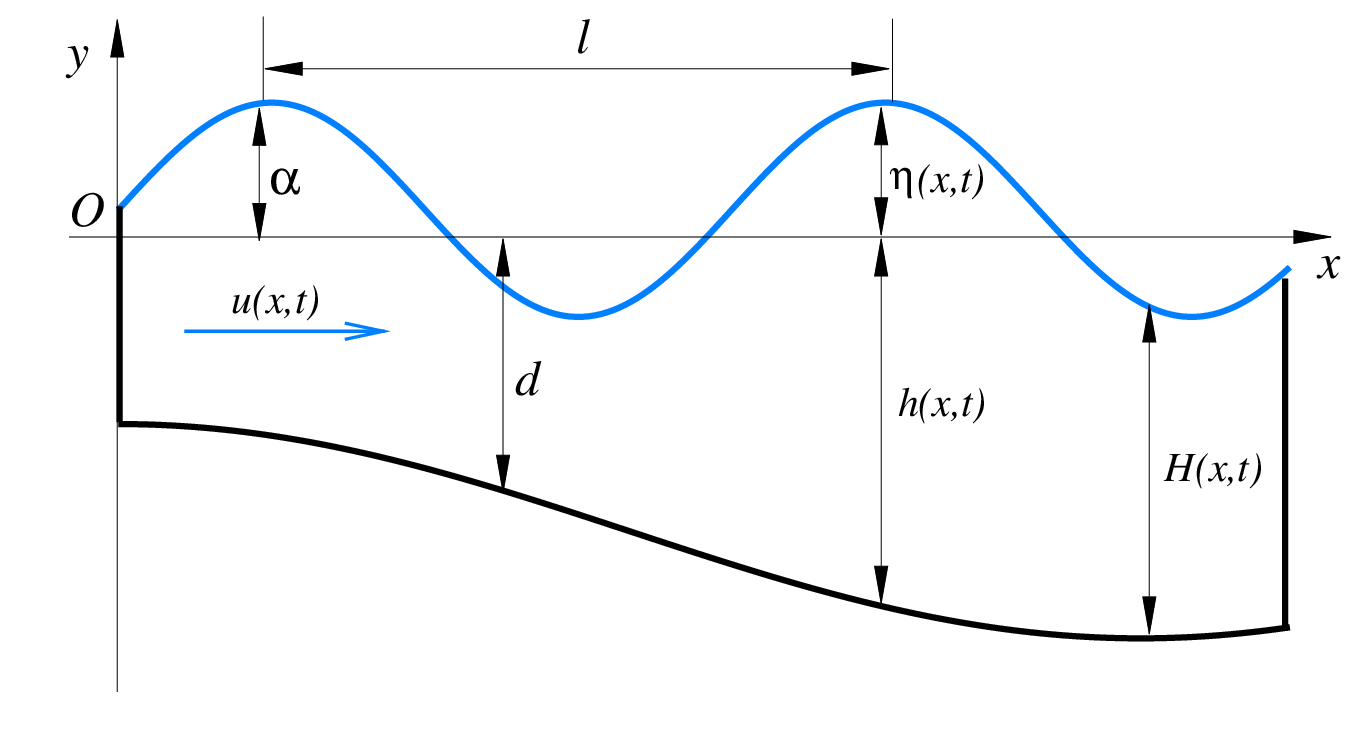}
  \caption{\small\em Sketch of the fluid domain in 2D.}
  \label{fig:sketch}
\end{figure}

In this study we consider the following system of the \textsc{Serre}--\textsc{Green}--\textsc{Naghdi} (SGN) equations, which describes the incompressible homogeneous fluid flow in a layer bounded from below by the impermeable bottom $y\ =\ -h(\x,\,t)$ and above by the free surface $y\ =\ \eta\,(\x,\,t)$, $\x\ =\ (x_1,\,x_2)\ \in\ \R^2$:
\begin{align}\label{eq:eq1}
  \H_t\ +\ \div\bigl[\,\H \u\,\bigr]\ &=\ 0\,, \\
  \u_t\ +\ (\u\scal\grad)\,\u\ +\ \frac{\grad \Pp}{\H}\ &=\ \frac{\pc}{\H}\;\grad h\,, \label{eq:eq2}
\end{align}
where for simplicity we drop the bars over the horizontal velocity variable $\u(\x,\,t)\ =\ \bigl(u_1(\x,\,t), u_2(\x,\,t)\bigr)$. Function $\H(\x,\,t)\,\ \eqdef\,\ h(\x,\,t)\ +\ \eta(\x,\,t)$ being the total water depth. The sketch of the fluid domain is schematically depicted in Figure~\ref{fig:sketch}. For the derivation of equations \eqref{eq:eq1}, \eqref{eq:eq2} we refer to the first part of the present series of papers \cite{Khakimzyanov2016c}. The depth-integrated pressure $\Pp(\u,\,\H)$ is defined as
\begin{equation*}
  \Pp\,(\u,\,\H)\,\ \eqdef\ \,\frac{g\, \H^2}{2}\ -\ \Pnh(\x,\,t)\,,
\end{equation*}
where $\Pnh(\x,\,t)$ is the non-hydrostatic part of the pressure:
\begin{equation}\label{eq:pnh}
  \Pnh\,(\x,\,t)\,\ \eqdef\,\ \frac{\H^{\,3}}{3}\; \Rr_1\ +\ \frac{\H^{\,2}}{2}\; \Rr_2\,,
\end{equation}
with
\begin{equation*}
  \Rr_1\,\ \eqdef\,\ \D(\div\u)\ -\ (\div\u)^2\,, \qquad
  \Rr_2\,\ \eqdef\,\ \D^{\,2}\, h\,, \qquad
  \D\,\ \eqdef\,\ \partial_t\ +\ \u\scal\grad\,.
\end{equation*}
Above, $\D$ is the total or material derivative operator. On the right hand side of equation \eqref{eq:eq2} we have the pressure trace at the bottom $\pc\,\ \eqdef\,\ \left.p\right|_{y\, =\, -h}$, which can be written as
\begin{equation*}
  \pc\,(\x,\,t)\ =\ g \H\ -\ \pb\,(\x,\,t)\,,
\end{equation*}
where $\pb(\x,\,t)$ is again the non-hydrostatic pressure contribution:
\begin{equation}\label{eq:pb}
  \pb\,(\x,\,t)\,\ \eqdef\,\ \frac{\H^{\,2}}{2}\;\Rr_1\ +\ \H\,\Rr_2\,.
\end{equation}

Equations above are much more complex comparing to the classical NSWE (or \textsc{Saint}-\textsc{Venant} equations) \cite{SV1871}, since they contain mixed derivatives up to the third order. From the numerical perspective these derivatives have to be approximated. However, the problem can be simplified if we `extract' a second order sub-problem for the non-hydrostatic component of the pressure. Indeed, it can be shown (see Appendix~\ref{app:press}) that function $\Pnh(\x,\,t)$ satisfies the following second order nonlinear elliptic equation with variable coefficients (by analogy with incompressible \textsc{Navier}--\textsc{Stokes} equations, where the pressure is found numerically by solving a \textsc{Poisson}-type problem \cite{Harlow1965, Chorin1968}):
\smallskip
\begin{equation}\label{eq:ell}
  \div\biggl[\,\frac{\grad\Pnh}{\H}\ -\ \frac{(\grad\Pnh\scal\grad h)\,\grad h}{\H\,\r}\,\biggr]\ -\ 6\,\underbrace{\biggl[\,\frac{2}{\H^{\,3}}\cdot\frac{\r\ -\ 3}{\r}\ +\ \div\Bigl(\frac{\grad h}{\H^{\,2}\,\r}\Bigr)\,\biggr]}_{(\Maltese)}\;\Pnh\ =\ \F\,,
\end{equation}
where $\r\,\ \eqdef\,\ 4\ +\ \abs{\grad h}^2$ and $\F$, $\Rr$ are defined as
\begin{align}\label{eq:F}
  \F\ &\eqdef\ \grad\scal\biggl[\,g\,\grad\eta\ +\ \frac{\Rr\,\grad h}{\r}\,\biggr]\ -\ \frac{6\,\Rr}{\H\,\r}\ +\ 2\,(\div\u)^{\,2}\ -\ 2\,\begin{vmatrix}
    u_{1\,x_1} & u_{1\,x_2} \\
    u_{2\,x_1} & u_{2\,x_2}
  \end{vmatrix}\,,\\
  \Rr\ &\eqdef\ -g\,\grad\eta\scal\grad h\ +\ \bigl[\,(\u\scal\grad)\grad h\,\bigr]\scal\u\ +\ h_{tt}\ +\ 2\,\u\scal\grad h_t \,. \label{eq:R}
\end{align}
Symbol $\abs{\cdot}$ in \eqref{eq:F} denotes the determinant of a $2\times 2$ matrix.

Equation \eqref{eq:ell} is uniformly elliptic and it does not contain time derivatives of the fluid velocity $\u$. If the coefficient ($\Maltese$) is positive (for instance, it is the case for a flat bottom $h(\x,\,t)\ \equiv\ \const$), we deal with a positive operator and stable robust discretizations can be proposed. Taking into account the fact that equation \eqref{eq:ell} is linear with respect to the variable $\Pnh(\x,\,t)$, its discrete counterpart can be solved by direct or iterative methods\footnote{In our implementation we use the direct \textsc{Thomas} algorithm, since in 1D the resulting linear system of equations is tridiagonal with the dominant diagonal.}. Well-posedness of this equation is discussed below (see Section~\ref{sec:well}). The boundary conditions for equation \eqref{eq:ell} will be discussed below in Section~\ref{sec:bcond} (in 1D case only, the generalization to 2D is done by projecting on the normal direction to the boundary).

Introduction of the variable $\Pnh(\x,\,t)$ allows to rewrite equation \eqref{eq:eq2} in the following equivalent form:
\begin{equation}\label{eq:nswe}
  \u_t\ +\ (\u\scal\grad)\u\ +\ g\,\grad\H\ =\ g\,\grad h\ +\ \frac{\grad\Pnh\ -\ \pb\,\grad h}{\H}\,.
\end{equation}
The non-hydrostatic pressure at the bottom $\pb(\x,\,t)$ can be expressed through $\Pnh$ in the following way:
\begin{equation}\label{eq:relpb}
  \pb(\x,\,t)\ =\ \frac{1}{\r}\;\Bigl[\,\frac{6\,\Pnh}{\H}\ +\ \H\,\Rr\ +\ \grad\Pnh\scal\grad h\,\Bigr]\,.
\end{equation}
The derivation of this equation \eqref{eq:relpb} is given in Appendix~\ref{app:press} as well. So, thanks to this relation \eqref{eq:relpb}, the usage of equation \eqref{eq:pb} is not necessary anymore. Once we found the function $\Pnh(\x,\,t)$, we can compute the bottom component from \eqref{eq:relpb}.

\begin{remark}
It can be easily seen that taking formally the limit $\Pnh\ \to\ 0$ and $\pb\ \to\ 0$ of vanishing non-hydrostatic pressures, allows us to recover the classical NSWE (or \textsc{Saint}-\textsc{Venant} equations) \cite{SV1871}. Thus, the governing equations verify the \textsc{Bohr} correspondence principle \cite{Bohr1920}.
\end{remark}

\subsection{Well-posedness conditions}
\label{sec:well}

In order to obtain a well-posed elliptic problem \eqref{eq:ell}, one has to ensure that coefficient ($\Maltese$) is positive. This coefficient involves the bathymetry function $h(\x,\,t)$ and the total water depth $\H(\x,\,t)$. In other words, the answer depends on local depth and wave elevation. It is not excluded that for some wave conditions the coefficient ($\Maltese$) may become negative. In the most general case the positivity condition is trivial and, thus, not very helpful, \ie
\begin{equation}\label{eq:cond}
  (\Maltese)\ \equiv\ \frac{2}{\H^{\,3}}\cdot\frac{\r\ -\ 3}{\r}\ +\ \div\Bigl(\frac{\grad h}{\H^{\,2}\,\r}\Bigr)\ \geq\ 0\,.
\end{equation}
On the flat bottom $h(\x,\,t)\ \to\ d\ =\ \const$ we know that the above condition is satisfied since $\r\ \to\ 4$ and $(\Maltese)\ \to\ \dfrac{1}{2\,\H^{\,3}}\ >\ 0$. Consequently, by continuity of the coefficient ($\Maltese$) we conclude that the same property will hold for some (sufficiently small) variations of the depth $h(\x,\,t)$, \ie $\abs{\grad h}\ \ll\ 1$. In practice it can be verified that bathymetry variations can be even finite so that condition \eqref{eq:cond} still holds.

\begin{remark}
It may appear that restrictions on the bathymetry variations are inherent to our formulation only. However, it is the case of all long wave models, even if this assumption does not appear explicitly in the derivation. For instance, bottom irregularities will inevitably generate short waves (\ie higher frequencies) during the wave propagation process. A priori, this part of the spectrum is not modeled correctly by approximate equations, unless some special care is taken.
\end{remark}

\subsubsection{Linear waves}

Let us take the limit of linear waves $\eta\ \to\ 0$ in expression ($\Maltese$). It will become then
\begin{equation*}
  (\Maltese)\ \to\ (\openbigstar)\ \eqdef\ \frac{2}{h^3}\cdot\frac{\r\ -\ 3}{\r}\ +\ \div\Bigl(\frac{\grad h}{h^{\,2}\,\r}\Bigr)\,.
\end{equation*}
The positivity\footnote{Non-negativity, to be more precise.} condition of ($\openbigstar$) then takes the following form:
\begin{equation*}
  2\,\r\ +\ h\,\bigl\{\,h_{x_1 x_1}\,\bigl(1 - h_{x_1}^2 + h_{x_2}^2\bigr)\ +\ h_{x_2 x_2}\,\bigl(1 + h_{x_1}^2 - h_{x_2}^2\bigr)\ -\ 4\, h_{x_1}\, h_{x_2}\, h_{x_1 x_2}\,\bigr\}\ \geq\ 0\,.
\end{equation*}
If we restrict our attention to the one-dimensional bathymetries (\ie $h_{x_2}\ \to\ 0$), then we obtain an even simpler condition:
\begin{equation*}
  h_{xx}\ \geq\ -\frac{2}{h}\cdot\frac{1\ +\ h_x^{\,2}}{1\ -\ h_x^{\,2}}\,,
\end{equation*}
where by $x$ we denote $x_1$ for simplicity. The last condition can be easily checked at the problem outset. A further simplification is possible if we additionally assume that 
\begin{equation}\label{eq:restr}
  \abs{\grad h}\ \equiv\ \abs{h_x}\ <\ 1\,,
\end{equation}
then we have the following elegant condition:
\begin{equation}\label{eq:simple}
  h_{xx}\ >\ -\frac{2}{h}\,.
\end{equation}

\subsection{Conservative form of equations}

Equations \eqref{eq:eq1}, \eqref{eq:eq2} admit an elegant conservative form, which is suitable for numerical simulations:
\begin{align}\label{eq:cons1}
  \H_t\ +\ \div\bigl[\,\H\u\,\bigr]\ &=\ 0\,, \\
  (\H\u)_t\ +\ \div\Fl\ &=\ g\, \H\, \grad h\ +\ \grad\Pnh\ -\ \pb\grad h\,, \label{eq:cons2}
\end{align}
where the flux matrix $\Fl(\H,\,\u)$ is the same as in NSWE (or \textsc{Saint}-\textsc{Venant} equations):
\begin{equation*}
  \Fl(H,\,\u)\,\ \eqdef\,\ \begin{bmatrix}
    \H\, u_1^2\ +\ \dfrac{g\, \H^{\,2}}{2} & \H\, u_1\cdot u_2 \\
    \H\, u_1\cdot u_2 & \H\, u_2^2\ +\ \dfrac{g\, \H^{\,2}}{2}
  \end{bmatrix}\,.
\end{equation*}
Notice that it is slightly different from the (fully-)conservative form given in Part~\textsc{I} \cite{Khakimzyanov2016c}. Conservative equations\footnote{It is not difficult to see that the mass conservation equation \eqref{eq:eq1} is already in a conservative form in the SGN model. Thus, equations \eqref{eq:eq1} and \eqref{eq:cons1} are obviously identical.} \eqref{eq:cons1}, \eqref{eq:cons2} can be supplemented by the energy conservation equation which can be used to check the accuracy of simulation (in conservative case, \ie $h_t\ \equiv\ 0$) and/or to estimate the energy of generated waves \cite{Dutykh2009b}:
\begin{equation}\label{eq:energyEq}
  (\H\,\E)_t\ +\ \div\Bigl[\,\H\,\u\,\Bigl(\E\ +\ \frac{\Pp}{\H}\Bigr)\,\Bigr]\ =\ -\pc\, h_t\,,
\end{equation}
where the total energy $\E$ is defined as
\begin{equation*}
  \E\ \eqdef\ \half\,\abs{\u}^{\,2}\ +\ \sixth\, \H^{\,2}\,(\div\u)^2\ +\ \half\,\H\,(\D h)\,(\div\u)\ +\ \half\,(\D h)^2\ +\ \frac{g}{2}\;(\H\ -\ 2h)\,.
\end{equation*}
Notice that equation \eqref{eq:energyEq} is not independent. It is a differential consequence of the mass and momentum conservations \eqref{eq:cons1}, \eqref{eq:cons2} (as it is the case for incompressible flows in general).

\subsubsection{Intermediate conclusions}

As a result, the system of nonlinear dispersive equations \eqref{eq:eq1}, \eqref{eq:eq2} was split in two main parts:
\begin{enumerate}
  \item Governing equations \eqref{eq:cons1}, \eqref{eq:cons2} in the form of (hyperbolic) balance laws with source terms
  \item A scalar nonlinear elliptic equation to determine the non-hydrostatic part of the pressure $\Pnh(\x,\,t)$ (and consequently $\pb(\x,\,t)$ as well)
\end{enumerate}
This splitting idea will be exploited below in the numerical algorithm in order to apply the most suitable and robust algorithm for each part of the solution process \cite{Barakhnin1999}.

\subsection{One-dimensional case}

In this study for the sake of simplicity we focus on the two-dimensional physical problem, \ie one horizontal and one vertical dimensions. The vertical flow structure being resolved using the asymptotic expansion (see Part~\textsc{I} \cite{Khakimzyanov2016c} of this series of papers), thus we deal with PDEs involving one spatial (horizontal) dimension ($x\ \eqdef\ x_1$) and one temporal variable $t\ \in\ \R^+$. The horizontal velocity variable $u(x,\,t)$ becomes a scalar function in this case. Below we provide the full set of governing equations (which follow directly from \eqref{eq:cons1}, \eqref{eq:cons2} and \eqref{eq:ell}):
\begin{align}\label{eq:mass}
  \H_t\ +\ [\,\H\, u\,]_x\ &=\ 0\,, \\
  (\H\, u)_t\ +\ \Bigl[\,\H\, u^2\ +\ \frac{g\, \H^2}{2}\,\Bigr]_x\ &=\ g\,\H\, h_x\ +\ \Pnh_x\ -\ \pb\, h_x\,,\label{eq:moment} \\
  4\biggl[\,\frac{\Pnh_x}{\H\,\r}\,\biggr]_x\ -\ 6\biggl[\,\frac{2}{\H^{\,3}}\cdot\frac{\r\ -\ 3}{\r}\ +\ \Bigl[\frac{h_x}{\H^{\,2}\,\r}\Bigr]_x\,\biggr]\;\Pnh\ &=\ \F\,, \label{eq:ell1}
\end{align}
where $\r\ \eqdef\ 4\ +\ h_x^2$ and
\begin{align*}
  \F\ &\eqdef\ \biggl[\,g\eta_x\ +\ \frac{\Rr\,h_x}{\r}\,\biggr]_x\ -\ \frac{6\,\Rr}{\H\,\r}\ +\ 2\,u_x^2\,, \\
  \Rr\ &\eqdef\ -g\,\eta_x h_x\ +\ u^2 h_{xx}\ +\ h_{tt}\ +\ 2\, u\, h_{xt}\,.
\end{align*}
The last equations can be trivially obtained from corresponding two-dimensional versions given in \eqref{eq:F}, \eqref{eq:R}. This set of equations will be solved numerically below (see Section~\ref{sec:num}).


\subsubsection{Boundary conditions on the elliptic part}
\label{sec:bcond}

First, we rewrite elliptic equation \eqref{eq:ell1} in the following equivalent form:
\begin{equation}\label{eq:ell2}
  \bigl[\,\K\,\Pnh_x\,\bigr]_x\ -\ \K_0\,\Pnh\ =\ \F\,,
\end{equation}
where
\begin{equation*}
  \K\ \eqdef\ \frac{4}{\H\,\r}\,,\, \qquad
  \K_0\ \eqdef\ 6\;\Bigl[\,\frac{2}{\H^{\,3}}\;\frac{\r\ -\ 3}{\r}\ +\ \Bigl(\frac{h_x}{\H^{\,2}\r}\Bigr)_x\,\Bigr]\,.
\end{equation*}
We assume that we have to solve an initial-boundary value problem for the system \eqref{eq:mass}--\eqref{eq:ell1}. If we have a closed numerical wave tank\footnote{Other possibilities have to be discussed separately.} (as it is always the case in laboratory experiments), then on vertical walls the horizontal velocity satisfies:
\begin{equation*}
  \left.u(x,\,t)\,\right|_{x\,=\,0}\ =\ \left.u(x,\,t)\,\right|_{x\,=\,\ell}\ \equiv\ 0, \qquad \forall t\ \in\ \R^+\,.
\end{equation*}
For the situation where the same boundary condition holds on both boundaries, we introduce a short-hand notation:
\begin{equation*}
  \left.u(x,\,t)\,\right|_{x\,=\,0}^{x\,=\,\ell}\ \equiv\ 0\,, \qquad \forall t\ \in\ \R^+\,.
\end{equation*}
Assuming that equation \eqref{eq:nswe} is valid up to the boundaries, we obtain the following boundary conditions for the elliptic equation \eqref{eq:ell1}:
\begin{equation*}
  \left.\Bigl(\,\frac{\Pnh_x\ -\ \pb\,h_x}{\H}\ -\ g\,\eta_x\,\Bigr)\,\right|_{x\,=\,0}^{x\,=\,\ell}\ =\ 0\,, \qquad \forall t\ \in\ \R^+\,.
\end{equation*}
Or in terms of equation \eqref{eq:ell2} we equivalently have:
\begin{equation}\label{eq:bvp}
  \left.\Bigl(\,\K\Pnh_x\ -\ \frac{6\,h_x}{\H^{\,2}\r}\;\Pnh\,\Bigr)\,\right|_{x\,=\,0}^{x\,=\,\ell}\ =\ \left.\Bigl(\,g\,\eta_x\ +\ \frac{\Rr\,h_x}{\r}\,\Bigr)\,\right|_{x\,=\,0}^{x\,=\,\ell}\,, \qquad \forall t\ \in\ \R^+\,.
\end{equation}
The boundary conditions for the non-hydrostatic pressure component $\Pnh$ are of the 3\up{rd} kind (sometimes they are referred to as of \textsc{Robin}-type). For the case where locally at the boundaries the bottom is flat (to the first order), \ie $\left.h_x\right|_{x\,=\,0}^{x\,=\,\ell}\ \equiv\ 0$, then we have the (non-homogeneous) Neumann boundary condition of the 2\up{nd} kind:
\begin{equation*}
  \left.\K\Pnh_x\,\right|_{x\,=\,0}^{x\,=\,\ell}\ =\ \left.g\,\eta_x\,\right|_{x\,=\,0}^{x\,=\,\ell}\,, \qquad \forall t\ \in\ \R^+\,.
\end{equation*}
For a classical \textsc{Poisson}-type equation this condition would not be enough to have a well-posed problem. However, we deal rather with a \textsc{Helmholtz}-type equation (if $\K_0\ >\ 0$). So, the flat bottom does not represent any additional difficulty for us and the unicity of the solution can be shown in this case as well.

\subsubsection{Unicity of the elliptic equation solution}

The mathematical structure of equation \eqref{eq:ell2} is very advantageous since it allows to show the following
\begin{theorem}\label{thm:unique}
Suppose that the Boundary Value Problem (BVP) \eqref{eq:bvp} for equation \eqref{eq:ell2} admits a solution and the following conditions are satisfied:
\begin{equation}\label{eq:conds}
  \K_0\ >\ 0,\, \quad \left.h_x\,\right|_{x\,=\,0}\ \geq\ 0,\, \quad \left.h_x\,\right|_{x\,=\,\ell}\ \leq\ 0\,,
\end{equation}
then this solution is unique.
\end{theorem}

\begin{proof}
Assume that there are two such solutions $\Pnh_1$ and $\Pnh_2$. Then, their difference $\Pnh\ \eqdef\ \Pnh_1\ -\ \Pnh_2$ satisfies the following homogeneous BVP:
\begin{align}\label{eq:hm1}
  \bigl[\,\K\,\Pnh_x\,\bigr]_x\ -\ \K_0\,\Pnh\ &=\ 0\,, \\
  \left.\Bigl(\,\K\Pnh_x\ -\ \frac{6\,h_x}{\H^{\,2}\r}\;\Pnh\,\Bigr)\,\right|_{x\,=\,0}^{x\,=\,\ell}\ &=\ 0\,. \label{eq:hm2}
\end{align}
Let us multiply the first equation \eqref{eq:hm1} by $\Pnh$ and integrate over the computational domain:
\begin{equation*}
  \underbrace{\int_{0}^{\ell}\bigl[\,\K\,\Pnh_x\,\bigr]_x\,\Pnh\;\ud x}_{(\blacklozenge)}\ -\ \int_0^{\ell}\K_0\,\Pnh^2\;\ud x\ =\ 0\,.
\end{equation*}
Integration by parts of the first integral ($\blacklozenge$) yields:
\begin{equation*}
  \left.\K\,\Pnh_x\,\Pnh\right|^{x\,=\,\ell}\ -\ \left.\K\,\Pnh_x\,\Pnh\right|_{x\,=\,0}\ -\ \int_0^{\ell}\K\,\Pnh_x^2\;\ud x\ -\ \int_0^{\ell}\K_0\,\Pnh^2\;\ud x\ =\ 0\,.
\end{equation*}
And using boundary conditions \eqref{eq:hm2} we finally obtain:
\begin{equation*}
  \left.\frac{6\,h_x}{\H^{\,2}\r}\;\Pnh^2\right|^{x\,=\,\ell}\ -\ \left.\frac{6\,h_x}{\H^{\,2}\r}\;\Pnh^2\right|_{x\,=\,0}\ -\ \int_0^{\ell}\K\,\Pnh_x^2\;\ud x\ -\ \int_0^{\ell}\K_0\,\Pnh^2\;\ud x\ =\ 0\,.
\end{equation*}
Taking into account this Theorem assumptions \eqref{eq:conds} and the fact that $\K\ >\ 0$, the last identity leads to a contradiction, since the left hand side is strictly negative. Consequently, the solution to equation \eqref{eq:ell2} with boundary condition \eqref{eq:bvp} is unique.
\end{proof}

\begin{remark}
Conditions in Theorem~\ref{thm:unique} are quite natural. The non-negativity of coefficient $\K_0$ has already been discussed in Section~\ref{sec:well}. Two other conditions mean that the water depth is increasing in the offshore direction ($\left.h_x\right|_{x\,=\,0}\ \geq\ 0$) and again it is decreasing ($\left.h_x\right|_{x\,=\,\ell}\ \leq\ 0$) when we approach the opposite shore.
\end{remark}


\subsection{Vector short-hand notation}

For the sake of convenience we shall rewrite governing equations \eqref{eq:mass}, \eqref{eq:moment} in the following vectorial form:
\begin{equation}\label{eq:hyp}
  \v_t\ +\ \bigl[\,\Fl(\v)\,\bigr]_x\ =\ \Gl(\v,\,\Pnh,\,\pb,\,h)\,,
\end{equation}
where we introduced the following vector-valued functions:
\begin{equation*}
  \v\ \eqdef\ \begin{pmatrix}
    \H \\
    \H u
  \end{pmatrix}\,, \qquad
  \Fl(\v)\ \eqdef\ \begin{pmatrix}
    \H u \\
    \H u^2\ +\ \dfrac{g\,\H^{\,2}}{2}
  \end{pmatrix}\,,
\end{equation*}
and the source term is defined as
\begin{equation*}
  \Gl(\v,\,\Pnh,\,\pb,\,h)\ \eqdef\ \begin{pmatrix}
    0 \\
    g\,\H\,h_x\ +\ \Pnh_x\ -\ \pb h_x
  \end{pmatrix}\,.
\end{equation*}
The point of view that we adopt in this study is to view the SGN equations as a system of hyperbolic equations \eqref{eq:hyp} with source terms $\Gl(\v,\,\Pnh_x,\,\pb,\,h)$. Obviously, one has to solve also the elliptic equation \eqref{eq:ell1} in order to compute the source term $\Gl$.

The Jacobian matrix of the advection operator coincides with that of classical NSWE equations:
\begin{equation*}
  \A(\v)\ \eqdef\ \od{\Fl(\v)}{\v}\ =\ \begin{pmatrix}
    0 & 1 \\
    -u^2\ +\ g\,\H & 2 u
  \end{pmatrix}\,.
\end{equation*}
Eigenvalues of the Jacobian matrix $\A(\v)$ can be readily computed:
\begin{equation}\label{eq:eigs}
  \lambda^{-}\ =\ u\ -\ s\,, \qquad
  \lambda^{+}\ =\ u\ +\ s\,, \qquad
  s\ \eqdef\ \sqrt{g\,\H}\,.
\end{equation}
The Jacobian matrix appears naturally in the non-divergent form of equations \eqref{eq:hyp}:
\begin{equation}\label{eq:ncons}
  \v_t\ +\ \A(\v)\cdot\v_x\ =\ \Gl\,,
\end{equation}
By multiplying both sides of the last equation by $\A(\v)$ we obtain the equations for the advection flux function $\Fl(\v)$:
\begin{equation}\label{eq:flux}
  \Fl_t\ +\ \A(\v)\cdot\Fl_x\ =\ \A\cdot\Gl\,.
\end{equation}
In order to study the characteristic form of equations one needs also to know the matrix of left and right eigenvectors correspondingly:
\begin{equation}\label{eq:leftright}
  \Lb\ \eqdef\ \frac{1}{s^2}\;\begin{pmatrix}
    -\lambda^{+} & 1 \\
    -\lambda^{-} & 1
  \end{pmatrix}\,, \qquad
  \Rb\ \eqdef\ \frac{s}{2}\;\begin{pmatrix}
    -1 & 1 \\
    -\lambda^{-} & \lambda^{+}
  \end{pmatrix}\,.
\end{equation}
If we introduce also the diagonal matrix of eigenvalues
\begin{equation*}
  \Lab\ \eqdef\ \begin{pmatrix}
    \lambda^{-} & 0 \\
    0 & \lambda^{+}
  \end{pmatrix}\,,
\end{equation*}
the following relations can be easily checked:
\begin{equation*}
  \Rb\cdot\Lab\cdot\Lb\ \equiv\ \A\,, \qquad
  \Rb\cdot\Lb\ =\ \Lb\cdot\Rb\ \equiv\ \Id\,,
\end{equation*}
where $\Id$ is the identity $2\times 2$ matrix.


\subsubsection{Flat bottom}
\label{sec:flat}

Equations above become particularly simple on the flat bottom. In this case the bathymetry functions is constant, \ie
\begin{equation*}
  h(x,\,t)\ \equiv\ d\ =\ \const\ >\ 0\,.
\end{equation*}
Substituting it into governing equations above, we straightforwardly obtain:
\begin{align*}
  \H_t\ +\ [\,\H\, u\,]_x\ &=\ 0\,, \\
  (\H\, u)_t\ +\ \Bigl[\,\H\, u^2\ +\ \frac{g\, \H^2}{2}\,\Bigr]_x\ &=\ \Pnh_x\,, \\
  \biggl[\,\frac{\Pnh_x}{\H}\,\biggr]_x\ -\ \frac{3}{\H^3}\;\Pnh\ &=\ g\,\eta_{xx}\ +\ 2\,u_x^2\,.
\end{align*}

\bigskip
\paragraph*{Solitary wave solution.}

Equations above admit an elegant analytical solution known as the \emph{solitary wave}\footnote{Solitary waves are to be distinguished from the so-called \emph{solitons} which interact elastically \cite{John}. Since the SGN equations are not integrable (for the notion of integrability we refer to \eg \cite{Zakharov1991}), the interaction of solitary waves is inelastic \cite{Mitsotakis2014}.}. It is given by the following expressions:
\begin{equation}\label{eq:sol}
  \eta(x,\,t)\ =\ \alpha\cdot\sech^2\biggl[\,\frac{\sqrt{3\,\alpha\, g}}{2 d \upsilon}\;\bigl(x\ -\ x_0\ -\ \upsilon\,t\bigr)\,\biggr]\,, \qquad
  u(x,\,t)\ =\ \frac{\upsilon\cdot\eta(x,\,t)}{d\ +\ \eta(x,\,t)}\,,
\end{equation}
where $\alpha$ is the wave amplitude, $x_0\ \in\ \R$ is the initial wave position and $\upsilon$ is the velocity defined as
\begin{equation*}
  \upsilon\ \eqdef\ \sqrt{g\,(d\ +\ \alpha)}\,.
\end{equation*}
The non-hydrostatic pressure under the solitary wave can be readily computed as well:
\begin{equation*}
  \Pnh(x,\,t)\ =\ \frac{g}{2}\;\bigl[\,\H^{\,2}(x,\,t)\ -\ d^{\,2}\,\bigr]\ -\ d\,\upsilon\, u(x,\,t)\,.
\end{equation*}

One can derive also periodic travelling waves known as \emph{cnoidal waves}. For their expressions we refer to \eg \cite{Dutykh2011a, Dutykh2016}.


\subsection{Linear dispersion relation}
\label{sec:disp}

The governing equations \eqref{eq:mass}--\eqref{eq:ell1} after linearizations take the following form:
\begin{align*}
  \eta_t\ +\ d\,u_x\ &=\ 0\,, \\
  u_t\ +\ g\,\eta_x\ &=\ \frac{\Pnh_x}{d}\,, \\
  \Pnh_{xx}\ -\ \frac{3}{d^{\,2}}\;\Pnh\ &=\ c^2\,\eta_{xx}\,,
\end{align*}
where $c\ \eqdef\ \sqrt{g\,d}$ is the linear gravity wave speed. By looking for plane wave solutions of form
\begin{equation*}
  \eta(x,\,t)\ =\ \alpha\,\ue^{\ui\, (k x\ -\ \upomega\, t)}\,, \quad
  u(x,\,t)\ =\ \upupsilon\,\ue^{\ui\, (k x\ -\ \upomega\, t)}\,, \quad
  \Pnh(x,\,t)\ =\ \uprho\,\ue^{\ui\, (k x\ -\ \upomega\, t)}\,,
\end{equation*}
where $k$ is the wave number, $\upomega(k)$ is the wave frequency and $\bigl\{\alpha,\, \upupsilon,\,\uprho\bigr\}\ \in\ \R$ are some (constant) real amplitudes. The necessary condition for the existence of plane wave solutions reads
\begin{equation}\label{eq:omega}
  \upomega(k)\ =\ \pm\;\frac{c k}{\sqrt{1\ +\ \dfrac{(kd)^2}{3}}}\,.
\end{equation}
By substituting the definition of $k\ =\ \dfrac{2\pi}{\lambda}$ into the last formula and dividing both sides by $k$ we obtain the relation between the phase speed $c_p$ and the wavelength $\lambda$:
\begin{equation*}
  c_p(\lambda)\ \eqdef\ \frac{\upomega\bigl(k(\lambda)\bigr)}{k\bigl(\lambda\bigr)}\ =\ \frac{c}{\sqrt{1\ +\ \dfrac{4\pi^2 d^{\,2}}{3\lambda^2}}}\,.
\end{equation*}
This dispersion relation is accurate to 2\up{nd} order at the limit of long waves $k d\ \to\ 0$. There are many other nonlinear dispersive wave models which share the same linear dispersion relation, see \eg \cite{Ertekin1986, Peregrine1967, Zheleznyak1985, Fedotova2009}. However, their nonlinear properties might be very different.


\section{Numerical method}
\label{sec:num}

The construction of numerical schemes for hyperbolic conservation laws on moving grids was described in our previous work \cite{Khakimzyanov2015a}. In the present manuscript we make an extension of this technology to dispersive PDEs illustrated on the example of the SGN equations \eqref{eq:mass}--\eqref{eq:ell1}. The main difficulty which arises in the dispersive case is handling of high order (possibly mixed) derivatives. The SGN system is an archetype of such systems with sufficient degree of nonlinearity and practically important applications in Coastal Engineering \cite{Kim2000}.


\subsection{Adaptive mesh construction}

In the present work we employ the method of moving grids initially proposed in early 60's by \textsc{Tikhonov} \& \textsc{Samarskii} \cite{Tikhonov1961, Tikhonov1962} and developed later by \textsc{Bakhvalov} (1969) \cite{Bakhvalov1969} and \textsc{Il'in} (1969) \cite{Ilin1969}. This technology was recently described by the authors to steady \cite{Khakimzyanov2015b} and unsteady \cite{Khakimzyanov2015a} problems. For more details we refer to our recent publications \cite{Khakimzyanov2015a, Khakimzyanov2015b}. An alternative recent approach can be found in \eg \cite{Arvanitis2006, Arvanitis2010}. In the present Section we just recall briefly the main steps of the method.

The main idea consists in assuming that there exists a (time-dependent) diffeomorphism from the reference domain $\Q\ \eqdef\ [0,\,1]$ to the computational domain $\I\ =\ [0,\,\ell]$:
\begin{equation*}
  x(q,\,t):\ \Q\ \mapsto\ \I\,.
\end{equation*}
It is natural to assume that boundaries of the domains correspond to each other, \ie
\begin{equation*}
  x(0,\,t)\ =\ 0\,, \qquad x(1,\,t)\ =\ \ell\,, \qquad \forall t\ \geq\ 0\,.
\end{equation*}
We shall additionally assume that the Jacobian of this map is bounded from below and above
\begin{equation}\label{eq:condJ}
  0\ <\ m\ \leq\ \Jj(q,\,t)\ \eqdef\ \pd{x}{q}\ \leq\ M\ < +\infty
\end{equation}
by some real constants $m$ and $M$.

The construction of this diffeomorphism $x(q,\,t)$ is the heart of the matter in the moving grid method. We employ the so-called equidistribution method. The required non-uniform grid $\I_h$ of the computational domain $\I$ is then obtained as the image of the uniformly distributed nodes $\Q_h$ under the mapping $x(q,\,t)$:
\begin{equation*}
  x_j\ =\ x(q_j,\,t)\,, \qquad q_j\ =\ j\,\h\,, \qquad \h\ =\ \frac{1}{N}\,,
\end{equation*}
where $N$ is the total number of grid points. Notice, that strictly speaking, we do not even need to know the mapping $x(q,\,t)$ in other points besides $\{q_j\}_{j=0}^{N}$. Under condition \eqref{eq:condJ} it easily follows that the maximal discretization step in the physical space vanishes when we refine the mesh in the reference domain $\Q_h$:
\begin{equation*}
  \max_{j=0\,\ldots\,,N-1}\abs{x_{j+1}\ -\ x_j}\ \leq\ M\,\h\ \to\ 0\,, \quad \mbox{ as }\quad \h\ \to\ 0\,.
\end{equation*}

\subsubsection{Initial grid generation}

Initially, the desired mapping $x(q,\,0)$ is obtained as a solution to the following nonlinear elliptic problem
\begin{equation}\label{eq:initgrid}
  \od{}{q}\;\biggl[\om(x)\,\od{x}{q}\biggr]\ =\ 0, \qquad x(0)\ =\ 0, \quad x(1)\ =\ \ell\,,
\end{equation}
where we drop in this Section the 2\up{nd} constant argument $0$. The function $\om(x)$ is the so-called \emph{monitor function}. Its choice will be specified below, but we can say that this functions has to be positive defined and bounded from below, \ie
\begin{equation*}
  \om(x)\ \geq\ C\ >\ 0\,, \qquad \forall x\ \in\ \R\,.
\end{equation*}
In practice the lower bound $C$ is taken for simplicity to be equal to $1$. A popular choice of the monitor function is, for example,
\begin{equation*}
  \om[\eta](x)\ =\ 1\ +\ \coef_0\,\abs{\eta}\,, \qquad \coef_0\ \in\ \R^+\,,
\end{equation*}
where $\eta$ is the free surface elevation. Another possibility consists in taking into account the free surface gradient:
\begin{equation*}
  \om[\eta](x)\ =\ 1\ +\ \coef_1\,\abs{\eta_x}\,, \qquad \coef_1\ \in\ \R^+\,,
\end{equation*}
or even both effects:
\begin{equation*}
  \om[\eta](x)\ =\ 1\ +\ \coef_0\,\abs{\eta}\ +\ \coef_1\,\abs{\eta_x}\,, \qquad \coef_{0,1}\ \in\ \R^+\,.
\end{equation*}

In some simple cases equation \eqref{eq:initgrid} can be solved analytically (see \eg \cite{Khakimzyanov2015b}). However, in most cases we have to solve the nonlinear elliptic problem \eqref{eq:initgrid} numerically. For this purpose we use an iterative scheme, where at every stage we have a linear three-diagonal problem to solve:
\begin{equation}\label{eq:discr}
  \frac{1}{\h}\;\biggl[\,\om(x_{j+1/2}^{(n)})\;\frac{x_{j+1}^{(n+1)}\ -\ x_{j}^{(n+1)}}{\h}\ -\ \om(x_{j-1/2}^{(n)})\;\frac{x_{j}^{(n+1)}\ -\ x_{j-1}^{(n+1)}}{\h}\,\biggr]\ =\ 0, \qquad n\ \in\ \N_0\,.
\end{equation}
The iterations are continued until the convergence is achieved to the prescribed tolerance parameter (typically $\propto\ 10^{-10}$).


\subsubsection{Grid motion}

In unsteady computations the grid motion is given by the following nonlinear parabolic equation:
\begin{equation}\label{eq:npar}
  \pd{}{q}\;\biggl[\om(x,\,t)\,\pd{x}{q}\biggr]\ =\ \beta\,\pd{x}{t}\,, \qquad \beta\ \in\ \R^+\,.
\end{equation}
The parameter $\beta$ plays the r\^ole of the diffusion coefficient here. It is used to control the smoothness of nodes trajectories. Equation \eqref{eq:npar} is discretized using an implicit scheme:
\begin{equation}\label{eq:discr2}
  \frac{1}{\h}\;\biggl\{\om^n_{j+1/2}\;\frac{x^{n+1}_{j+1}\ -\ x^{n+1}_j}{\h}\ -\ \om^n_{j-1/2}\;\frac{x^{n+1}_j\ -\ x^{n+1}_{j-1}}{\h}\biggr\}\ =\ \beta\;\frac{x^{n+1}_j - x^{n}_j}{\tau}\,,
\end{equation}
with boundary conditions $x_0^{n+1}\ =\ 0$, $x_N^{n+1}\ =\ \ell$ as above. We would like to reiterate that at every time step we solve only one additional (tridiagonal) linear system. Nonlinear iterative computations are performed only once when we project the initial condition on the ad-hoc non-uniform grid. So, the additional overhead due to the mesh motion is linear in complexity, \ie $\O(N)$.

Similarly to the elliptic case \eqref{eq:initgrid}, equation \eqref{eq:npar} admits smooth solutions provided that the monitor function $\om(x,\,t)$ is bounded from below by a positive constant. In numerical examples shown below we always take monitor functions which satisfy the condition $\om(x,\,t)\ \geq\ 1$, $\forall x\ \in\ \I\,$, $\forall t\ \geq\ 0\,$. Thus, for any $t\ >\ 0$ equation \eqref{eq:npar} provides us the required diffeomorphism between the reference domain $\Q$ and the computational domain $\I$.


\subsection{The SGN equations on a moving grid}

Before discretizing the SGN equations \eqref{eq:mass}--\eqref{eq:ell1}, we have to pose them on the reference domain $\Q$. The composed functions will be denoted as:
\begin{equation*}
  \uc(q,\,t)\ \eqdef\ (u\circ x)\,(q,\,t)\ \equiv\ u\bigl(x(q,\,t),\,t\bigr)\,.
\end{equation*}
And we introduce similar notations for all other variables, \eg $\Hc(q,\,t) \ \eqdef\ \H\bigl(x(q,\,t),\,t\bigr)\,$. The conservative \eqref{eq:hyp} and non-conservative \eqref{eq:ncons}, \eqref{eq:flux} forms of hyperbolic equations read:
\begin{align}\label{eq:cons}
  (\Jj\,\vc)_t\ +\ \bigl[\,\Flc\ -\ x_t\,\vc\,\bigr]_q\ &=\ \Glc\,, \\
  \vc_t\ +\ \frac{1}{\Jj}\;\bigl[\,\Flc_q\ -\ x_t\,\vc_q\,\bigr]\ &=\ \frac{1}{\Jj}\;\Glc\,, \label{eq:nconsQ} \\
  \Flc_t\ +\ \frac{1}{\Jj}\;\Ac\cdot\bigl[\,\Flc_q\ -\ x_t\,\vc_q\,\bigr]\ &=\ \frac{1}{\Jj}\;\Ac\cdot\Glc\,, \label{eq:fluxQ}
\end{align}
where the terms on the right-hand sides are defined similarly as above:
\begin{equation*}
  \Glc\ \eqdef\ \begin{pmatrix}
    0 \\
    g\,\Hc\,\hc_q\ +\ \Pnhc_q\ -\ \pbc\,\hc_q
  \end{pmatrix}\,.
\end{equation*}
The non-hydrostatic pressure on the bottom is computed in $\Q$ space as:
\begin{equation}\label{eq:pbc}
  \pbc\ \eqdef\ \frac{1}{\rc}\;\biggl[\,\frac{6\,\Pnhc}{\Hc}\ +\ \Hc\,\Rrc\ +\ \frac{\Pnhc_q\,\hc_q}{\Jj^2}\,\biggr]\,, \qquad
  \rc\ \eqdef\ 4\ +\ \dfrac{\hc_q^{\,2}}{\Jj^2}
\end{equation}
Finally, we just have to specify the expression for $\Rrc$:
\begin{equation*}
  \Rrc\ \eqdef\ -g\,\frac{\etac_q\,\hc_q}{\Jj^2}\ +\ \frac{\uc^{\,2}}{\Jj}\;\Bigl[\,\frac{\hc_q}{\Jj}\,\Bigr]_q\ +\ \Bigl(\hc_t\ -\ \frac{x_t}{\Jj}\;\hc_q\Bigr)_t\ +\ \frac{2\,\uc\ -\ x_t}{\Jj}\cdot\Bigl[\,\hc_t\ -\ \frac{x_t}{\Jj}\;\hc_q\,\Bigr]_q\,.
\end{equation*}
We have to specify also the equations which allow us to find the non-hydrostatic part of the pressure field $\Pnhc$. Equation \eqref{eq:ell2} posed on the reference domain $\Q$ reads:
\begin{equation}\label{eq:pnhcq}
  \bigl[\,\Kc\,\Pnhc_q\,\bigr]_q\ -\ \Kc_0\,\Pnhc\ =\ \Fc\,,
\end{equation}
where the coefficients and the right-hand side are defined as
\begin{equation*}
  \Kc\ \eqdef\ \frac{4}{\Jj\,\Hc\,\rc}\,, \qquad
  \Kc_0\ \eqdef\ 6\;\biggl[\,\frac{2\,\Jj}{\Hc^{\,3}}\cdot\frac{\rc\ -\ 3}{\rc}\ +\ \Bigl(\frac{\hc_q}{\Jj\,\Hc^{\,2}\,\rc}\Bigr)_q\,\biggr]\,,
\end{equation*}
\begin{equation*}
  \Fc\ \eqdef\ \Bigl[\,g\,\frac{\etac_q}{\Jj}\ +\ \frac{\Rrc\,\hc_q}{\Jj\,\rc}\,\Bigr]_q\ -\ \frac{6\,\Rrc\,\Jj}{\Hc\,\rc}\ +\ 2\,\frac{\uc_q^{\,2}}{\Jj}\,.
\end{equation*}
Finally, the boundary conditions are specified now at $q\ =\ 0$ and $q\ =\ 1$. For the hyperbolic part of the equations they are
\begin{equation*}
  \uc(0,\,t)\ =\ 0\, \qquad
  \uc(1,\,t)\ =\ 0\, \qquad
  \forall t\ \geq\ 0\,.
\end{equation*}
For the elliptic part we have the following mixed-type boundary conditions:
\begin{equation}\label{eq:bcc}
  \left.\biggl[\,\frac{4}{\Jj\,\Hc\,\rc}\;\Pnhc_q\ -\ \frac{6\,\hc_q}{\Jj\,\Hc^{\,2}\,\rc}\;\Pnhc\,\biggr]\right|_{q\, =\, 0}^{q\, =\, 1}\ =\ \left.\frac{1}{\Jj}\;\biggl[\,g\etac_q\ +\ \frac{\Rrc}{\rc}\;\hc_q\,\biggr]\right|_{q\, =\, 0}^{q\, =\, 1}\,.
\end{equation}


\subsection{Predictor--corrector scheme on moving grids}
\label{sec:prcor}

In this Section we describe the numerical finite volume discretization of the SGN equations on a moving grid. We assume that the reference domain $\Q$ is discretized with a uniform grid $\Q_h\ \eqdef\ \bigl\{q_j\ =\ j\,\h\bigr\}_{j\,=\,0}^{N}\,$, with the uniform spacing $\h\ =\ \frac{1}{N}\,$. Then, the grid $\I_h^n$ in the physical domain $\I$ at every time instance $t\ =\ t^{\,n}\ \geq\ 0$ is given by the image of the uniform grid $\Q_h$ under the mapping $x(q,\,t)\,$, \ie $x_j^n\ =\ x(q_j,\,t^{\,n})\,$, $j\ =\ 0,\,1,\,\ldots,\,N\,$ or simply $\I_h^n\ =\ x(\Q_h,\,t^{\,n})$. We assume that we know the discrete solution\footnote{With symbol $\sharp$ we denote the set of solution values at discrete spatial grid nodes.} $\vc_\sharp^{\,n}\ \eqdef\ \bigl\{\vc_j^{\,n}\bigr\}_{j\,=\,0}^{N}\,$, $\Pnhc_\sharp^n\ \eqdef\ \bigl\{\Pnhc_j^n\bigr\}_{j\,=\,0}^{N}$ at the current time $t\ =\ t^n$ and we already constructed the non-uniform grid $x_\sharp^{n+1}\ \eqdef\ \bigl\{x_j^{n+1}\bigr\}_{j\,=\,0}^{N}$ at the following time layer $t^{\,n+1}$ using the equidistribution method described above. We remind that the non-uniform grid at the following layer is constructed based only on the knowledge of $\vc_\sharp^{\,n}$.


\subsubsection{Predictor step}

In the nonlinear case, during the predictor step the hyperbolic part of equations is solved two times:
\begin{itemize}
  \item First, using equation \eqref{eq:nconsQ} we compute the discrete solution values $\vc^{\ast}_{\sharp,\,c}\ \eqdef\ \bigl\{\vc^\ast_{j+1/2}\bigr\}_{j\,=\,0}^{N-1}$ in the cell centers $\Q_{h,\,c}\ \eqdef\ \bigl\{q_{j+1/2}\ =\ q_j\ +\ \frac{\h}{2}\bigr\}_{j\,=\,0}^{N-1}\,$.
  \item Then, using equation \eqref{eq:fluxQ} we compute the values of the flux vector equally in the cell centers $\Flc^{\ast}_{\sharp,\,c}\ \eqdef\ \bigl\{\Flc^\ast_{j+1/2}\bigr\}_{j\,=\,0}^{N-1}\,$.
\end{itemize}
We rewrite equations \eqref{eq:nconsQ}, \eqref{eq:fluxQ} in the characteristic form by multiplying them on the left by the matrix $\Lbc$ (of left eigenvectors of the Jacobian $\Ac$):
\begin{align*}
  \Lbc\cdot\vc_t\ +\ \frac{1}{\Jj}\;\Lbc\cdot\bigl[\,\Flc_q\ -\ x_t\,\vc_q\,\bigr]\ &=\ \frac{1}{\Jj}\;\Lbc\cdot\Glc\,, \\
  \Lbc\cdot\Flc_t\ +\ \frac{1}{\Jj}\;\Labc\cdot\Lbc\cdot\bigl[\,\Flc_q\ -\ x_t\,\vc_q\,\bigr]\ &=\ \frac{1}{\Jj}\;\Labc\cdot\Lbc\cdot\Glc\,,
\end{align*}
The discretization of last equations reads:
\begin{align}\label{eq:pred1}
  \bigl(\Di^{-1}\cdot\Lbc\bigr)_{j+1/2}^n\cdot\frac{\vc_{j+1/2}^\ast\ -\ \vc_{j+1/2}^n}{\tau/2}\ +\ \Bigl(\frac{1}{\Jj}\;\Lbc\cdot\bigl[\,\Flc_q\ -\ x_t\,\vc_q\,\bigr]\Bigr)_{j+1/2}^n\ &=\ \Bigl(\frac{1}{\Jj}\;\Lbc\cdot\Glc\Bigr)_{j+1/2}^n\,, \\
  \bigl(\Di^{-1}\cdot\Lbc\bigr)_{j+1/2}^n\cdot\frac{\Flc_{j+1/2}^\ast\ -\ \Flc_{j+1/2}^n}{\tau/2}\ +\ \Bigl(\frac{1}{\Jj}\;\Labc\cdot\Lbc\cdot\bigl[\,\Flc_q\ -\ x_t\,\vc_q\,\bigr]\Bigr)_{j+1/2}^n\ &=\ \Bigl(\frac{1}{\Jj}\;\Labc\cdot\Lbc\cdot\Glc\Bigr)_{j+1/2}^n\,,\label{eq:pred2}
\end{align}
where $\tau$ is the time step, $\Lbc_{j+1/2}^n$ is an approximation of matrix $\Lbc$ in the cell centers $\Q_{h,\,c}$ (it will be specified below). The matrix $\Di$ is composed of cell parameters for each equation:
\begin{equation*}
  \Di_{j+1/2}^{\,n}\ \eqdef\ \begin{pmatrix}
    1\ +\ \theta_{j+1/2}^{1,\,n} & 0 \\
    0 & 1\ +\ \theta_{j+1/2}^{2,\,n}
  \end{pmatrix}\,, 
  \qquad
  \Labc_{j+1/2}^{\,n}\ \eqdef\ \begin{pmatrix}
    1\ +\ \lambda_{j+1/2}^{-,\,n} & 0 \\
    0 & 1\ +\ \lambda_{j+1/2}^{+,\,n}
  \end{pmatrix}\,,
\end{equation*}
with $\lambda_{j+1/2}^{\pm,\,n}$ being the approximations of eigenvalues \eqref{eq:eigs} in the cell centers $\Q_{h,\,c}$ (it will be specified below). On the right-hand side the source term is
\begin{equation*}
  \Glc_{j+1/2}^n\ \eqdef\ \begin{pmatrix}
    0 \\
    \bigl(g\,\Hc\,\hc_q\ +\ \Pnhc_q\ -\ \pbc\,\hc_q\bigr)_{j+1/2}^n
  \end{pmatrix}\,,
\end{equation*}
where derivatives with respect to $q$ are computed using central differences:
\begin{equation*}
  \Pnhc_{q,\,j+1/2}^n\ \eqdef\ \frac{\Pnhc_{j+1}^n\ -\ \Pnhc_{j}^n}{\h}\,, \qquad
  \hc_{q,\,j+1/2}^n\ \eqdef\ \frac{\hc_{j+1}^n\ -\ \hc_{j}^n}{\h}\,.
\end{equation*}
The value of the non-hydrostatic pressure trace at the bottom $\pbc_{j+1/2}^{\,n}$ is computed according to formula \eqref{eq:pbc}. Solution vector $\vc^n_{\sharp,\,c}$ and the fluxes $\Flc^n_{\sharp,\,c}$ in cell centers are computed as:
\begin{equation*}
  \vc_{j+1/2}^n\ \eqdef\ \frac{\vc_{j+1}^n\ +\ \vc_{j}^n}{2}\,, \qquad
  \Flc_{j+1/2}^n\ \eqdef\ \frac{\Flc_{j+1}^n\ +\ \Flc_{j}^n}{2}\,.
\end{equation*}
The derivatives of these quantities are estimated using simple finite differences:
\begin{equation*}
  \vc_{q,\,j+1/2}^n\ \eqdef\ \frac{\vc_{j+1}^n\ -\ \vc_{j}^n}{\h}\,, \qquad
  \Flc_{q,\,j+1/2}^n\ \eqdef\ \frac{\Flc_{j+1}^n\ -\ \Flc_{j}^n}{\h}\,.
\end{equation*}
Finally, we have to specify the computation of some mesh-related quantities:
\begin{equation*}
  x_{t,\,j}^{n}\ \eqdef\ \frac{x_{j}^{n+1}\ -\ x_{j}^{n}}{\tau}\,, \quad
  x_{t,\,j+1/2}^{n}\ \eqdef\ \frac{x_{t,\,j}^{n}\ +\ x_{t,\,j}^{n}}{2}\,, \quad
  \Jj_{j+1/2}^n\ \equiv\ x_{q,\,j+1/2}^n\ \eqdef\ \frac{x_{j+1}^n\ -\ x_{j}^n}{\h}\,.
\end{equation*}
The approximation of the matrix of left eigenvectors $\Lbc_{j+1/2}^n$ and eigenvalues $\lambda_{j+1/2}^{\pm,\,n}$ depends on the specification of the Jacobian matrix $\Ac_{j+1/2}^n$. Our approach consists in choosing the discrete approximation in order to have at discrete level
\begin{equation}\label{eq:jacond}
  \Flc_{q,\,j+1/2}^n\ \equiv\ \bigl(\Ac\cdot\vc_q\bigr)_{j+1/2}^n\,,
\end{equation}
which is the discrete analogue of the continuous identity $\Fc_q\ \equiv\ \Ac\cdot\vc_q$. Basically, our philosophy consists in preserving as many as possible continuous properties at the discrete level. For example, the following matrix satisfies the condition \eqref{eq:jacond}:
\begin{equation*}
  \Ac_{j+1/2}^n\ =\ \begin{pmatrix}
    0 & 1 \\
    -u_{j}^{n}\,u_{j+1}^{n}\ +\ g\,\H_{j+1/2}^{\,n} & 2\,u_{j+1/2}^{n}
  \end{pmatrix}\ =\ \bigl(\Rbc\cdot\Labc\cdot\Lbc\bigr)_{j+1/2}^n
\end{equation*}
The matrices $\Lb_{j+1/2}^n$ and $\Rb_{j+1/2}^n\ =\ (\Lb_{j+1/2}^n)^{-1}$ are computed by formulas \eqref{eq:leftright}. The Jacobian matrix $\Ac_{j+1/2}^n$ eigenvalues can be readily computed:
\begin{equation*}
  \lambda_{j+1/2}^{\pm,\,n}\ \eqdef\ (u\ \pm\ s)_{j+1/2}^n\,, \quad
  s_{j+1/2}^n\ \eqdef\ \sqrt{(u_{j+1/2}^n)^2\ -\ u_{j}^{n}\,u_{j+1}^n\ +\ g\,\H_{j+1/2}^n}\ \geq\ \sqrt{g\,\H_{j+1/2}^n}\ >\ 0\,.
\end{equation*}
Thanks to the discrete differentiation rule \eqref{eq:jacond}, we can derive elegant formulas for the predicted values $\vc^{\ast}_{\sharp,\,c}$, $\Flc^{\ast}_{\sharp,\,c}$ by drastically simplifying the scheme \eqref{eq:pred1}, \eqref{eq:pred2}:
\begin{align}\label{eq:predi1}
  \vc_{j+1/2}^{\ast}\ &=\ \Bigl[\,\vc\ -\ \frac{\tau}{2\,\Jj}\;\Rbc\cdot\Di\cdot\bigl(\Labbc\cdot\Pc\ -\ \Lbc\cdot\Glc\bigr)\,\Bigr]_{j+1/2}^{n}\,, \\
  \Flc_{j+1/2}^{\ast}\ &=\ \Bigl[\,\Flc\ -\ \frac{\tau}{2\,\Jj}\;\Rbc\cdot\Di\cdot\Labc\cdot\bigl(\Labbc\cdot\Pc\ -\ \Lbc\cdot\Glc\bigr)\,\Bigr]_{j+1/2}^{n}\,,\label{eq:predi2}
\end{align}
where we introduced two matrices:
\begin{equation*}
  \Labbc_{j+1/2}^n\ \eqdef\ \Labc_{j+1/2}^n\ -\ x_{t,\,j+1/2}^{\,n}\cdot\Id\,, \qquad
  \Pc_{j+1/2}^{\,n}\ \eqdef\ \bigl(\Lbc\cdot\vc_q\bigr)_{j+1/2}^n\,.
\end{equation*}
Finally, the scheme parameters $\theta_{j+1/2}^{1,2}$ are chosen as it was explained in our previous works \cite{Shokin2006, Khakimzyanov2015a} for the case of Nonlinear Shallow Water Equations. This choice guarantees the TVD property of the resulting scheme.

\bigskip
\paragraph*{Non-hydrostatic pressure computation.} 

Once we determined the predicted values $\vc^{\ast}_{\sharp,\,c}\,$, $\Flc^{\ast}_{\sharp,\,c}\,$, we have to determine also the predicted value for the non-hydrostatic pressure components $\Pnhc^{\ast}_{\sharp,\,c}\,$ located in cell centers $\Q_{h,\,c}$. In order to discretize the elliptic equation \eqref{eq:pnhcq} we apply the same finite volume philosophy. Namely, we integrate equation \eqref{eq:pnhcq} over one cell $[q_j,\,q_{j+1}]$. Right now for simplicity we consider an interior element. The approximation near boundaries will be discussed below. The integral form of equation \eqref{eq:pnhcq} reads
\begin{equation}\label{eq:intEq}
  \int_{q_j}^{\,q_{j+1}} \bigl[\,\Kc\,\Pnhc^{\ast}_q\,\bigr]_q\;\ud q\ -\ \int_{q_j}^{\,q_{j+1}}\Kc_0\,\Pnhc^{\ast}\;\ud q\ =\ \int_{q_j}^{\,q_{j+1}}\Fc\;\ud q\,.
\end{equation}
The coefficients $\Kc$, $\Kc_0$ are evaluated using the predicted value of the total water depth $\Hc^{\,\ast}_{\sharp,\,c}\,$. If the scheme parameter $\theta_{j+1/2}^{\,n}\ \equiv\ 0\,$, $\forall j\ =\ 0,\,\ldots,\,N-1$, then the predictor value would lie completely on the middle layer $t\ =\ t^{\,n}\ +\ \frac{\tau}{2}\,$. However, this simple choice of $\{\theta_{j+1/2}^{\,n}\}_{j\,=\,0}^{\,N-1}$ does not ensure the desired TVD property \cite{Barth2004, Shokin2006}.

The solution of this integral equation will give us the predictor value for the non-hydrostatic pressure $\Pnhc^{\ast}_{\sharp,\,c}\,$. The finite difference scheme for equation \eqref{eq:pnhcq} is obtained by applying the following quadrature formulas to all the terms in integral equation \eqref{eq:intEq}:
\begin{multline*}
  \int_{q_j}^{\,q_{j+1}} \bigl[\,\Kc\,\Pnhc^{\ast}_q\,\bigr]_q\;\ud q\ \simeq\ \frac{\Kc_{j+3/2}\ +\ \Kc_{j+1/2}}{2}\cdot\frac{\Pnhc^{\ast}_{j+3/2}\ -\ \Pnhc^{\ast}_{j+1/2}}{\h}\\ -\ \frac{\Kc_{j+1/2}\ +\ \Kc_{j-1/2}}{2}\cdot\frac{\Pnhc^{\ast}_{j+1/2}\ -\ \Pnhc^{\ast}_{j-1/2}}{\h}\,,
\end{multline*}
\begin{multline*}
  \int_{q_j}^{\,q_{j+1}}\Kc_0\,\Pnhc^{\ast}\;\ud q\ \simeq\ \biggl[\,\h\cdot\Bigl[\,\frac{12\,\Jj^n}{(\Hc^{\,\ast})^3}\cdot\frac{\rc\ -\ 3}{\rc}\,\Bigr]_{j+1/2}\\ +\ \Bigl[\,\frac{3\,\hc_q^n}{\rc\,\Jj^n\,(\Hc^{\,\ast})^2}\,\Bigr]_{j+3/2}\ -\ \Bigl[\,\frac{3\,\hc_q^n}{\rc\,\Jj^n\,(\Hc^{\,\ast})^2}\,\Bigr]_{j-1/2}\,\biggr]\,\Pnhc_{j+1/2}^{\ast}\,,
\end{multline*}
\begin{equation*}
  \int_{q_j}^{\,q_{j+1}}\Fc\;\ud q\ \simeq\ \h\cdot\Bigl(2\;\frac{(\uc_q^\ast)^2}{\Jj^n}\ -\ \frac{6\,\Rrc\,\Jj^n}{\rc\,\Hc^{\,\ast}}\Bigr)_{j+1/2}\ +\ \Bigl(g\;\frac{\etac_q^\ast}{\Jj^n}\ +\ \frac{\Rrc\,\hc_q^n}{\rc\,\Jj^n}\Bigr)_{j+1}\ -\ \Bigl(g\;\frac{\etac_q^\ast}{\Jj^n}\ +\ \frac{\Rrc\,\hc_q^n}{\rc\,\Jj^n}\Bigr)_{j}\,.
\end{equation*}
In approximation formulas above we introduced the following notations:
\begin{equation*}
  \Kc_{j+1/2}\ \eqdef\ \Bigl[\,\frac{4}{\rc\,\Jj^n\,\Hc^{\,\ast}}\,\Bigr]_{j+1/2}\,, \qquad
  \rc_{j+1/2}\ \eqdef\ 4\ +\ \Bigl(\,\frac{\hc_q^n}{\Jj^n}\,\Bigr)_{j+1/2}^2\,, \qquad
  \Jj_j^n\ \eqdef\ \frac{\Jj_{j+1/2}^n\ +\ \Jj_{j-1/2}^n}{2}\,.
\end{equation*}
In this way we obtain a three-point finite difference approximation of the elliptic equation \eqref{eq:pnhcq} in interior of the domain, \ie $j\ =\ 1,\,\ldots,\,N-2\,$:
\begin{multline}\label{eq:pred}
  \frac{\Kc_{j+3/2}\ +\ \Kc_{j+1/2}}{2}\cdot\frac{\Pnhc^{\ast}_{j+3/2}\ -\ \Pnhc^{\ast}_{j+1/2}}{\h}\ -\ \frac{\Kc_{j+1/2}\ +\ \Kc_{j-1/2}}{2}\cdot\frac{\Pnhc^{\ast}_{j+1/2}\ -\ \Pnhc^{\ast}_{j-1/2}}{\h}\\ -\ \biggl[\,\h\cdot\Bigl[\,\frac{12\,\Jj^n}{(\Hc^{\,\ast})^3}\cdot\frac{\rc\ -\ 3}{\rc}\,\Bigr]_{j+1/2}\ +\ \Bigl[\,\frac{3\,\hc_q^n}{\rc\,\Jj^n\,(\Hc^{\,\ast})^2}\,\Bigr]_{j+3/2}\ -\ \Bigl[\,\frac{3\,\hc_q^n}{\rc\,\Jj^n\,(\Hc^{\,\ast})^2}\,\Bigr]_{j-1/2}\,\biggr]\ =\\ \h\cdot\Bigl(2\;\frac{(\uc_q^\ast)^2}{\Jj^n}\ -\ \frac{6\,\Rrc\,\Jj^n}{\rc\,\Hc^{\,\ast}}\Bigr)_{j+1/2}\ +\ \Bigl(g\;\frac{\etac_q^\ast}{\Jj^n}\ +\ \frac{\Rrc\,\hc_q^n}{\rc\,\Jj^n}\Bigr)_{j+1}\ -\ \Bigl(g\;\frac{\etac_q^\ast}{\Jj^n}\ +\ \frac{\Rrc\,\hc_q^n}{\rc\,\Jj^n}\Bigr)_{j}\,.
\end{multline}
Two missing equations are obtained by approximating the integral equation \eqref{eq:intEq} in intervals adjacent to the boundaries. As a result, we obtain a linear system of equations where unknowns are $\{\Pnhc_{j+1/2}^{\ast}\}_{j\,=\,0}^{N-1}$. The approximation in boundary cells will be illustrated on the left boundary $[q_0\ \equiv\ 0,\, q_1]$. The right-most cell $[q_{N-1}\,, q_N\ \equiv\ 1]$ can be treated similarly. Let us write down one-sided quadrature formulas for the first cell:
\begin{multline*}
  \frac{\Kc_{3/2}\ +\ \Kc_{1/2}}{2}\cdot\frac{\Pnhc_{3/2}^\ast\ -\ \Pnhc_{1/2}^\ast}{\h}\ -\ 
  \underbrace{\left.\frac{4\,\Pnhc_q^\ast}{\Jj\,\Hc^{\,\ast}\,\rc}\right|_{q\,=\,0}}_{\mathlarger\logof_1}\ -\ 
  \Pnhc_{1/2}^\ast\,\biggl[\,\h\cdot\Bigl[\,\frac{12\,\Jj^n}{(\Hc^{\,\ast})^3}\cdot\frac{\rc\ -\ 3}{\rc}\,\Bigr]_{1/2}\\ +\ \Bigl[\,\frac{3\,\hc_q^n}{\rc\,\Jj^n\,(\Hc^{\,\ast})^2}\,\Bigr]_{3/2}\ +\ \Bigl[\,\frac{3\,\hc_q^n}{\rc\,\Jj^n\,(\Hc^{\,\ast})^2}\,\Bigr]_{1/2}\,\biggr]\ +\ 
  \underbrace{\left.\frac{6\,\hc_q^n\,\Pnhc^\ast}{\Jj\,(\Hc^{\,\ast})^2\,\rc}\right|_{q\,=\,0}}_{\mathlarger\logof_2}\\ =\ 
  \h\cdot\Bigl(2\;\frac{(\uc_q^\ast)^2}{\Jj^n}\ -\ \frac{6\,\Rrc\,\Jj^n}{\rc\,\Hc^{\,\ast}}\Bigr)_{1/2}\ +\ \Bigl(g\;\frac{\etac_q^\ast}{\Jj^n}\ +\ \frac{\Rrc\,\hc_q^n}{\rc\,\Jj^n}\Bigr)_{1}\ -\ 
  \underbrace{\left.\Bigl(g\;\frac{\etac_q^\ast}{\Jj^n}\ +\ \frac{\Rrc\,\hc_q^n}{\rc\,\Jj^n}\Bigr)\right|_{q\,=\,0}}_{\mathlarger\logof_3}\,.
\end{multline*}
It can be readily noticed that terms $\mathlarger{\logof_1\ +\ \logof_2\ +\ \logof_3}$ vanish thanks to the boundary condition \eqref{eq:bcc} (the part at $q\ =\ 0$). The same trick applies to the right-most cell $[q_{N-1}\,, q_N\ \equiv\ 1]$. We reiterate on the fact that in our scheme the boundary conditions are taken into account \emph{exactly}. Consequently, in two boundary cells we obtain a two-point finite difference approximation to equation \eqref{eq:pnhcq}. The resulting linear system of equations can be solved using \eg the direct \textsc{Thomas} algorithm with linear complexity $\O(N)$. Under the conditions $\Kc_0\ >\ 0\,$, $\left.\hc_q\right|_{q\,=\,0}\ \geq\ 0\,$, $\left.\hc_q\right|_{q\,=\,1}\ \leq\ 0\,$ the numerical solution exists, it is unique and stable \cite{Samarskii2001}.


\subsubsection{Corrector step}

During the corrector step we solve again separately the hyperbolic and elliptic parts of the SGN equations. In order to determine the vector of conservative variables $\vc^{n+1}_{\sharp}$ we use an explicit finite volume scheme based on the conservative equation \eqref{eq:cons}:
\begin{equation}\label{eq:corr}
  \frac{(\Jj\vc)_{j}^{n+1}\ -\ (\Jj\vc)_{j}^{n}}{\tau}\ +\ \frac{\bigl(\Flc^\ast - x_t\cdot\vc^\ast\bigr)_{j+1/2}\ -\ \bigl(\Flc^\ast - x_t\cdot\vc^\ast\bigr)_{j-1/2}}{\h}\ =\ \Glc_j^\ast\,,
\end{equation}
where 
\begin{equation*}
  \Glc_j^\ast\ \eqdef\ \begin{pmatrix}
    0 \\
    \Bigl(\bigl(g\,\Hc^{\,n\,+\,\flat}\ -\ \pbc^\ast\bigr)\,\hc_q^{n\,+\,\flat}\ +\ \Pnhc_q^\ast\Bigr)_j
  \end{pmatrix}\,, \quad
  \Pnhc_{q,\,j}^\ast\ \eqdef\ \frac{\Pnhc_{j+1/2}^\ast\ -\ \Pnhc_{j-1/2}^\ast}{\h}\,,
\end{equation*}
and
\begin{align}\label{eq:def1}
  \Hc_j^{\,n\,+\,\flat}\ &\eqdef\ \frac{\Hc_{j+1}^{\,n+1}\ +\ \Hc_{j-1}^{\,n+1}\ +\ 2\,\Hc_{j}^{\,n+1}\ +\ 2\,\Hc_{j}^{\,n} +\ \Hc_{j+1}^{\,n}\ +\ \Hc_{j-1}^{\,n}}{8}\,, \\
  \hc_q^{n\,+\,\flat}\ &\eqdef\ \frac{\hc_{j+1}^{n+1}\ -\ \hc_{j-1}^{n+1}\ +\ \hc_{j+1}^{n}\ -\ \hc_{j-1}^{n}}{4\,\h}\,. \label{eq:def2}
\end{align}
The algorithm of the corrector scheme can be summarized as follows:
\begin{enumerate}
  \item From the mass conservation equations (the first component in \eqref{eq:corr}) we find the total water depth $\Hc_{\sharp}^{\,n+1}$ in interior nodes of the grid
  \item Using the method of characteristics and the boundary conditions $\uc_0^{n+1}\ =\ \uc_{N}^{n+1}\ \equiv\ 0$ we determine the total water depth $\Hc_0^{\,n+1}$, $\Hc_N^{\,n+1}$ in boundary points $q_0\ \equiv\ 0$ and $q_N\ \equiv\ 1$
  \item Then, using the momentum conservation equation (the second component in \eqref{eq:corr}) we find the momentum values $(\Hc\,\uc)_{\sharp}^{n+1}$ on the next time layer.
\end{enumerate}
In this way, we obtain an explicit scheme despite the fact that the right hand side $\Glc_{\sharp}^\ast$ depends on the water depth $\Hc_{\sharp}^{\,n+1}\,$ at the new time layer $t\ =\ t^{n+1}\,$.

\bigskip
\paragraph*{Non-hydrostatic pressure correction.} The non-hydrostatic pressure correction $\Pnhc_{\sharp}^{n+1}$ is computed by integrating locally the elliptic equation \eqref{eq:pnhcq} around each grid point:
\begin{equation*}
  \int_{q_{j-1/2}}^{\,q_{j+1/2}} \bigl[\,\Kc\,\Pnhc^{n+1}_q\,\bigr]_q\;\ud q\ -\ \int_{q_{j-1/2}}^{\,q_{j+1/2}} \Kc_0\,\Pnhc^{n+1}\;\ud q\ =\ \int_{q_{j-1/2}}^{\,q_{j+1/2}} \Fc^{\,n+1}\;\ud q\,, \qquad j\ =\ 1,\,\ldots,\,N-1\,,
\end{equation*}
The details of integrals approximations are similar to the predictor step described above. Consequently, we provide directly the difference scheme in interior nodes:
\begin{multline}\label{eq:Korr}
  \K_{j+1/2}\;\frac{\Pnhc_{j+1}^{n+1}\ -\ \Pnhc_{j}^{n+1}}{\h}\ -\ \K_{j-1/2}\;\frac{\Pnhc_{j}^{n+1}\ -\ \Pnhc_{j-1}^{n+1}}{\h}\ =\\ -6\,\Pnhc_{j}^{n+1}\biggl[\,\Bigl(\h\;\frac{(\rc - 3)\,\Jj}{\rc\,\Hc^{\,3}}\ -\ \frac{\hc_q}{\rc\,\Jj\,\Hc^{\,2}}\Bigr)_{j-1/2}^{n+1}\ +\ \Bigl(\h\;\frac{(\rc - 3)\,\Jj}{\rc\,\Hc^{\,3}}\ +\ \frac{\hc_q}{\rc\,\Jj\,\Hc^{\,2}}\Bigr)_{j+1/2}^{n+1}\,\biggr]\ =\\ \h\;\Bigl(2\;\frac{\uc_q^2}{\Jj}\ -\ \frac{6\,\Rrc\,\Jj}{\rc\,\Hc}\Bigr)_{j}^{n+1}\ +\ \Bigl(g\;\frac{\etac_q}{\Jj}\ +\ \frac{\Rrc\,\hc_q}{\rc\,\Jj}\Bigr)_{j+1/2}^{n+1}\ -\ \Bigl(g\;\frac{\etac_q}{\Jj}\ +\ \frac{\Rrc\,\hc_q}{\rc\,\Jj}\Bigr)_{j-1/2}^{n+1}\,,
\end{multline}
where
\begin{equation*}
  \K_{j+1/2}\ \eqdef\ \frac{4}{\bigl(\rc\,\Jj\,\Hc\bigr)_{j+1/2}^{n+1}}\,, \quad
  \rc_{j+1/2}^{n+1}\ \eqdef\ 4\ +\ \Bigl[\,\frac{\hc_{j+1}^{n+1}\ -\ \hc_{j}^{n+1}}{x_{j+1}^{n+1}\ -\ x_j^{n+1}}\,\Bigr]^2\,, \quad
  \Jj_{j+1/2}^{n+1}\ \eqdef\ \frac{x_{j+1}^{n+1}\ -\ x_{j}^{n+1}}{\h}\,.
\end{equation*}
In order to complete the scheme description, we have to specify the discretization of the elliptic equation \eqref{eq:pnhcq} in boundary cells. To be specific we take again the left-most cell $[q_0\ \equiv\ 0,\, q_{1/2}]$. The integral equation in this cell reads:
\begin{equation*}
  \int_{q_{0}}^{\,q_{1/2}} \bigl[\,\Kc\,\Pnhc^{n+1}_q\,\bigr]_q\;\ud q\ -\ \int_{q_{0}}^{\,q_{1/2}} \Kc_0\,\Pnhc^{n+1}\;\ud q\ =\ \int_{q_{0}}^{\,q_{1/2}} \Fc^{\,n+1}\;\ud q\,.
\end{equation*}
And the corresponding difference equation is
\begin{multline*}
  \K_{1/2}\;\frac{\Pnhc_1^{n+1}\ -\ \Pnhc_0^{n+1}}{\h}\ -\ \underbrace{\left.\frac{4\,\Pnhc_q}{\Jj\,\Hc\,\rc}\right|_{q\,=\,0}^{n+1}}_{\mathlarger\Diamond_1}\ -\ 6\,\Pnhc_0^{n+1}\;\biggl[\,\h\;\frac{(\rc - 3)\,\Jj}{\rc\,\Hc^{\,3}}\ +\ \frac{\hc_q}{\Jj\,\Hc^{\,2}\,\rc}\,\biggr]_{1/2}^{n+1}\ +\ \underbrace{\left.\frac{6\,\hc_q\,\Pnhc}{\Jj\,\Hc^{\,2}\,\rc}\right|_{q\,=\,0}^{n+1}}_{\mathlarger\Diamond_2}\ =\\
  \biggl[\,g\,\frac{\etac_q}{\Jj}\ +\ \frac{\Rrc\,\hc_q}{\rc\,\Jj}\ +\ \h\;\Bigl(\frac{\uc_q^2}{\Jj}\ -\ \frac{3\,\Rrc\,\Jj}{\rc\,\Hc}\Bigr)\,\biggr]_{1/2}^{n+1}\ -\ \underbrace{\left.\Bigl(g\,\frac{\etac_q}{\Jj}\ +\ \frac{\Rrc\,\hc_q}{\rc\,\Jj}\Bigr)\right|_{q\,=\,0}^{n+1}}_{\mathlarger\Diamond_3}\,.
\end{multline*}
By taking into account the boundary condition \eqref{eq:bcc} we obtain that three under-braced terms vanish:
\begin{equation*}
  \mathlarger\Diamond_1\ +\ \mathlarger\Diamond_2\ +\ \mathlarger\Diamond_3\ \equiv\ 0\,.
\end{equation*}
A similar two-point approximation can be obtained by integrating over the right-most cell $\bigl[\,q_{N-1/2},\,q_N\,\bigr]\ \equiv\ \Bigl[\,1\ -\ \dfrac{\h}{2},\,1\,\Bigr]\,$. In this way we obtain again a three-diagonal system of linear equations which can be efficiently solved with the \textsc{Thomas} algorithm \cite{Higham2002}.

\bigskip
\paragraph*{Stability of the scheme.}

In order to ensure the stability of (nonlinear) computations, we impose a slightly stricter restriction on the time step $\tau$ than the linear analysis given below predicts (see Section~\ref{sec:linstab}). Namely, at every time layer we apply the same restriction as for hyperbolic (non-dispersive) Nonlinear Shallow Water Equations \cite{Khakimzyanov2015a}:
\begin{equation*}
  \max_{j}\{\,\Cs_{j+1/2}^{\,n,\,\pm}\,\}\ \leq\ 1\,,
\end{equation*}
where $\Cs_{j+1/2}^{n,\,\pm}$ are local \textsc{Courant} numbers \cite{Courant1928} which are defined as follows
\begin{equation*}
  \Cs_{j+1/2}^{\,n,\,\pm}\ \eqdef\ \frac{\tau}{\h}\;\Bigl[\,\frac{\abs{\lambda^\pm\ -\ x_t}}{\Jj}\,\Bigr]_{j+1/2}^n\,.
\end{equation*}

\subsubsection{Well-balanced property}

It can be easily established that the predictor--corrector scheme presented above preserves exactly the so-called states `lake-at-rest':

\begin{lemma}
Assume that the bottom is stationary (\ie $h_t\ \equiv\ 0\,$, but not necessary flat) and initially the fluid is at the `lake-at-rest' state, \ie
\begin{equation}\label{eq:assume}
  \etac_{j}^{\,0}\ \equiv\ 0\,, \qquad \uc_{j}^{\,0}\ \equiv\ 0\, \qquad j\ =\ 0,\,1,\,2,\,\ldots,\,N\,.
\end{equation}
Then, the predictor--corrector scheme will preserve this state at all time layers.
\end{lemma}
\begin{proof}
In order to prove this Lemma, we employ the mathematical induction \cite{Hermes1973}. First, we have to discuss the generation of the initial grid and how it will be transformed to the next time layer along with the discrete numerical solution:
\begin{equation*}
  x_\sharp^{0}\ \hookrightarrow\ x_\sharp^{1}\,, \qquad
  \vc_{\sharp}^{0}\ \hookrightarrow\ \vc_{c,\,\sharp}^{\ast}\ \hookrightarrow\ \vc_{\sharp}^{1}\,.
\end{equation*}
Then, by assuming that our statement is true at the $n$\up{th} time layer, we will have to show that it is true on the upcoming $(n+1)$\up{th} layer. This will complete the proof \cite{Hermes1973}.

If the monitoring function $\om(x,\,t)$ depends only on the free surface elevation $\eta(x,\,t)$ and fluid velocity $u(x,\,t)$, then the monitoring function $\om(x,\,t)\ \equiv\ 1$ thanks to Lemma assumption \eqref{eq:assume}. And the equidistribution principle \eqref{eq:initgrid} will give us the uniform mesh. However, in most general situations one can envisage the grid adaptation upon the bathymetry profile\footnote{In the present study we do not consider such example. However, the idea of grid adaptation upon the bathymetry function certainly deserves to be studied more carefully.} $h(x,\,t)$. Consequently, in general we can expect that the mesh will be non-uniform even under condition \eqref{eq:assume}, since $h_x\ \neq\ 0\,$. However, we know that the initial grid satisfies the fully converged discrete equidistribution principle \eqref{eq:discr}. From now on we assume that the initial grid is generated and it is not necessarily uniform. In order to construct the grid at the next layer, we solve just one linear equation \eqref{eq:discr2}. Since, system \eqref{eq:discr2} is diagonally dominant, its solution exists and it is unique \cite{Samarskii2001}. It is not difficult to check that the set of values $\{x_j^{\,1}\ \equiv\ x_j^{\,0}\}_{j\,=\,0}^{N}$ solves the system \eqref{eq:discr2}. It follows from two observations:
\begin{itemize}
  \item The right-hand side of \eqref{eq:discr2} vanishes when $x_j^{1}\ \equiv\ x_j^{0}$, $\forall j\ =\ 0,\,\ldots,\,N\,$.
  \item The monitor function $\{\om_{j+1/2}^{\,0}\}_{j\,=\,0}^{N-1}$ is evaluated on the previous time layer $t\ =\ 0\,$.
\end{itemize}
Thus, we obtain that $x_{\sharp}^{\,1}\ \equiv\ x_{\sharp}^{\,0}\,$. Consequently, we have $x_{t,\,j}^{\,0}\ \equiv\ 0\,$ and $\Jj_j^{\,1}\ =\ \Jj_j^{\,0}\,$, $\forall j\ =\ 0,\,\ldots,\,N\,$.

In order to complete the predictor step we need to determine the quantities $\Pnhc_{\sharp}^{\,0}$ and $\pbc_{\sharp}^{\,0}\,$ on which depends the source term $\Glc_{j+1/2}^{\,0}\,$. These quantities are uniquely determined by prescribed initial conditions. For instance, $\Pnhc_{\sharp}^{\,0}$ are obtained by solving linear equations \eqref{eq:Korr}. We showed above also that the solution to this equation is unique. We notice also that the right-hand side in equation \eqref{eq:Korr} vanishes under conditions of this Lemma. Consequently, we obtain $\Pnhc_{\sharp}^{\,0}\ \equiv\ \vO\,$. By applying a finite difference analogue of equation \eqref{eq:pbc} we obtain also that $\pb_{\sharp}^{\,0}\ \equiv\ \vO\,$. As the result, at the `lake-at-rest' state the right-hand side of predictor equations \eqref{eq:pred1}, \eqref{eq:pred2} reads
\begin{equation*}
  \Glc_{j+1/2}^{\,0}\ =\ \begin{pmatrix}
    0 \\
    (g\,\hc\,\hc_q)_{j+1/2}^{\,0}
  \end{pmatrix}\,.
\end{equation*}
Taking into account the fact that the mesh does not evolve $x_\sharp^{0}\ \hookrightarrow\ x_\sharp^{1}\ \equiv x_\sharp^{0}\,$, we obtain $x_{t,\,j}^{\,0}\ \equiv\ 0$ and thus $\Labbc_{j+1/2}^{\,0}\ \equiv\ \Labc_{j+1/2}^{\,0}\,$, $s_{j+1/2}^{\,0}\ \equiv\ \sqrt{g\,\hc_{j+1/2}}\,$,
\begin{equation*}
  \bigl(\Labbc\cdot\Pc\bigr)_{j+1/2}^{\,0}\ \equiv\ \begin{pmatrix}
    \hc_{q,\,j+1/2} \\
    \hc_{q,\,j+1/2}
  \end{pmatrix}\,, \qquad
  \bigl(\Lbc\cdot\Glc\bigr)_{j+1/2}^{\,0}\ \equiv\ \begin{pmatrix}
    \hc_{q,\,j+1/2} \\
    \hc_{q,\,j+1/2}
  \end{pmatrix}\,.
\end{equation*}
Consequently, the predictor step \eqref{eq:predi1}, \eqref{eq:predi2} gives us the following values:
\begin{equation*}
  \vc_{j+1/2}^{\ast}\ \equiv\ \vc_{j+1/2}^{\,0}\,, \qquad
  \Flc_{j+1/2}^{\ast}\ \equiv\ \Flc_{j+1/2}^{\,0}\,.
\end{equation*}
For the sake of clarity, we rewrite the last predictions in component-wise form:
\begin{equation*}
  \vc_{j+1/2}^{\ast}\ \equiv\ \begin{pmatrix}
    \hc_{j+1/2} \\
    0
  \end{pmatrix}\,, \qquad
  \Flc_{j+1/2}^{\ast}\ \equiv\ \begin{pmatrix}
    0 \\
    \frac{g\,\hc_{j+1/2}^{\,2}}{2}
  \end{pmatrix}\,.
\end{equation*}
Thus, $\Hc_{j+1/2}^{\ast}\ \equiv\ \hc_{j+1/2}$. As an intermediate conclusion of the predictor step we have:
\begin{equation*}
  \eta_{j+1/2}^{\ast}\ \equiv\ 0\,, \qquad \uc_{j+1/2}^{\ast}\ \equiv\ 0\,,
\end{equation*}
and all dispersive corrections $\Pnhc_{\sharp}^{\,\ast}$, $\pbc_{\sharp}^{\,\ast}\,$ vanish as well by applying similar arguments to equation \eqref{eq:pred}.

The corrector step \eqref{eq:corr}, written component-wise reads:
\begin{align*}
  \frac{(\Jj\,\Hc)_{j}^{\,1}\ -\ (\Jj\,\Hc)_j^{\,0}}{\tau}\ &=\ 0\,, \\
  \frac{(\Jj\,\uc\,\Hc)_{j}^{\,1}\ -\ (\Jj\,\uc\,\Hc)_j^{\,0}}{\tau}\ +\ \frac{g\,\hc_{j+1/2}^{\,2}\ -\ g\,\hc_{j-1/2}^{\,2}}{2\,\h}\ &=\ g\,\bigl(\Hc\,\hc_q\bigr)_{j}^{\,\flat}\,
\end{align*}
From the first equation above taking into account that $\Jj_j^{\,1}\ \equiv\ \Jj_{j}^{\,0}$ and $\Hc_j^{\,0}\ =\ \hc_{j}$ we obtain $\Hc_j^{\,1}\ =\ \hc_{j}\,$. And thus, by the definition of the total water depth we obtain $\etac_{j}^{\,1}\ \equiv\ 0\,$. In the second equation above by condition \eqref{eq:assume} we have that $\uc_j^{\,0}\ \equiv\ 0\,$. Moreover, in the left-hand side:
\begin{equation}\label{eq:comp1}
  \frac{g\,\hc_{j+1/2}^{\,2}\ -\ g\,\hc_{j-1/2}^{\,2}}{2\,\h}\ =\ g\;\frac{(\hc_{j+1}\ -\ \hc_{j-1})\cdot(\hc_{j+1}\ +\ 2\,\hc_{j}\ +\ \hc_{j-1})}{8\,\h}\,.
\end{equation}
The right-hand side of the same corrector equation can be rewritten using definitions \eqref{eq:def1}, \eqref{eq:def2} as
\begin{equation}\label{eq:comp2}
  g\,\bigl(\Hc\,\hc_q\bigr)_{j}^{\,\flat}\ =\ g\;\frac{2\,\hc_{j+1}\ +\ 4\,\hc_{j}\ +\ 2\,\hc_{j-1}}{8}\cdot\frac{2\,\hc_{j+1}\ -\ 2\,\hc_{j-1}}{4\,\h}\,.
\end{equation}
Comparing equation \eqref{eq:comp1} with \eqref{eq:comp2} yields the desired well-balanced property of the predictor--corrector scheme and thus $\uc_j^{\,1}\ \equiv\ 0\,$.

By assuming that \eqref{eq:assume} is verified at the time layer $t\ =\ t^{\,n}$ and repeating precisely the same reasoning as above (by substituting superscripts $0\ \leftarrow\ n$ and $1\ \leftarrow\ n\,+\,1$) we obtain that \eqref{eq:assume} is verified at the next time layer $t\ =\ t^{\,n+1}\,$. It completes the proof of this Lemma.
\end{proof}

We would like to mention that the well-balanced property of the proposed scheme was checked also in numerical experiments on various configurations of general uneven bottoms (not reported here for the sake of manuscript compactness) --- in all cases we witnessed the preservation of the `lake-at-rest' state up to the machine precision. This validates our numerical implementation of the proposed algorithm.


\subsection{Numerical scheme for linearized equations}
\label{sec:linear}

In order to study the numerical scheme stability and its dispersive properties, we consider the discretization of the linearized SGN equations on a uniform unbounded grid (for simplicity we consider an IVP without boundary conditions). The governing equations after linearization can be written as (we already gave these equations in Section~\ref{sec:disp})
\begin{align*}
  \eta_t\ +\ d\,u_x\ &=\ 0\,, \\
  u_t\ +\ g\,\eta_x\ &=\ \frac{1}{d}\;\Pnh_x\,, \\
  \Pnh_{xx}\ -\ \frac{3}{d^2}\;\Pnh\ &=\ c^2\,\eta_{xx}\,,
\end{align*}
where $c\ =\ \sqrt{g\,d}$ is the speed of linear gravity waves. We shall apply to these PDEs precisely the same scheme as described above. Since the grid is uniform, we can return to the original notation, \ie $\vc\ \equiv\ \v$, \etc\ Let $\dx$ be the discretization step in the computational domain $\I_h$ and $\tau$ is the local time step. We introduce the following finite difference operators (illustrated on the free surface elevation $\eta_{\,\sharp}^{\,n}$):
\begin{equation*}
  \eta_{t,\,j}^{\,n}\ \eqdef\ \frac{\eta_j^{\,n+1}\ -\ \eta_j^{\,n}}{\tau}\,, \qquad
  \eta_{x,\,j}^{\,n}\ \eqdef\ \frac{\eta_{j+1}^{\,n}\ -\ \eta_{j}^{\,n}}{\dx}\,, \qquad
  \eta_{(x),\,j}^{\,n}\ \eqdef\ \frac{\eta_{j+1}^{\,n}\ -\ \eta_{j-1}^{\,n}}{2\,\dx}\,,
\end{equation*}
\begin{equation*}
  \eta_{xx,\,j}^{\,n}\ \eqdef\ \frac{\eta_{j+1}^{\,n} - 2\,\eta_j^{\,n} + \eta_{j-1}^{\,n}}{\dx^2}\,, \quad \eta_{xx,\,j+1/2}^{\,n}\ \eqdef\ \frac{\eta_{xx,\,j}^{\,n} + \eta_{xx,\,j+1}^{\,n}}{2}\,, \quad \eta_{xxx,\,j}^{\,n}\ \eqdef\ \frac{\eta_{xx,\,j+1}^{\,n} - \eta_{xx,\,j}^{\,n}}{\dx}\,.
\end{equation*}
Then, at the predictor step we compute auxiliary quantities $\{\eta_{j+1/2}^{\ast}\}_{j\,=\,-\infty}^{+\infty}$, $\{u_{j+1/2}^{\ast}\}_{j\,=\,-\infty}^{+\infty}$ and $\{\Pnh_{j+1/2}^{\ast}\}_{j\,=\,-\infty}^{+\infty}\,$. First, we solve the hyperbolic part of the linearized SGN equations:
\begin{align*}
  \frac{\eta_{j+1/2}^\ast\ -\ \frac{1}{2}\,\bigl(\eta_{j+1}^{\,n}\ +\ \eta_j^{\,n}\bigr)}{\tau^\ast_{j+1/2}}\ +\ d\,u_{x,\,j}^n\ &=\ 0\,, \\
  \frac{u_{j+1/2}^\ast\ -\ \frac{1}{2}\,\bigl(u_{j+1}^{n}\ +\ u_j^{n}\bigr)}{\tau^\ast_{j+1/2}}\ +\ g\,\eta_{x,\,j}^{\,n}\ &=\ \frac{1}{d}\;\Pnh_{x,\,j}^n\,,
\end{align*}
and then we solve the elliptic equation to find $\{\Pnh_{j+1/2}^{\ast}\}_{j\,=\,-\infty}^{+\infty}\,$:
\begin{equation*}
  \frac{\Pnh_{j+3/2}^\ast\ -\ 2\,\Pnh_{j+1/2}^\ast\ +\ \Pnh_{j-1/2}^\ast}{\dx^2}\ -\ \frac{3}{d^2}\;\Pnh_{j+1/2}^\ast\ =\ c^2\,\frac{\eta_{j+3/2}^\ast\ -\ 2\,\eta_{j+1/2}^\ast\ +\ \eta_{j-1/2}^\ast}{\dx^2}\,,
\end{equation*}
where $\tau_{j+1/2}^\ast\ \eqdef\ \dfrac{\tau}{2}\;(1\ +\ \theta_{j+1/2}^{\,n})$ and $\theta_{j+1/2}^{\,n}$ is the numerical scheme parameter \cite{Khakimzyanov2015a}, whose choice guarantees the TVD property (strictly speaking the proof was done for scalar hyperbolic equations only).

Then, the predicted values are used on the second --- corrector step, to compute all physical quantities $\{\eta_{j}^{\,n+1}\}_{j\,=\,-\infty}^{+\infty}$, $\{u_{j}^{n+1}\}_{j\,=\,-\infty}^{+\infty}$ and $\{\Pnh_{j}^{n+1}\}_{j\,=\,-\infty}^{+\infty}\,$ on the next time layer $t\ =\ t^{n+1}$:
\begin{align}
  \eta_{t,\,j}^{\,n}\ +\ d\,\frac{u_{j+1/2}^\ast\ -\ u_{j-1/2}^{\ast}}{\dx}\ &=\ 0\,, \\
  u_{t,\,j}^{\,n}\ +\ g\,\frac{\eta_{j+1/2}^\ast\ -\ \eta_{j-1/2}^\ast}{\dx}\ &=\ \frac{1}{d}\;\frac{\Pnh_{j+1/2}^\ast\ -\ \Pnh_{j-1/2}^\ast}{\dx}\,, \\
  \Pnh_{xx,\,j}^{n+1}\ -\ \frac{3}{d^2}\;\Pnh_{j}^{n+1}\ &=\ c^2\,\eta_{xx,\,j}^{\,n+1}\,.\label{eq:103}
\end{align}

It can be easily checked that the scheme presented above has the first order accuracy if $\theta_{j+1/2}^{\,n}\ =\ \const\,$, $\forall j$ and the second order if $\theta_{j+1/2}^{\,n}\ \equiv\ 0\,$, $\forall j\,$. However, the last condition can be somehow relaxed. There is an interesting case of quasi-constant values of the scheme parameter:
\begin{equation*}
  \theta_{j+1/2}^{\,n}\ =\ \O\,(\tau\ +\ \dx)\,.
\end{equation*}
In this case the scheme is second order accurate as well. In the present Section we perform a theoretical analysis of the scheme and we shall assume for simplicity that $\theta_{j+1/2}^{\,n}\ \equiv\ \const$. Consequently, from now on we shall drop the index $j+1/2$ in the intermediate time step $\tau_{j+1/2}^\ast\,$.


\subsubsection{Linear stability of the scheme}
\label{sec:linstab}

In this Section we apply the so-called \textsc{von Neumann} stability analysis to the predictor--corrector scheme described above \cite{Charney1950}. In order to study the scheme stability, first we exclude algebraically the predicted values $\{\eta_{j+1/2}^{\ast}\}_{j\,=\,-\infty}^{+\infty}$, $\{u_{j+1/2}^{\ast}\}_{j\,=\,-\infty}^{+\infty}$ from difference equations. The resulting system reads:
\begin{align}\label{eq:105a}
  \eta_{t,\,j}^{\,n}\ +\ d\,u_{(x),\,j}^{n}\ &=\ \tau^\ast\,c^2\,\eta_{xx,\,j}^{\,n}\ -\ \tau^\ast\,\Pnh_{xx,\,j}^{n}\,, \\
  u_{t,\,j}^n\ +\ g\,\eta_{(x),\,j}^{\,n}\ &=\ \tau^\ast\,c^2\,u_{xx,\,j}^{n}\ +\ \frac{1}{d}\;\frac{\Pnh_{j+1/2}^\ast\ -\ \Pnh_{j-1/2}^\ast}{\dx}\,,\label{eq:105b} \\
  \frac{\Pnh_{j+3/2}^\ast - 2\,\Pnh_{j+1/2}^\ast + \Pnh_{j-1/2}^\ast}{\dx^{\,2}} - \frac{3}{d^2}\;\Pnh_{j+1/2}^\ast\ &= c^2\,\bigl(\eta_{xx,\,j+1/2}^{\,n}\ -\ \tau^\ast\,d\,u_{xxx,\,j}^n\bigr)\,.\label{eq:104}
\end{align}
We substitute in all difference relations above the following elementary harmonics
\begin{equation}\label{eq:pwave}
  \eta_j^{\,n}\ =\ \Lambda_0\,\uprho^n\,\ue^{\,\ui\, j\, \xi}\,, \quad
  u_j^n\ =\ \Psi_0\,\uprho^n\,\ue^{\,\ui\, j\, \xi}\,, \quad
  \Pnh_j^n\ =\ \Phi_0\,\uprho^n\,\ue^{\ui\, j\, \xi}\,, \quad
  \Pnh_{j+1/2}^\ast\ =\ \Phi_0^\ast(\uprho)\,\ue^{\,\ui\, (j+1/2)\, \xi}\,,
\end{equation}
where $\xi\ \eqdef\ k\cdot\dx\ \in\ [0,\,\pi]$ is the \emph{scaled} wavenumber and $\uprho$ is the transmission factor between the time layers $t^n$ and $t^{n+1}$. As a result, from equations \eqref{eq:103} and \eqref{eq:104} we obtain the following expressions for $\Phi_0$ and $\Phi_0^\ast$:
\begin{equation*}
  \Phi_0\ =\ \frac{4\,c^2\,d^2}{3\,\hslash\,\dx^{\,2}}\;\gimel^2\,\Lambda_0\,, \qquad
  \Phi_0^\ast(\uprho)\ =\ \uprho^n\,\frac{2\,c^2\,d^2}{3\,\hslash\,\dx^{\,2}}\;\gimel\;\Bigl[\,\Lambda_0\,\sin(\xi)\ -\ \ui\,\tau^\ast\,\frac{4\,d}{\dx}\;\gimel^{\,2}\,\Psi_0\,\Bigr]\,,
\end{equation*}
where we introduced some short-hand notations:
\begin{equation*}
  \aleph\ \eqdef\ \frac{\tau}{\dx}\,, \quad 
  \gimel\ \eqdef\ \sin\Bigl(\frac{\xi}{2}\Bigr)\,, \quad
  \daleth\ \eqdef\ 4\,c^2\,\aleph^{\,2}\,\gimel^{\,2}\,, \quad
  \beth\ \eqdef\ c\,\aleph\,\sin(\xi)\,, \quad
  \hslash\ \eqdef\ 1\ +\ \frac{4\,d^{\,2}}{3\,\dx^{\,2}}\;\gimel^{\,2}\,.
\end{equation*}
By substituting just obtained expressions for $\Phi_0$ and $\Phi_0^\ast$ into equations \eqref{eq:105a}, \eqref{eq:105b} we obtain two linear equations with respect to amplitudes $\Lambda_0$ and $\Psi_0$:
\begin{align*}
  \Bigl[\,\uprho\ -\ 1\ +\ \frac{2\,c^2\,\aleph^2(1 + \theta)}{\hslash}\;\gimel^{\,2}\,\Bigr]\;\Lambda_0\ +\ \ui\,\aleph\,d\,\sin(\xi)\,\Psi_0\ &=\ 0\,, \\
  \ui\,\frac{g\,\aleph\,\sin(\xi)}{\hslash}\;\Lambda_0\ +\ \Bigl[\,\uprho\ -\ 1\ +\ \frac{2\,c^2\,\aleph^2(1 + \theta)}{\hslash}\;\gimel^{\,2}\,\Bigr]\;\Psi_0\ &=\ 0\,.
\end{align*}
The necessary condition to have non-trivial solutions gives us an algebraic equation for the transmission factor $\uprho$:
\begin{equation*}
  (\uprho\ -\ 1)^2\ +\ \frac{\daleth\,(1\ +\ \theta)}{\hslash}\;(\rho\ -\ 1)\ +\ \frac{\daleth^2\,(1\ +\ \theta)^2}{4\,\hslash^2}\ +\ \frac{\beth^2}{\hslash}\ =\ 0\,.
\end{equation*}
This quadratic equation admits two distinct roots:
\begin{equation}\label{eq:roots}
  \uprho^\pm\ =\ 1\ -\ \frac{\daleth\,(1\ +\ \theta)}{2\,\hslash}\ \pm\ \ui\,\frac{\beth}{\sqrt{\hslash}}\,.
\end{equation}
The necessary stability condition $\abs{\uprho}\ \leq\ 1$ is equivalent to the following condition on quadratic equation coefficients:
\begin{equation}\label{eq:stab}
  c^2\,\aleph^2\,(1\ +\ \theta)^2\,\zeta\ -\ \Bigl[\,1\ +\ \frac{4\,\varsigma^2}{3}\,\zeta\,\Bigr]\,(\zeta\ +\ \theta)\ \leq\ 0\,, \qquad \varsigma\ \eqdef\ \frac{d}{\dx}\,,
\end{equation}
which has to be fulfilled for all $\zeta\ \eqdef\ \gimel^{\,2}\ \in\ [0,\,1]$. The parameter $\varsigma$ characterizes the grid resolution relative to the mean water depth. This parameter appears in stability condition along with the \textsc{Courant} ratio $\aleph$. It is one of the differences with non-dispersive equations whose discretization stability depends only on $\aleph$.

\bigskip
\paragraph*{Further thoughts about stability.}

When long waves travel towards the shoreline, their shoaling process is often accompanied with the formation of undular bores \cite{Peregrine1966, Grue2008}. Undular bores have dispersive nature and cannot be correctly described by dispersionless models. In \cite{Grue2008} it was shown that satisfactory description of dispersive effects in shallow water environments is obtained for $\varsigma\ \eqdef\ \dfrac{d}{\dx}\ =\ 2\ \sim\ 4\,$. In another study \cite{Glimsdal2006} it was shown that for satisfactory modeling of trans-oceanic wave propagation it is sufficient to choose $\varsigma\ \approx\ 4$ in deep ocean and $\varsigma\ \approx\ 1$ in shallow coastal areas. In other words, it is sufficient to choose the grid size equal to water depth in shallow waters and in deep areas --- four times smaller than the water depth. On coarser grids the numerical dispersion may dominate over the physical one \cite{Glimsdal2006}. In the present study we shall assume that parameter $\varsigma\ \geq\ \dfrac{\sqrt{3}}{2}\,$.

Substituting into equation \eqref{eq:stab} the value $\zeta\ \equiv\ 0$ we obtain that for stability reasons necessarily the scheme parameter $\theta\ \geq\ 0\,$. Since the predictor layer should be in between time layers $t\ =\ t^n$ and $t\ =\ t^{n+1}$ we have $\theta\ \leq\ 1\,$. Then, for fixed values of parameters $\varsigma$ and $\theta$ the stability condition \eqref{eq:stab} takes the following form:
\begin{equation*}
  c\,\aleph\ \leq\ \frac{\sqrt{1\ +\ \dfrac{4}{3}\,\varsigma^2\,\theta}}{1\ +\ \theta}\,.
\end{equation*}
For $\theta\ \equiv\ 0$ the last condition simply becomes:
\begin{equation*}
  c\,\aleph\ \leq\ 1\,,
\end{equation*}
and it does not depend on parameter $\varsigma$. However, when $\theta\ >\ 0$, then the scheme stability depends on the mesh refinement $\varsigma$ relative to the mean water depth. Surprisingly, more we refine the grid, less stringent becomes the stability barrier. In the asymptotic limit $\varsigma\ \gg\ 1$ we obtain the following restriction on the time step $\tau$:
\begin{equation*}
  \tau\ \leq\ \frac{2\,\sqrt{\theta}}{\sqrt{3}\,(1\ +\ \theta)}\;\tau_0\ <\ \frac{1}{\sqrt{3}}\;\tau_0\ \approx\ 0.58\,\tau_0\,,
\end{equation*}
where $\tau_0\ \eqdef\ \frac{d}{c}\ \equiv\ \frac{d}{\sqrt{gd}}$ is the characteristic time scale of gravity waves. Above we used the following obvious inequality:
\begin{equation*}
  1\ +\ \theta\ \geq\ 2\,\sqrt{\theta}\,, \qquad \forall\,\theta\ \in\ \R^+\,.
\end{equation*}

So, in practice for \emph{sufficiently refined} grids the stability condition \emph{de facto} does not involve the grid spacing $\dx$ anymore. This property is very desirable for numerical simulations. For the sake of comparison we give here (without underlying computations) the stability restriction of the same predictor--corrector scheme for NSWE equations:
\begin{equation*}
  c\,\aleph\ \leq\ \frac{1}{\sqrt{1\ +\ \theta}}\,.
\end{equation*}
So, another surprising conclusion obtained from this linear stability analysis is that the SGN equations require \emph{in fine} a less stringent condition on the time step than corresponding dispersionless NSWE. Most probably, this conclusion can be explained by the regularization effect of the dispersion. Indeed, the NSWE \emph{bores} are replaced by smooth \emph{undular bores} whose regularity is certainly higher. The smoothness of solutions allows to use a larger time step $\tau$ to propagate the numerical solution. This conclusion was checked in (fully nonlinear) numerical experiments (not reported here) where the time step $\tau$ was artificially pushed towards the stability limits. In general, the omission of dispersive effects yields a stricter stability condition. The authors of \cite{Glimsdal2013} came experimentally to similar conclusions about the time step limit in dispersive and hydrostatic simulations. Our theoretical analysis reported above may serve as a basis of rational explanation of this empirical fact.

This result is to be compared with a numerical scheme proposed in \cite{Tkalich2007} for a weakly nonlinear weakly dispersive water wave model. They used splitting technique and solved an elliptic equation to determine the non-hydrostatic pressure correction. The main drawback of the scheme proposed in \cite{Tkalich2007} is the stability condition:
\begin{equation*}
  \dx\ \geq\ 1.5\,d\,.
\end{equation*}
One can easily see that a numerical computation with a sufficiently refined grid is simply impossible with that scheme. Our method is free of such drawbacks.


\subsubsection{Discrete dispersion relation}

The dispersion relation properties are crucial to understand and explain the behaviour of the numerical solution \cite{LeRoux2012}. In this Section we perform the dispersion relation analysis of the proposed above predictor--corrector scheme. This analysis is based on the study of elementary plane-wave solutions \eqref{eq:pwave}. The continuous case was already analyzed in Section~\ref{sec:disp}. Dispersive properties of the scheme can be completely characterized by the phase error $\Delta\varphi\ \eqdef\ \phi\ -\ \varphi$ committed during solution transfer from time layer $t\ =\ t^n$ to $t\ =\ t^{n+1}\ =\ t^n\ +\ \tau$. Here we denote by $\phi$ the phase shift due to the SGN equations dynamics and $\varphi$ is its discrete counterpart. From equations \eqref{eq:omega} and \eqref{eq:roots} we obtain correspondingly:
\begin{align}\label{eq:118}
  \phi\ &=\ \arg(\ue^{-\ui\,\omega\,\tau})\ \equiv\ -\omega\,\tau\ =\ \pm\,\frac{c\,\aleph\,\xi}{\sqrt{1\ +\ \dfrac{\varsigma^2\,\xi^2}{3}}}\,, \quad \xi\ \in\ [0,\,\pi]\,, \\
  \varphi\ &=\ \arg\uprho\ =\ \pm\,\arccos\Bigl[\,\Bigl(1\ -\ \frac{\daleth\,(1\ +\ \theta)}{2\,\hslash}\Bigr)/\abs{\uprho}\,\Bigr]\,,\label{eq:119}
\end{align}
In other words, the phase change $\phi$ is predicted by the `exact' SGN equations properties, while $\varphi$ comes from the approximate dynamics as predicted by the predictor--corrector scheme. Since we are interested in long wave modelling, we can consider \textsc{Taylor} expansions of the phase shifts in the limit $\xi\ \to\ 0$ (assuming that $\varsigma$ and $\aleph$ are kept constant):
\begin{align*}
  \phi\ &=\ \pm\,\Bigl[\,c\,\aleph\,\xi\ -\ \frac{c\aleph}{6}\;\varsigma^2\,\xi^3\ +\ \O(\xi^4)\,\Bigr]\,, \\
  \varphi\ &=\ \pm\,\Bigl[\,c\,\aleph\,\xi\ +\ \frac{c\,\aleph}{6}\,\bigl((c\,\aleph)^2\,(3\,\theta\ +\ 1)\ -\ 1\ -\ \varsigma^2\bigr)\,\xi^3\ +\ \O(\xi^4)\,\Bigr]\,.
\end{align*}
The asymptotic expression for the phase error is obtained by subtracting above expressions:
\begin{equation*}
  \Delta\varphi\ =\ \mp\,\frac{c\,\aleph}{6}\;\bigl[\,(c\,\aleph)^2\,(3\,\theta\ +\ 1)\ -\ 1\,\bigr]\,\xi^3\ +\ \O(\xi^4)\,.
\end{equation*}
From the last relation one can see that the leading part of the phase error has the same asymptotic order as the `physical' dispersion of the SGN equations. In general, this result is not satisfactory. However, this situation can be improved if for the given scheme parameter $\theta\ \geq\ 0$, the \textsc{Courant} ratio $\aleph$ is chosen according to the following formula:
\begin{equation*}
  c\,\aleph\ =\ \frac{1}{\sqrt{1\ +\ 3\,\theta}}\,.
\end{equation*}
In this case the numerical phase error will be one order lower than the physical dispersion of the SGN system.

\begin{figure}
  \centering
  \includegraphics[width=0.69\textwidth]{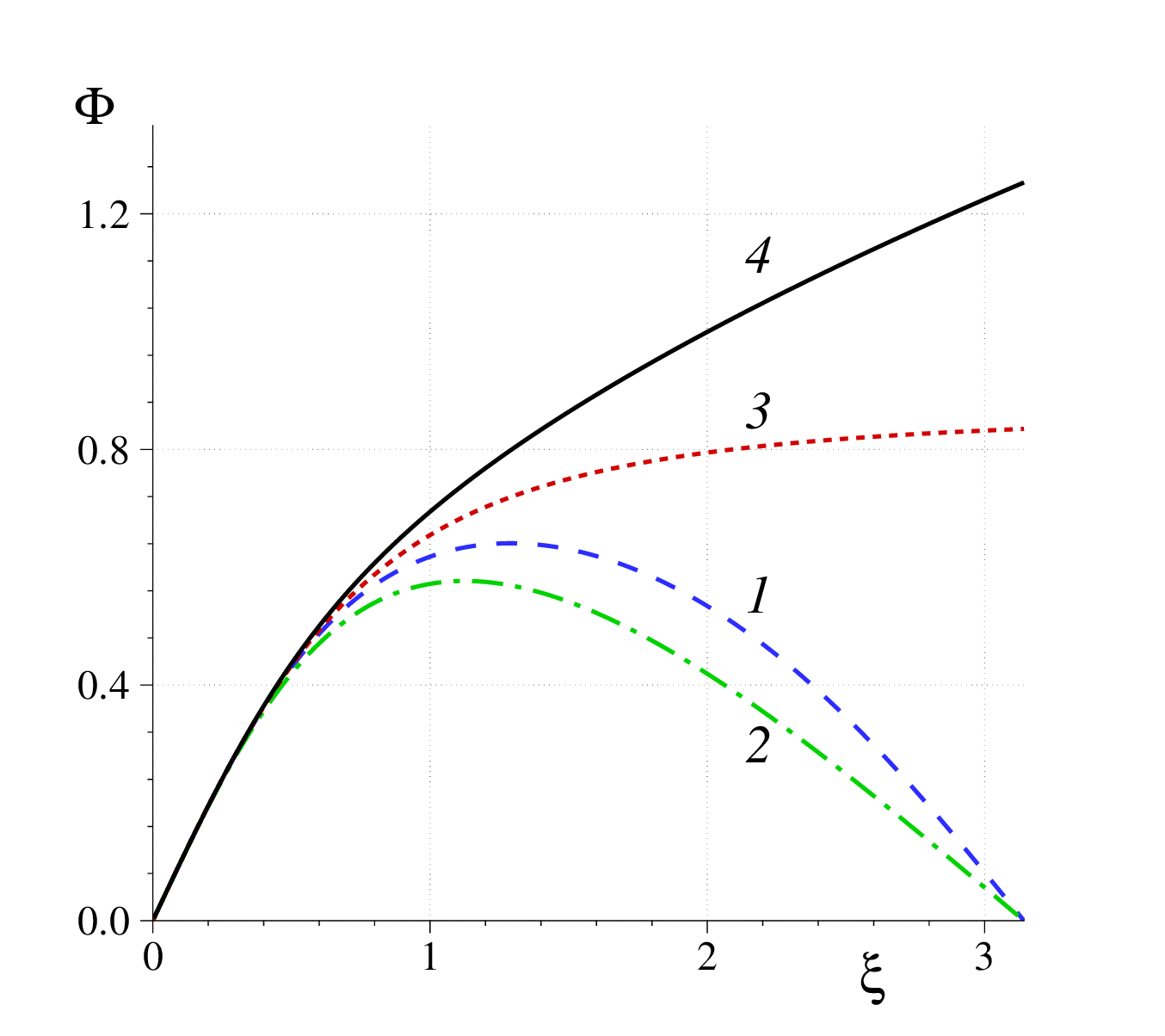}
  \caption{\small\em Phase shifts in different models: (1) predictor--corrector scheme; (2) \textsc{Peregrine}'s numerical scheme \cite{Peregrine1966}; (3) the SGN equations; (4) full \textsc{Euler} equations.}
  \label{fig:disp}
\end{figure}

In Figure~\ref{fig:disp} we represent graphically phase shifts predicted by various models. The dashed line (1) is the phase shift of the predictor--corrector scheme given by equation \eqref{eq:119} (taken with $+$ sign) for the parameters values $\theta\ =\ 0\,$, $c\,\aleph\ =\ 1\,$, $\varsigma\ =\ 2\,$. The continuous dispersion relation are shown with the dotted line (3) (the SGN equations, formula \eqref{eq:118}) and the solid line (4) (full \textsc{Euler} equations):
\begin{equation*}
  \phi_{\mathrm{Euler}}\ =\ \pm\, c\,\aleph\,\xi\;\sqrt{\frac{\tanh(\varsigma\,\xi)}{\varsigma\,\xi}}\,.
\end{equation*}
It can be seen that our predictor--corrector scheme provides a better approximation to the dispersion relation than the scheme proposed by \textsc{Peregrine} \cite{Peregrine1966} (dash-dotted line (2) in Figure~\ref{fig:disp}). The analysis of the discrete dispersion relation of \textsc{Peregrine}'s scheme is not given here, but we provide only the final result for the phase change:
\begin{equation*}
  \phi_{\mathrm{Peregrine}}\ =\ \pm\,\arccos\Bigl(1\ -\ \frac{\beth^{\,2}}{2\,\hslash}\Bigr)\,.
\end{equation*}
In Figure~\ref{fig:disp} one can see that the predictor--corrector scheme (curve (1)) approximates well the dispersion relation of the SGN equations (curve (3)) up to $\xi\ =\ k\cdot \dx\ \apprle \frac{\pi}{4}$. In terms of the wave length $\lambda$ we obtain that $\lambda\ \apprge\ 8\,\dx$ and for $\varsigma\ =\ 2$ we obtain the inequality $\lambda\ \apprge\ 4\,d$. So, as the main result of the present analysis we conclude that our scheme is able to propagate accurately water waves whose length is four times longer than the mean water depth $d$.

\section{Numerical results}
\label{sec:res}

Below we present a certain number of test cases which aim to validate and illustrate the performance of the numerical scheme described above along with our implementation of this method.

\subsection{Solitary wave propagation over the flat bottom}

As we saw above in Section~\ref{sec:flat}, in a special case of constant water depth $h(x,\,t)\ =\ d$ the SGN equations admit solitary wave solutions (given by explicit simple analytical formulas) which propagate with constant speed without changing their shapes.

\subsubsection{Uniform grid}

These analytical solutions can be used to estimate the accuracy of the fully discrete numerical scheme. Consequently, we take a sufficiently large domain $[0,\,\ell]$ with $\ell\ =\ 80$. In this Section all lengths are relative to the water depth $d$, and time is scaled with $\sqrt{g/d}$. For instance, if the solitary wave amplitude $\alpha\ =\ 0.7$, then $\alpha_{\,d}\ =\ 0.7\,d$ in dimensional variables. So, the solitary wave is initially located at $x_0\ =\ 40$. In computations below we take a solitary wave of amplitude $\alpha\ =\ 0.4$. In general, the SGN travelling wave solutions approximate fairly well those of the full \textsc{Euler} model up to amplitudes $\alpha\ \apprle\ \frac{1}{2}$ (see \cite{Duran2013} for comparisons).

\begin{figure}
  \centering
  \subfigure[]{\includegraphics[width=0.485\textwidth]{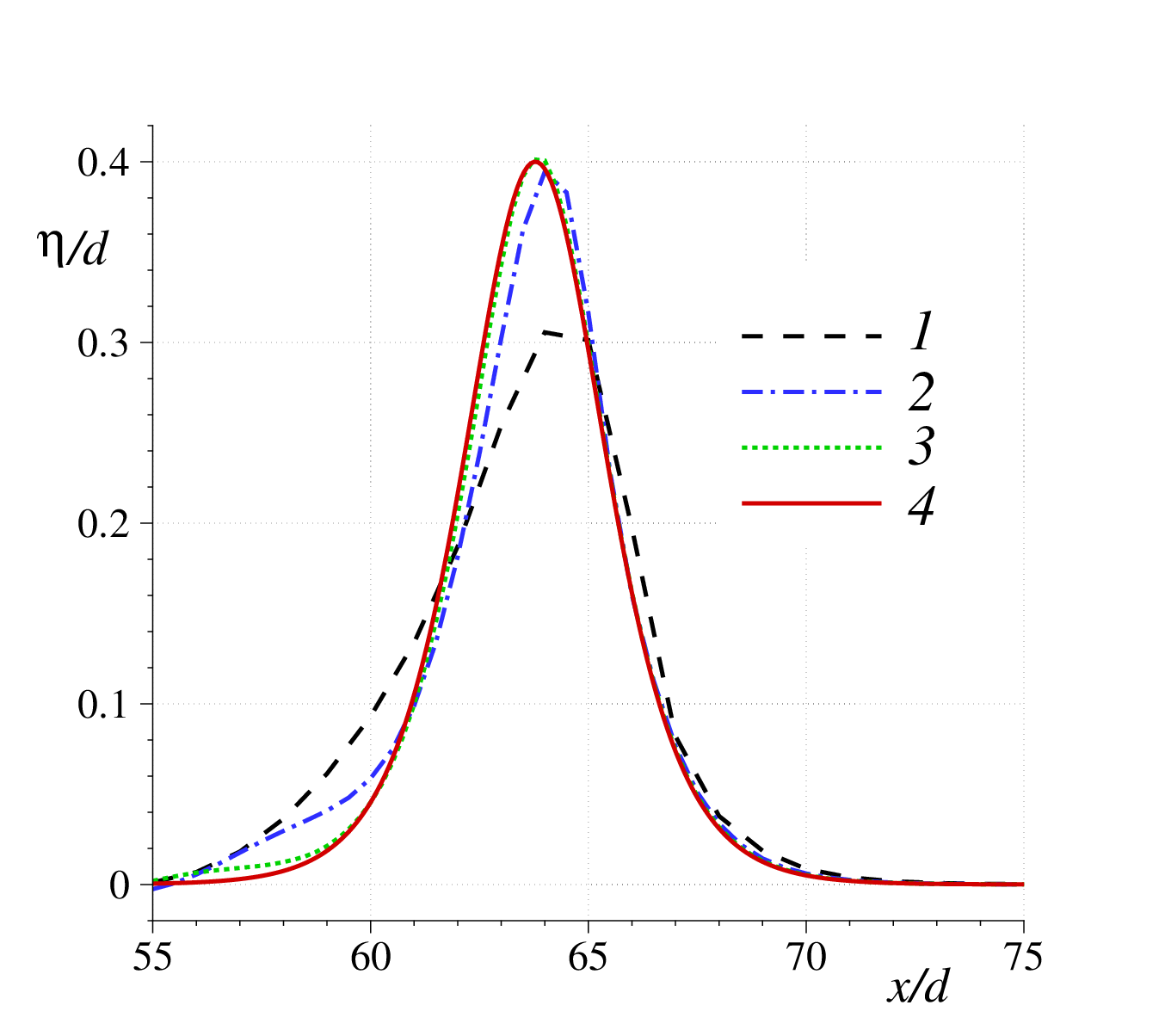}}
  \subfigure[]{\includegraphics[width=0.485\textwidth]{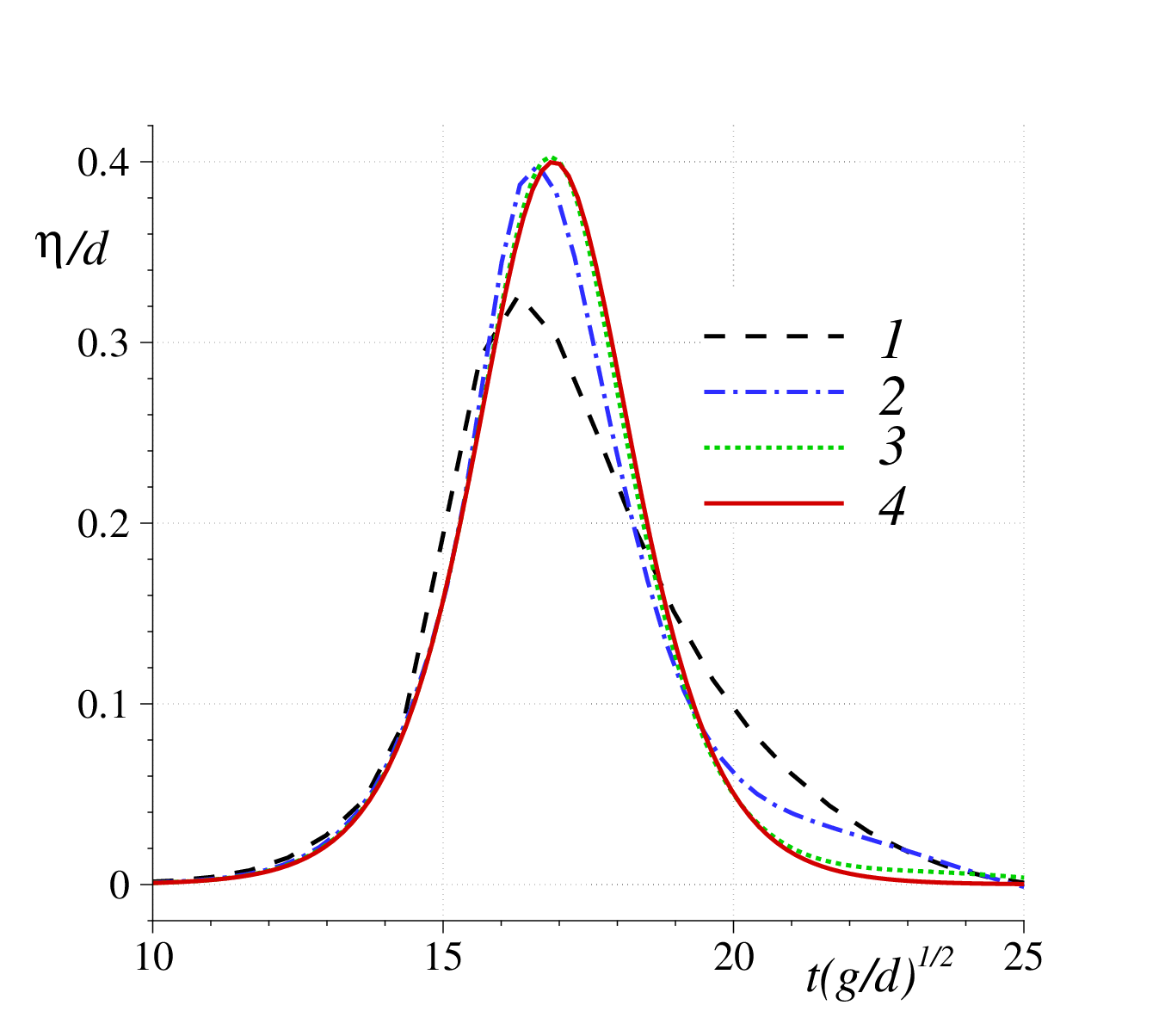}}
  \caption{\small\em Propagation of a solitary wave over the flat bottom: (a) free surface profile at $t\ =\ 20$; (b) wave gauge data at $x\ =\ 60$. Various lines denote: (1) --- $N\ =\ 80$, (2) --- $N\ =\ 160$, (3) --- $N\ =\ 320$, (4) --- the exact analytical solution given by formula \eqref{eq:sol}.}
  \label{fig:sw1}
\end{figure}

In Figure~\ref{fig:sw1} we show a zoom on free surface profile (\textit{a}) at $t\ =\ 20$ and wave gauge data (\textit{b}) in a fixed location $x\ =\ 60$ for various spatial (and uniform) resolutions. By this time, the solitary wave propagated the horizontal distance of $20$ mean water depths. It can be seen that the numerical solution converges to the analytical one.

In order to quantify the accuracy of the numerical solution we measure the relative $l_{\infty}$ discrete error:
\begin{equation*}
  \norm{\eps_h}_{\infty}\ \eqdef\ \alpha^{-1}\,\norm{\eta_h\ -\ \eta}_{\infty}\,,
\end{equation*}
where $\eta_h$ stands for the numerical and $\eta$ -- for the exact free surface profiles. The factor $\alpha^{-1}$ is used to obtain the dimensionless error. Then, the order of convergence $\p$ can be estimated as
\begin{equation*}
  \p\ \simeq\ \log_2\biggl\{\,\frac{\norm{\eps_h}_{\infty}}{\norm{\eps_{h/2}}_{\infty}}\,\biggr\}\,.
\end{equation*}
The numerical results in Table~\ref{tab:conv} indicate that $\p\ \to\ 2$, when $N\ \to\ +\infty$. This validates the proposed scheme and the numerical solver.

\begin{table}
  \centering
  \setlength{\tabcolsep}{2em}
  \begin{tabular}{c|c|c|c}
  \hline\hline
  $N$ & $\varsigma$ & $\norm{\eps_h}_{\infty}$ & $\p$ \\
  \hline\hline
  80      & 1   & $0.2442$               & --- \\
  160     & 2   & $0.1277$               & 0.94 \\
  320     & 4   & $0.3344\times 10^{-1}$ & 1.93 \\
  640     & 8   & $0.8639\times 10^{-2}$ & 1.95 \\
  1280    & 16  & $0.2208\times 10^{-2}$ & 1.97 \\
  2560    & 32  & $0.5547\times 10^{-3}$ & 1.99 \\
  \hline\hline
  \end{tabular}
  \bigskip
  \caption{Numerical estimation of the convergence order for the analytical solitary wave propagation test case. The parameter $\varsigma\ =\ \frac{d}{\dx}$ characterizes the mesh resolution relative to the mean water depth $d$.}
  \label{tab:conv}
\end{table}

\subsubsection{Adaptive grid}

\begin{figure}
  \centering
  \subfigure[]{%
  \adjustbox{trim={0.025\width} {0.02\height} {0.10\width} {0.05\height}, clip}%
  {\includegraphics[width=0.33\textwidth]{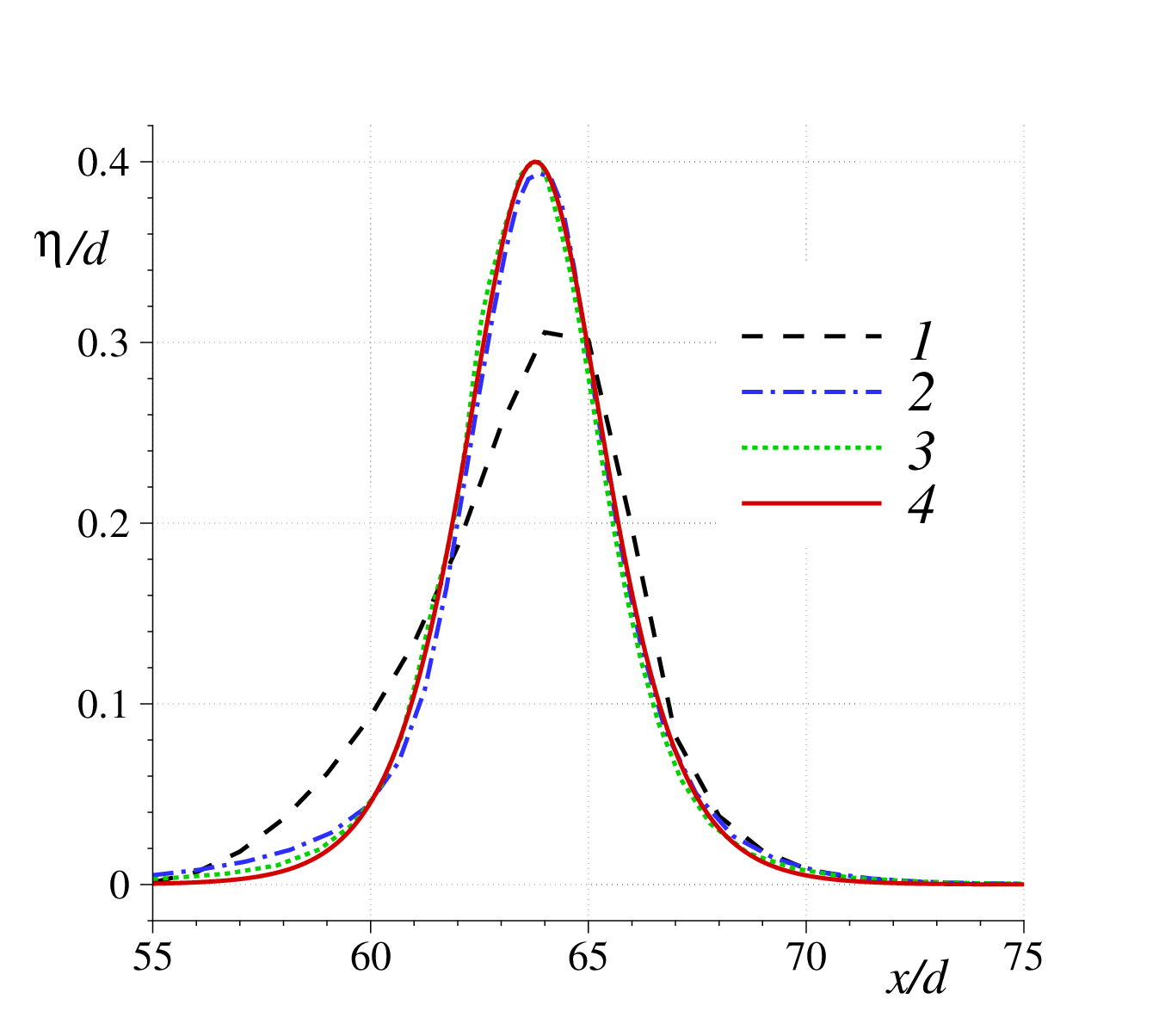}}}
  \subfigure[]{%
  \adjustbox{trim={0.025\width} {0.02\height} {0.10\width} {0.05\height}, clip}%
  {\includegraphics[width=0.325\textwidth]{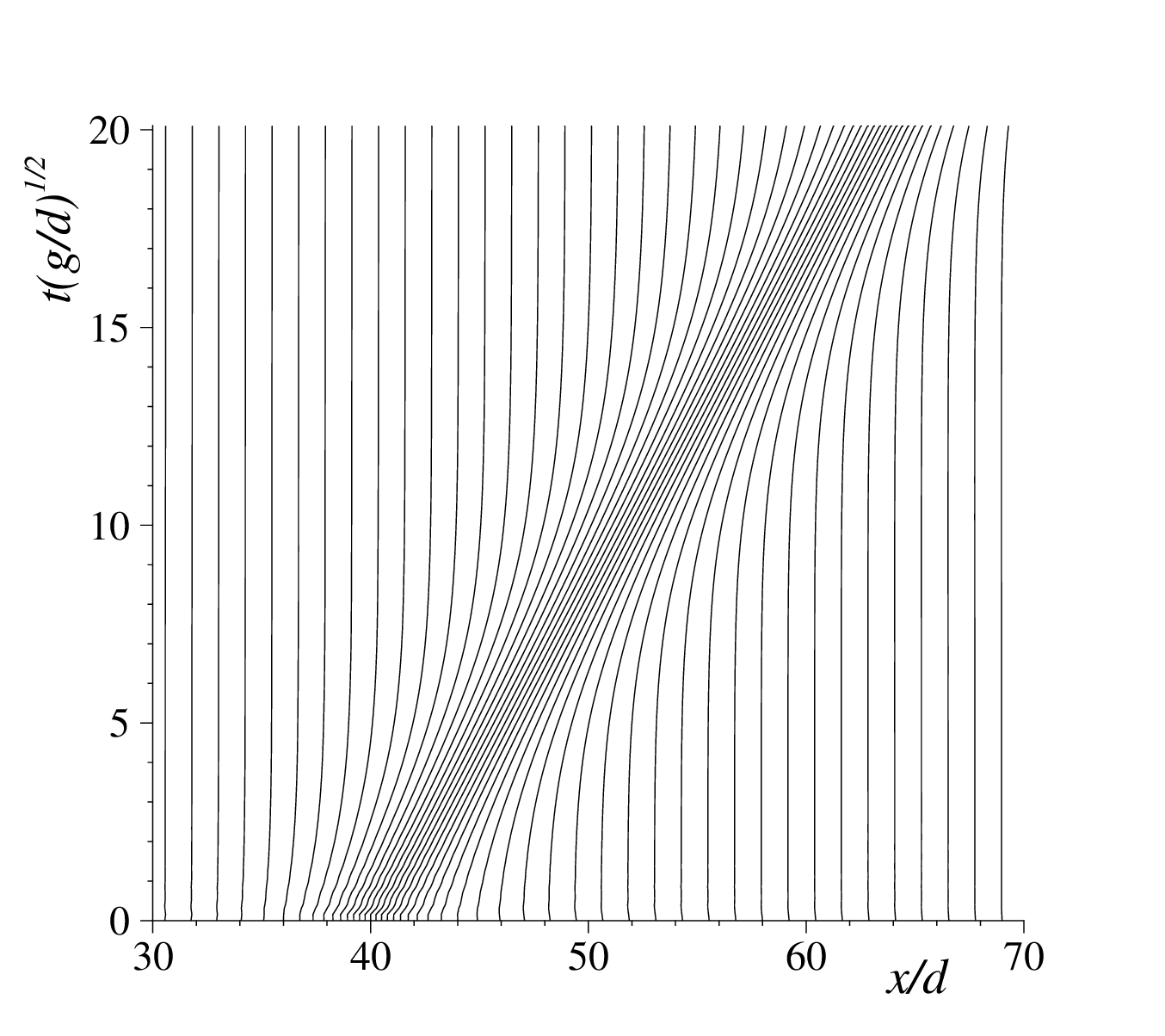}}}
  \subfigure[]{%
  \adjustbox{trim={0.025\width} {0.02\height} {0.10\width} {0.05\height}, clip}%
  {\includegraphics[width=0.33\textwidth]{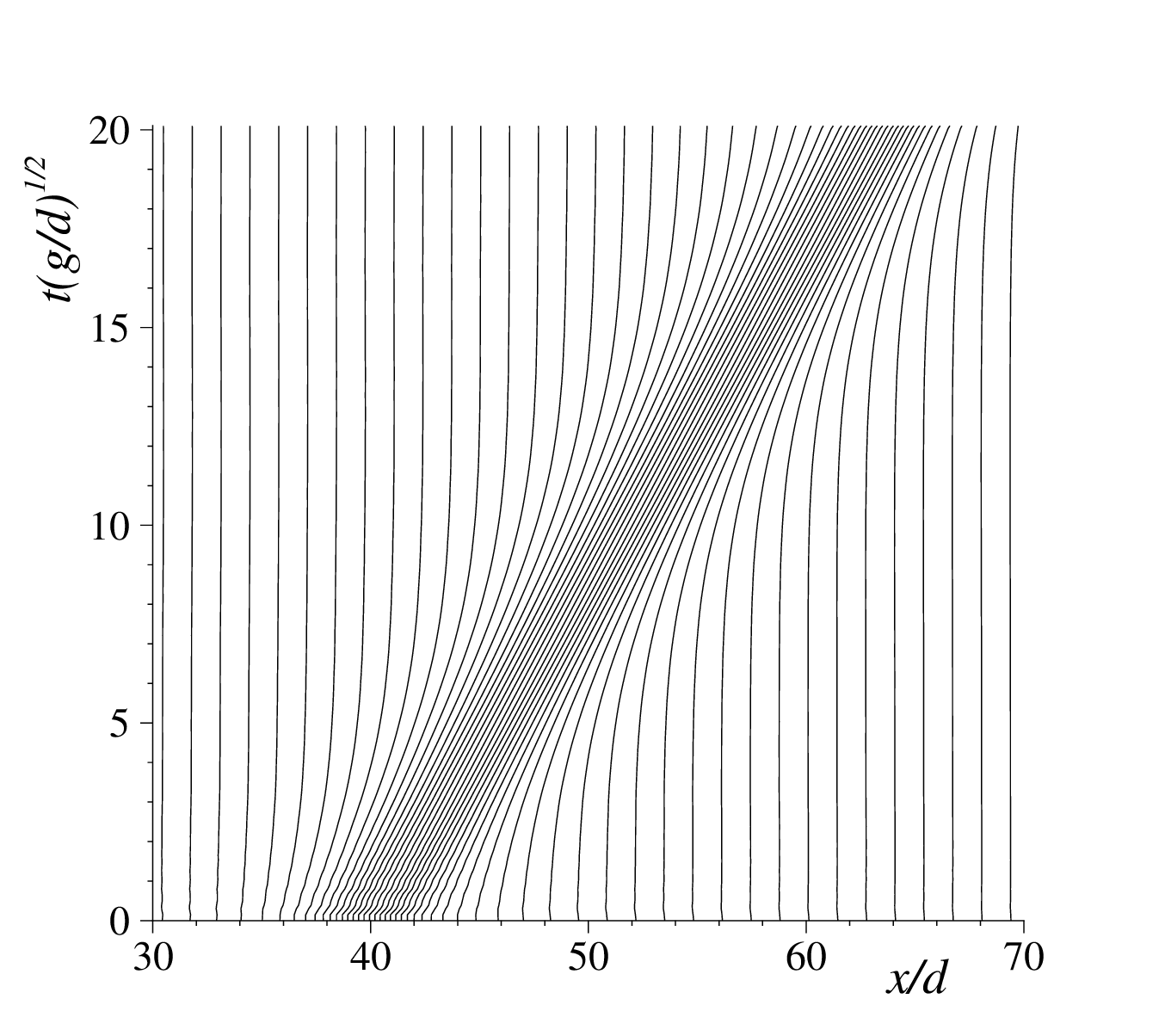}}}
  \caption{\small\em Propagation of a solitary wave over the flat bottom simulated with moving adapted grids: (a) free surface profile at $t\ =\ 20$; (b) trajectory of some grid points predicted with monitor function $\om_0(x,\,t)\,$; (c) the same but with monitor function $\om_1(x,\,t)\,$. On panel (a) the lines are defined as: (1) --- numerical solution on a uniform fixed grid; (2) --- numerical solution predicted with monitor function $\om_0(x,\,t)\,$; (3) --- the same with $\om_1(x,\,t)\,$; (4) --- exact analytical solution.}
  \label{fig:sw2}
\end{figure}

In order to show the performance of the adaptive algorithm, we adopt two monitor functions in our computations:
\begin{align}\label{eq:om0}
  \om_0[\,\eta\,]\,(x,\,t)\ &=\ 1\ +\ \coef_0\,\abs{\eta(x,\,t)}\,, \\
  \om_1[\,\eta\,](x,\,t)\ &=\ 1\ +\ \coef_0\,\abs{\eta(x,\,t)}\ +\ \coef_1\,\abs{\eta_x(x,\,t)}\,, \label{eq:om1}
\end{align}
where $\coef_{\,0,\,1}\ \geq\ 0$ are some positive constants. In numerical simulations we use $\coef_{\,0}\ =\ \coef_{\,1}\ =\ 10$ and only $N\ =\ 80$ grid points. Above we showed that numerical results are rather catastrophic when these $80$ grid points are distributed uniformly (see  Figure~\ref{fig:sw1}). Numerical results on adaptive moving grids obtained with monitor functions $\om_{\,0,\,1}(x,\,t)$ are shown in Figure~\ref{fig:sw2}. The monitor function $\om_{\,0}(x,\,t)$ ensures that points concentrate around the wave crest, leaving the areas in front and behind relatively rarefied. The visual comparison of panels \ref{fig:sw2}(\textit{b}) and \ref{fig:sw2}(\textit{c}) shows that the inclusion of the spatial derivative $\eta_x$ into the monitor function $\om_{1}(x,\,t)$ yields the increase of dense zones around the wave crest. With an adaptive grid involving only $N\ =\ 80$ points we obtain a numerical solution of quality similar to the uniform grid with $N\ =\ 320$ points.

\subsection{Solitary wave/wall interaction}

For numerous practical purposes in Coastal Engineering it is important to model correctly wave/structure interaction processes \cite{Peregrine2003}. In this Section we apply the above proposed numerical algorithm to the simulation of a simple solitary wave/vertical wall interaction. The reason is two-fold:
\begin{enumerate}
  \item Many coastal structures involve vertical walls as building elements,
  \item This problem is well studied by previous investigators and, consequently, there is enough available data/results for comparisons
\end{enumerate}
We would like to underline that this problem is equivalent to the head-on collision of two equal solitary waves due to simple symmetry considerations. This `generalized' problem was studied in the past using experimental \cite{Maxworthy1976, Seabra-Santos1989}, numerical \cite{Chan1970a, Fenton1982} and analytical techniques \cite{Su1980, Mirie1982, Byatt-Smith1988}. More recently this problem gained again some interest of researchers \cite{Cooker1997, Madsen2002, Chambarel2009, Dutykh2011e, Dutykh2010e, Dutykh2011a, Carbone2013, Touboul2014}. Despite the simple form of the obstacle, the interaction process of sufficiently large solitary waves with it takes a highly non-trivial character as it will be highlighted below.

\begin{figure}
  \centering
  \subfigure[]{\includegraphics[width=0.495\textwidth]{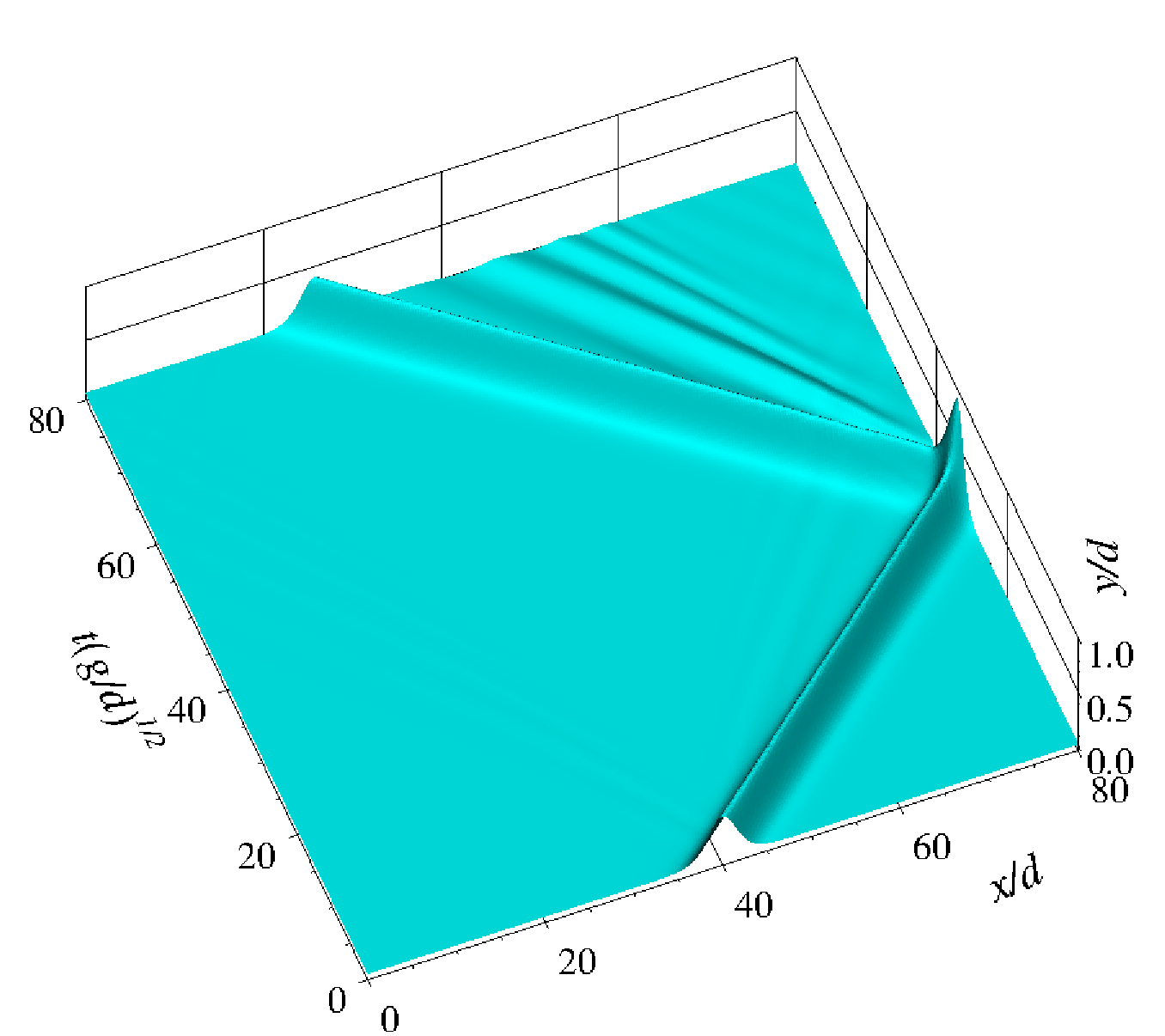}}
  \subfigure[]{\includegraphics[width=0.495\textwidth]{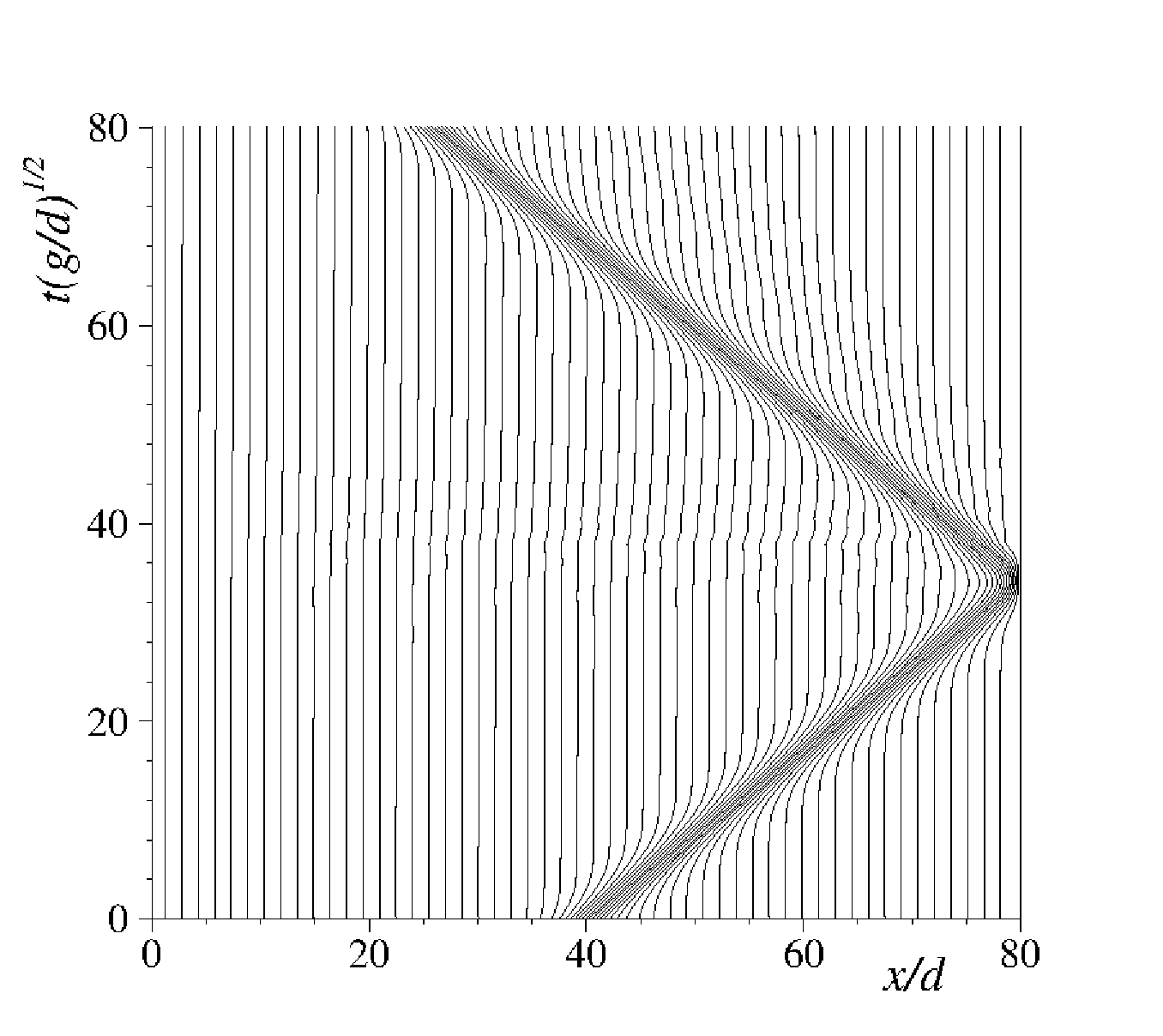}}
  \caption{\small\em Solitary wave (amplitude $\alpha\ =\ 0.4$)/vertical wall interaction in the framework of the SGN equations: (a) space-time plot of the free surface elevation; (b) nodes trajectories. For the sake of clarity every 5\up{th} node is shown only, the total number of nodes $N\ =\ 320$.}
  \label{fig:wall}
\end{figure}

Figure~\ref{fig:wall}(\textit{a}) shows the free surface dynamics as it is predicted by the SGN equations solved numerically using the moving grid with $N\ =\ 320$ nodes. The initial condition consists of an exact analytical solitary wave \eqref{eq:sol} of amplitude $\alpha\ =\ 0.4$ moving rightwards to the vertical wall (where the wall boundary condition $u\ =\ 0$ is imposed\footnote{The same condition is imposed on the left boundary as well, even if during our simulation time there are no visible interactions with the left wall boundary.} on the velocity, for the pressure see Section~\ref{sec:bcond}). The computational domain is chosen to be sufficiently large $[0,\,\ell]\ =\ [0,\,80]$, so there is no interaction with the boundaries at $t\ =\ 0$. Initially the solitary wave is located at $x_0\ =\ 40$ (right in the middle). The bottom is flat $h(x,\,t)\ =\ d\ =\ \const$ in this test case. From Figure~\ref{fig:wall}(\textit{a}) it can be clearly seen that the reflection process generates a train of weakly nonlinear waves which propagate with different speeds in agreement with the dispersion relation. The moving grid was constructed using the monitor function $\om_1(x,\,t)$ from the previous Section (see the definition in equation \eqref{eq:om1}).
with $\coef_0\ =\ \coef_1\ =\ 10$. The resulting trajectories of mesh nodes are shown in Figure~\ref{fig:wall}(\textit{b}). The grid is clearly refined around the solitary wave and nodes follow it. Moreover, we would like to note also a slight mesh refinement even in the dispersive tail behind the reflected wave (it is not clearly seen in Figure~\ref{fig:wall}(\textit{b}) since we show only every 5\up{th} node).

\begin{figure}
  \centering
  \includegraphics[width=0.76\textwidth]{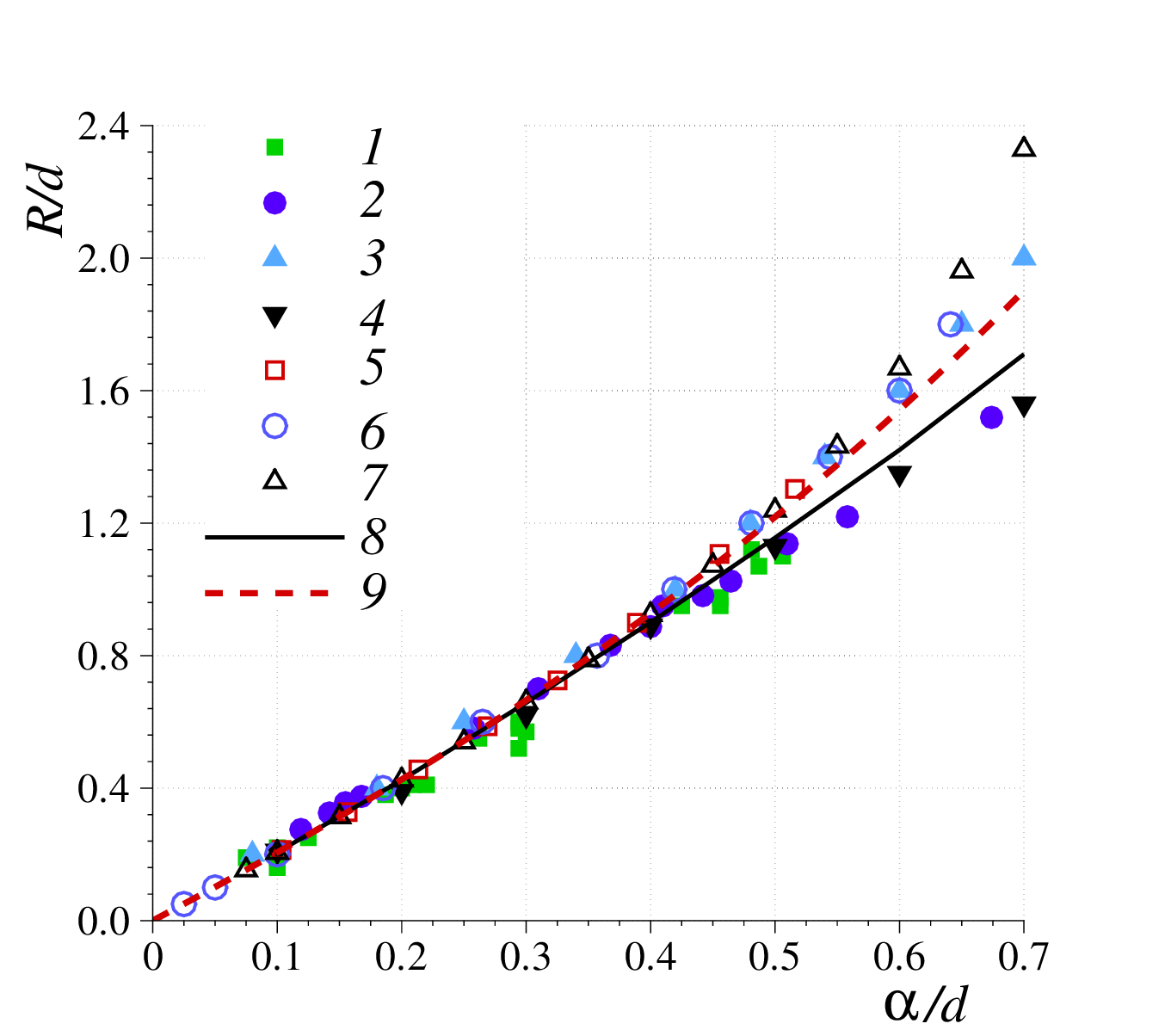}
  \caption{\small\em Dependence of the maximal run-up $\Ru$ on the amplitude $\alpha$ of the incident solitary wave. Experimental data: (1) --- \cite{Zagryadskaya1980}, (2) --- \cite{Maxworthy1976}, (3) --- \cite{Davletshin1984}, (4) --- \cite{Manoylin1989}. Numerical data: (5) --- \cite{Fenton1982}, (6) --- \cite{Chan1970a}, (7) --- \cite{Cooker1997}. The solid line (8) --- our numerical results, the dashed line (9) --- the analytical prediction \eqref{eq:predict}.}
  \label{fig:compar}
\end{figure}

One of the main interesting characteristics that we can compute from these numerical experiments is the \emph{maximal} wave run-up $\Ru$ on the vertical wall:
\begin{equation*}
  \Ru\ \eqdef\ \sup_{0\ \leq\ t\ \leq\ T} \{\eta(\ell,\,t)\}\,.
\end{equation*}
The $\sup$ is taken in some time window when the wave/wall interaction takes place. For the class of incident solitary wave solutions it is clear that \emph{maximal} run-up $\Ru$ will depend on the (dimensionless) solitary wave amplitude $\alpha$. In \cite{Su1980} the following asymptotic formula was derived in the limit $\alpha\ \to\ 0$:
\begin{equation}\label{eq:predict}
  \Ru\,(\alpha)\ =\ 2\,\alpha\,\bigl[\,1\ +\ \fourth\,\alpha\ +\ \threeeights\,\alpha^2\,\bigr]\ +\ \O(\alpha^4)\,.
\end{equation}
The last approximation was already checked against full the \textsc{Euler} simulations \cite{Fenton1982, Cooker1997} and even laboratory experiments \cite{Maxworthy1976}. Figure~\ref{fig:compar} shows the dependence of the maximal run-up $\Ru$ on the incident solitary wave amplitude $\alpha$ as it is predicted by our numerical model, by formula \eqref{eq:predict} and several other experimental \cite{Zagryadskaya1980, Maxworthy1976, Davletshin1984, Manoylin1989} and numerical \cite{Fenton1982, Chan1970a, Cooker1997} studies. In particular, one can see that almost all models agree fairly well up to the amplitudes $\alpha\ \apprle\ 0.4$. Then, there is an apparent `separation' of data in two branches. Again, our numerical model gives a very good agreement with experimental data from \cite{Maxworthy1976, Manoylin1989, Zagryadskaya1980} up to the amplitudes $\alpha\ \apprle\ 0.7$.

\subsubsection{Wave action on the wall}

The nonlinear dispersive SGN model can be used to estimate also the wave force exerted on the vertical wall. Moreover, we shall show below that this model is able to capture the non-monotonic behaviour of the force when the incident wave amplitude is increased. This effect was first observed experimentally \cite{Zagryadskaya1980} and then numerically \cite{Zheleznyak1985a}.

\begin{figure}
  \centering
  \subfigure[]{\includegraphics[width=0.495\textwidth]{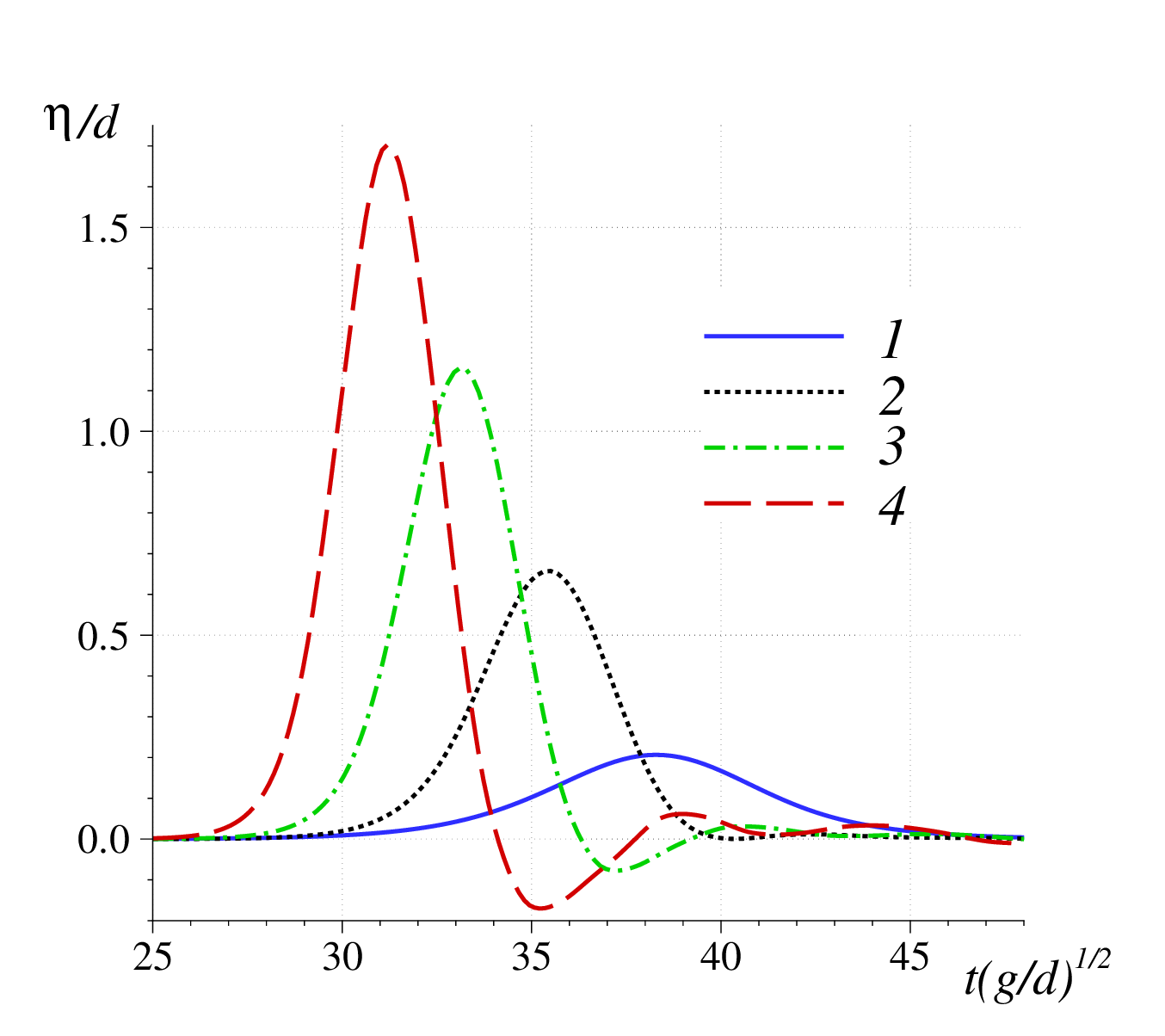}}
  \subfigure[]{\includegraphics[width=0.495\textwidth]{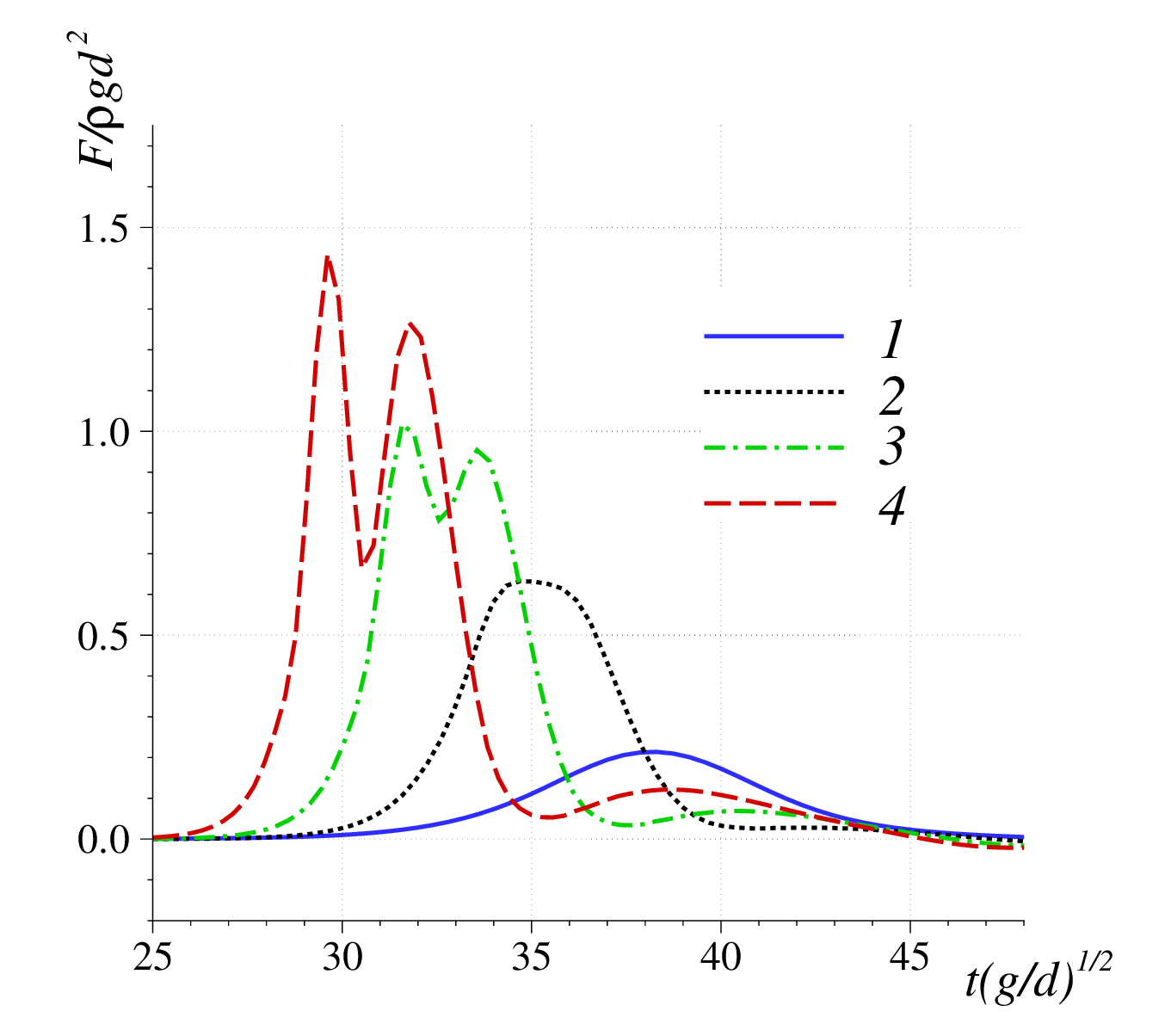}}
  \caption{\small\em Solitary wave/vertical wall interaction: (a) time series of wave run-up on the wall; (b) dynamic wave loading on the wall. Different lines correspond to different incident solitary wave amplitudes: (1) --- $\alpha\ =\ 0.1$, (2) --- $\alpha\ =\ 0.3$, (3) --- $\alpha\ =\ 0.5$, (4) --- $\alpha\ =\ 0.7$.}
  \label{fig:force}
\end{figure}

For the 2D case with flat bottom the fluid pressure $p(x,\,y,\,t)$ can be expressed:
\begin{equation}\label{eq:flatP}
  \frac{p(x,\,y,\,t)}{\rho}\ =\ g\bigl(\,\H\ -\ (y + d)\,\bigr)\ -\ \Bigl[\,\frac{\H^{\,2}}{2}\ -\ \frac{(y + d)^2}{2}\,\Bigr]\,\Rr_1\,, \quad -d\ \leq\ y\ \leq\ \eta(x,\,t)\,,
\end{equation}
with $\Rr_1\ \eqdef\ u_{xt}\ +\ u\,u_{xx}\ -\ u_x^{2}$. The \emph{horizontal} wave loading exerted on the vertical wall located at $x\ =\ \ell$ is given by the following integral:
\begin{equation*}
  \frac{\Ff_0 (t)}{\rho}\ =\ \int_{-d}^{\eta(\ell,t)} p(\ell,\,y,\,t)\;\ud y\ =\ \frac{g\,\H^{\,2}}{2}\ -\ \frac{\H^{\,3}}{3}\;\bar{\Rr}_1\,,
\end{equation*}
where due to boundary conditions $\bar{\Rr}_1\ =\ u_{xt}\ -\ u_x^{2}$. After removing the hydrostatic force, we obtain the dynamic wave loading computed in our simulations:
\begin{equation*}
  \frac{\Ff(t)}{\rho}\ =\ g\,\Bigl[\,\frac{\,\H^{\,2}}{2}\ -\ \frac{d^2}{2}\,\Bigr]\ -\ \frac{\H^{\,3}}{3}\;\bar{\Rr}_1\,.
\end{equation*}
The expression for corresponding tilting moment can be found in \cite[Remark~3]{Dutykh2011a}. Figure~\ref{fig:force} shows the wave elevation (\textit{a}) and the dynamic wave loading (\textit{b}) on the vertical wall. From Figure~\ref{fig:force}(\textit{b}) it can be seen that the force has one maximum for small amplitude solitary waves. However, when we gradually increase the amplitude (\ie $\alpha\ \apprge\ 0.4$), the second (local) maximum appears. For such large solitary waves a slight run-down phenomenon can be noticed in Figure~\ref{fig:force}(\textit{a}). We reiterate that this behaviour is qualitatively and quantitatively correct comparing to the full \textsc{Euler} equations \cite{Cooker1997, Chambarel2009}. However, the complexity of the nonlinear dispersive SGN model and, consequently, the numerical algorithm to solve it, is much lower.

\subsection{Solitary wave/bottom step interaction}
\label{sec:step}

Water waves undergo continuous changes while propagating over general uneven bottoms. Namely, the wave length and wave amplitude are modified while propagating over bottom irregularities. Such transformations have been studied in the literature \cite{Dingemans1997, Kurkin2015}. In the present Section we focus on the process of a Solitary Wave (SW) transformation over a bottom step. In the early work by \textsc{Madsen} \& \textsc{Mei} (1969) \cite{Madsen1969} it was shown using long wave approximation that a solitary wave can be disintegrated into a finite number of SWs with decreasing amplitudes while passing over an underwater step. This conclusion was supported in \cite{Madsen1969} by laboratory data as well. This test case was used later in many works, see \eg \cite{Dutykh2007, Chazel2007, Lai2010}.

We illustrate the behaviour of the adaptive numerical algorithm as well as the SGN model on the solitary wave/bottom interaction problem. The bottom bathymetry is given by the following discontinuous function:
\begin{equation*}
  y\ =\ -h(x)\ =\ \begin{dcases}
    -h_0\,, & 0\ \leq\ x\ \leq\ x_s\,, \\
    -h_s\,, & x_s\ <\ x\ \leq\ \ell\,, \\
  \end{dcases}
\end{equation*}
where $\ell$ is the numerical wave tank length, $h_0$ (respectively $h_s$) are the still water depths on the left (right) of the step located at $x\ =\ x_s$. We assume also that $0\ <\ h_s\ <\ h_0$. The initial condition is a solitary wave located at $x\ =\ x_0$ and propagating rightwards. For the experiment cleanliness we assume that initially the solitary wave does not `feel' the step. In other words it is located sufficiently far from the abrupt change in bathymetry. In our experiment we choose $x_0$ so that $\eta(x_s)\ \apprle\ 0.01\,\alpha$, where $\alpha$ is the SW amplitude. The main parameters in this problem are the incident wave amplitude $\alpha$ and the bottom step jump $\Delta b_s\ =\ h_0\ -\ h_s$. Various theoretical and experimental studies show that a solitary wave undergoes a splitting into a reflected wave and a finite number of solitary waves after passing over an underwater step. See \cite{Kurkin2015} for a recent review on this topic. Amplitudes and the number of solitary waves over the step were determined in \cite{Kabbaj1985} in the framework of the shallow water theory. These expressions were reported later in \cite{Seabra-Santos1987} and this result was improved recently in \cite{Pelinovsky2010}. However, in the vicinity of the step, one may expect important vertical accelerations of fluid particles, which are simplified (or even neglected) in shallow water type theories. Nevertheless, in \cite{Pelinovsky2010} a good agreement of this theory with numerical and experimental data was reported.

There is also another difficulty inherent to the bottom step modelling. In various derivations of shallow water models there is an implicit assumption that the bathymetry gradient $\grad h$ is bounded (or even small $\abs{\grad h}\ \ll\ 1$, \eg in the \textsc{Boussinesq}-type equations \cite{Bristeau2011}). On the other hand, numerical tests and comparisons with the full (\textsc{Euler} and even \textsc{Navier}--\textsc{Stokes}) equations for finite values of $\abs{\grad h}\ \sim\ \O(1)$ show that resulting approximate models have a larger applicability domain than it was supposed at the outset \cite{Bristeau2011}. In the case of a bottom step, the bathymetry function is even discontinuous which is an extreme case we study in this Section.

There are two main approaches to cope with this problem. One consists in running the approximate model directly on discontinuous bathymetry, and the resulting eventual numerical instabilities are damped out by ad-hoc dissipative terms (see \eg references in \cite{Pelinovsky2010}). The magnitude of these terms allows to increase the scheme dissipation, and overall computation appears to be stable. The difficulty of this approach consists in the fine tuning of dissipation, since
\begin{itemize}
  \item Insufficient dissipation will make the computation unstable,
  \item Excessive dissipation will yield unphysical damping of the solution.
\end{itemize}
An alternative approach consists in replacing the discontinuous bathymetry by a smoothed version over certain length $\bigl[\,x_s\ -\ \dfrac{\ell_s}{2},\, x_s\ +\ \dfrac{\ell_s}{2}\,\bigr]\,$, where $\ell_s$ is the smoothing length on which the jump from $h_0$ to $h_s$ is replaced by a smooth variation. For instance, in all numerical computations reported in \cite{Seabra-Santos1987} the smoothing length was chosen to be $\ell_s\ =\ 60$ $\cm$ independently of the water depths before $h_0$ and after $h_s$ the step. In another work \cite{Fedotova1997} the smoothing length was arbitrarily set to $\ell_s\ =\ 20$ $\cm$ independently of other parameters. Unfortunately, in a recent work \cite{Zhang2013} the smoothing procedure was not described at all. Of course, this method is not perfect since the bathymetry is slightly modified. However, one can expect that sufficiently long waves will not `notice' this modification. This assumption was confirmed by the numerical simulations reported in \cite{Fedotova1997, Seabra-Santos1987, Chubarov2000}.

In the present work we also employ the bottom smoothing procedure. However, the smoothing length $\ell_s$ is chosen in order to have a well-posed problem for the elliptic operator \eqref{eq:ell}. For simplicity, we use the sufficient condition \eqref{eq:simple} (obtained under restriction \eqref{eq:restr}), which is not necessarily optimal, but it allows us to invert stably the nonlinear elliptic operator \eqref{eq:ell1}. Namely, the smoothed step has the following analytical expression:
\begin{equation}\label{eq:bsmooth}
  y\ =\ -h(x)\ =\ \begin{dcases}
    -h_0\,, & 0\ \leq\ x\ \leq\ x_s\ -\ \frac{\ell_s}{2}\,, \\
    -h_0\ +\ \frac{\Delta b_s}{2}\cdot\bigl(1\ +\ \sin\upzeta\bigr)\,, & x_s\ -\ \frac{\ell_s}{2}\ \leq\ x\ \leq\ x_s\ +\ \frac{\ell_s}{2}\,, \\
    -h_s\,, & x_s\ +\ \frac{\ell_s}{2}\ \leq\ x\ \leq\ \ell\,,
  \end{dcases}
\end{equation}
where $\upzeta\ \eqdef\ \dfrac{\pi(x - x_s)}{\ell_s}$. For this bottom profile, the inequalities \eqref{eq:restr}, \eqref{eq:simple} take the form:
\begin{align*}
  \frac{\pi\,\Delta b_s}{2\,\ell_s}\;\cos\upzeta\ <&\ 1\,, \quad \forall\,\upzeta\ \in\ \bigl[\,-\frac{\pi}{2},\,\frac{\pi}{2}\,\bigr]\,, \\
  \frac{\pi^2\,\Delta b_s}{2\,\ell_s^{\,2}}\;\sin\upzeta\ >&\ -\,\frac{2}{\dfrac{h_0 + h_s}{2}\ -\ \dfrac{\Delta b_s}{2}\;\sin\upzeta}\,.
\end{align*}
These inequalities have corresponding solutions:
\begin{equation*}
  \ell_s\ >\ \frac{\pi\,\Delta b_s}{2}\,, \qquad
  \ell_s\ >\ \frac{\pi}{2}\;\sqrt{h_0\,\Delta b_s}\,.
\end{equation*}
The last inequalities are verified simultaneously if the second inequality is true. If we assume that the bottom step height $\Delta b_s$ is equal to the half of the water depth before it, then we obtain the following condition:
\begin{equation*}
  \ell_s\ >\ \frac{\pi}{2\,\sqrt{2}}\;h_0\ \approx\ 1.11\, h_0\,.
\end{equation*}
We underline that the last condition is only sufficient and stable numerical computations can most probably be performed even for shorter smoothing lengths $\ell_s$. For instance, we tested the value $\ell_s\ =\ h_0$ and everything went smoothly.

In \cite{Seabra-Santos1987} the results of $80$ experiments are reported for various values of $\alpha$ and $h_0$ (for fixed values of the bottom jump $\Delta b_s\ =\ 10$ $\cm$). In our work we repeated all experiments from \cite{Seabra-Santos1987} using the SGN equations solved numerically with the adaptive predictor--corrector algorithm described above. In general, we obtained a very good agreement with experimental data from \cite{Seabra-Santos1987} in terms of the following control parameters:
\begin{itemize}
  \item number of solitary waves moving over the step,
  \item amplitudes of solitary waves over the step,
  \item amplitude of the (main) reflected wave.
\end{itemize}
We notice that the amplitude of the largest solitary wave over the step corresponds perfectly to the measurements. However, the variation in the amplitude of subsequent solitary waves over the step could reach in certain cases $20\%$.

\begin{remark}
The conduction of laboratory experiments on the solitary wave/bottom step interaction encounters a certain number of technical difficulties \cite{Seabra-Santos1987, Chang2001} that we would like to mention. First of all, the wave maker generates a solitary wave with some dispersive components. Moreover, one has to take the step sufficiently long so that the transmitted wave has enough time to develop into a finite number of visible well-separated solitary waves. Finally, the reflections of the opposite wave flume's wall are to be avoided as well in order not to pollute the measurements. Consequently, the successful conduction of experiments and accurate measurement of wave characteristics requires a certain level of technique. We would like to mention the exemplary experimental work \cite{Hammack2004} on the head-on collision of solitary waves.
\end{remark}

\begin{table}
  \centering
  \begin{tabular}{l|c}
    \hline\hline
    \textit{Parameter} & \textit{Value} \\
    \hline\hline
    Wave tank length, $\ell$ & $35$ $\m$ \\
    Solitary wave amplitude, $\alpha$ & $3.65$ $\cm$ \\
    Solitary wave initial position, $x_0$ & $11$ $\m$ \\
    Water depth before the step, $h_0$ & $20$ $\cm$ \\
    Water depth after the step, $h_s$ & $10$ $\cm$ \\
    Water depth used in scaling, $d$ & $h_0$ \\
    Bottom step jump, $\Delta b_s$ & $10$ $\cm$ \\
    Bottom step location, $x_s$ & $14$ $\m$ \\
    Number of grid points, $N$ & $350$ \\
    Simulation time, $T$ & $17.6$ $\s$ \\
    \hline\hline
  \end{tabular}
  \bigskip
  \caption{\small\em Values of various numerical parameters used in the solitary wave/bottom step interaction test case.}
  \label{tab:params}
\end{table}

\begin{figure}
  \centering
  \subfigure[]{
  \includegraphics[width=0.48\textwidth]{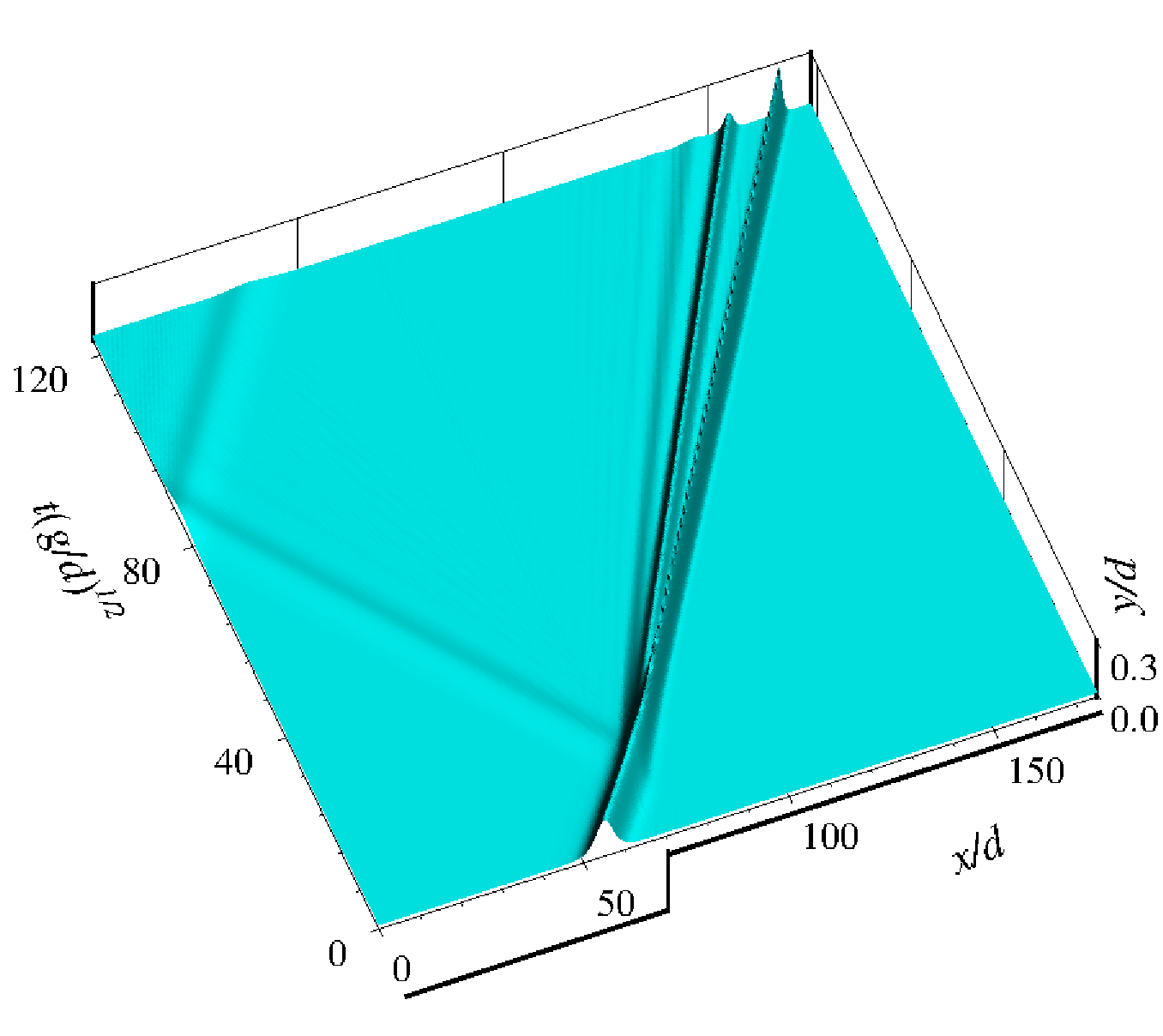}}
  \subfigure[]{
  \includegraphics[width=0.489\textwidth]{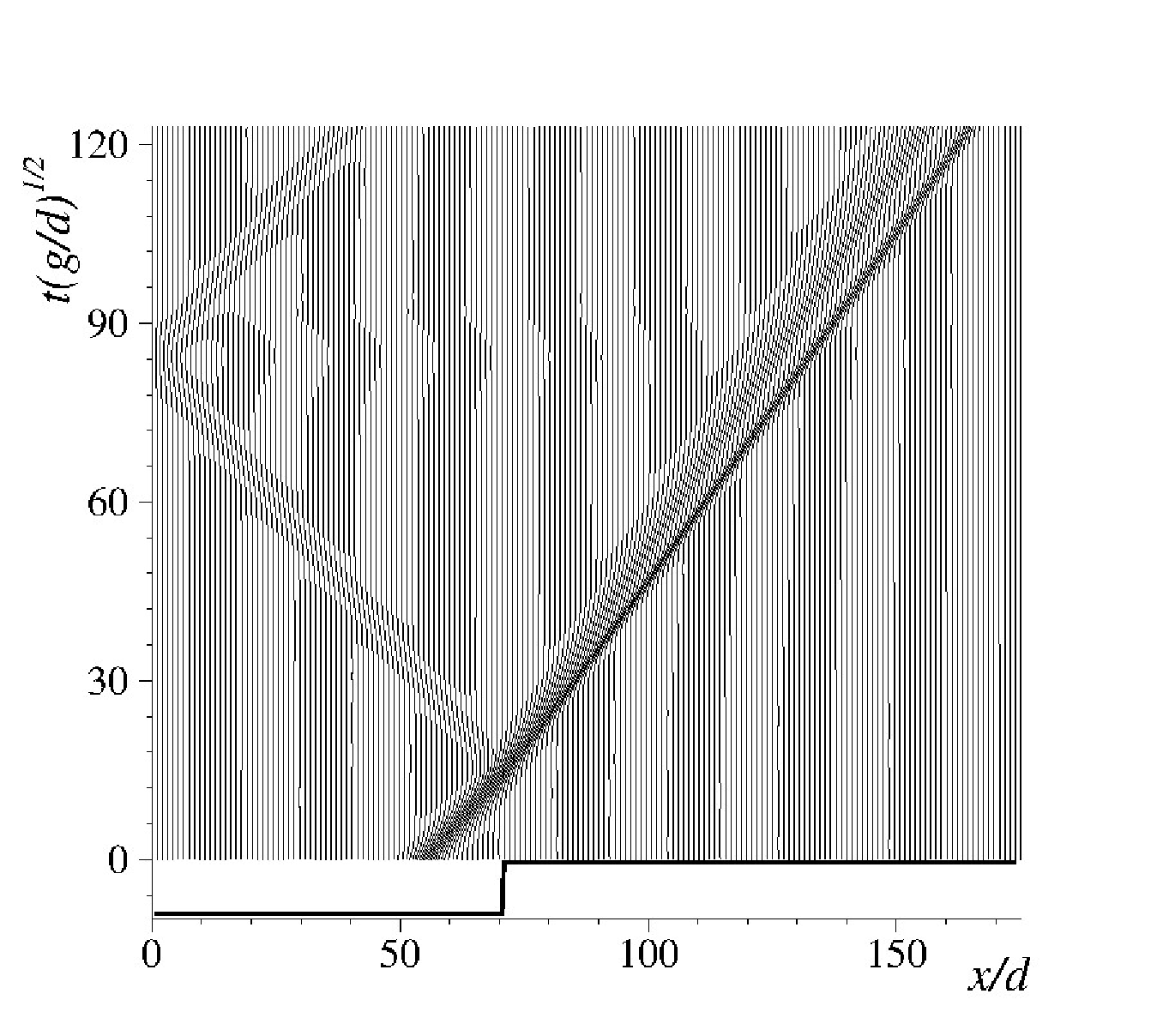}}
  \caption{\small\em Interaction of a solitary wave with an underwater step: (a) space-time plot of the free surface elevation $y\ =\ \eta(x,\,t)$ in the dimensional time interval $[0\; \s,\, 17.6\; \s]$; (b) trajectories of every second grid node. Numerical parameters are provided in Table~\ref{tab:params}.}
  \label{fig:ssurface}
\end{figure}

\begin{figure}
  \centering
  \subfigure[]{
  \includegraphics[width=0.48\textwidth]{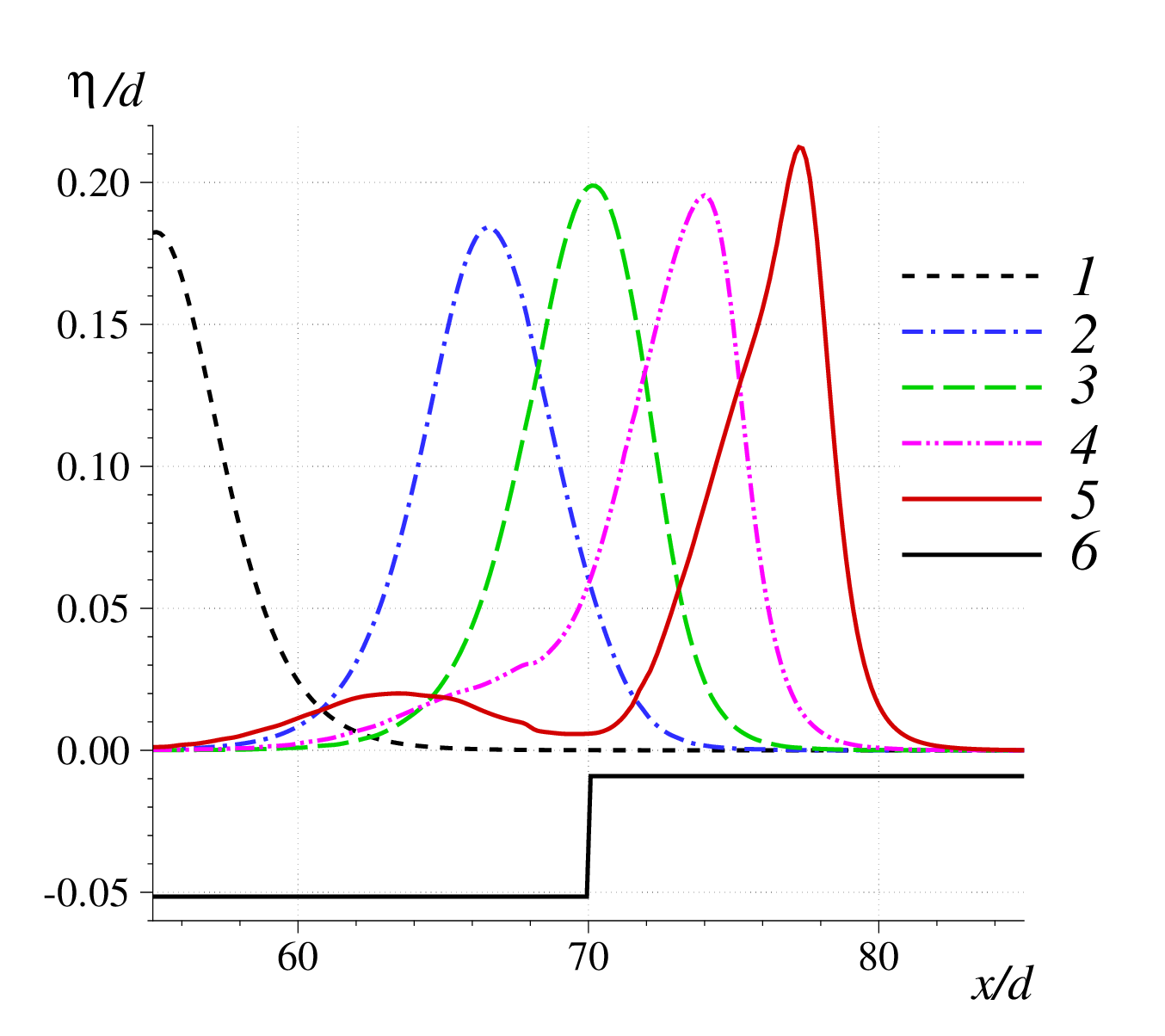}}
  \subfigure[]{
  \includegraphics[width=0.48\textwidth]{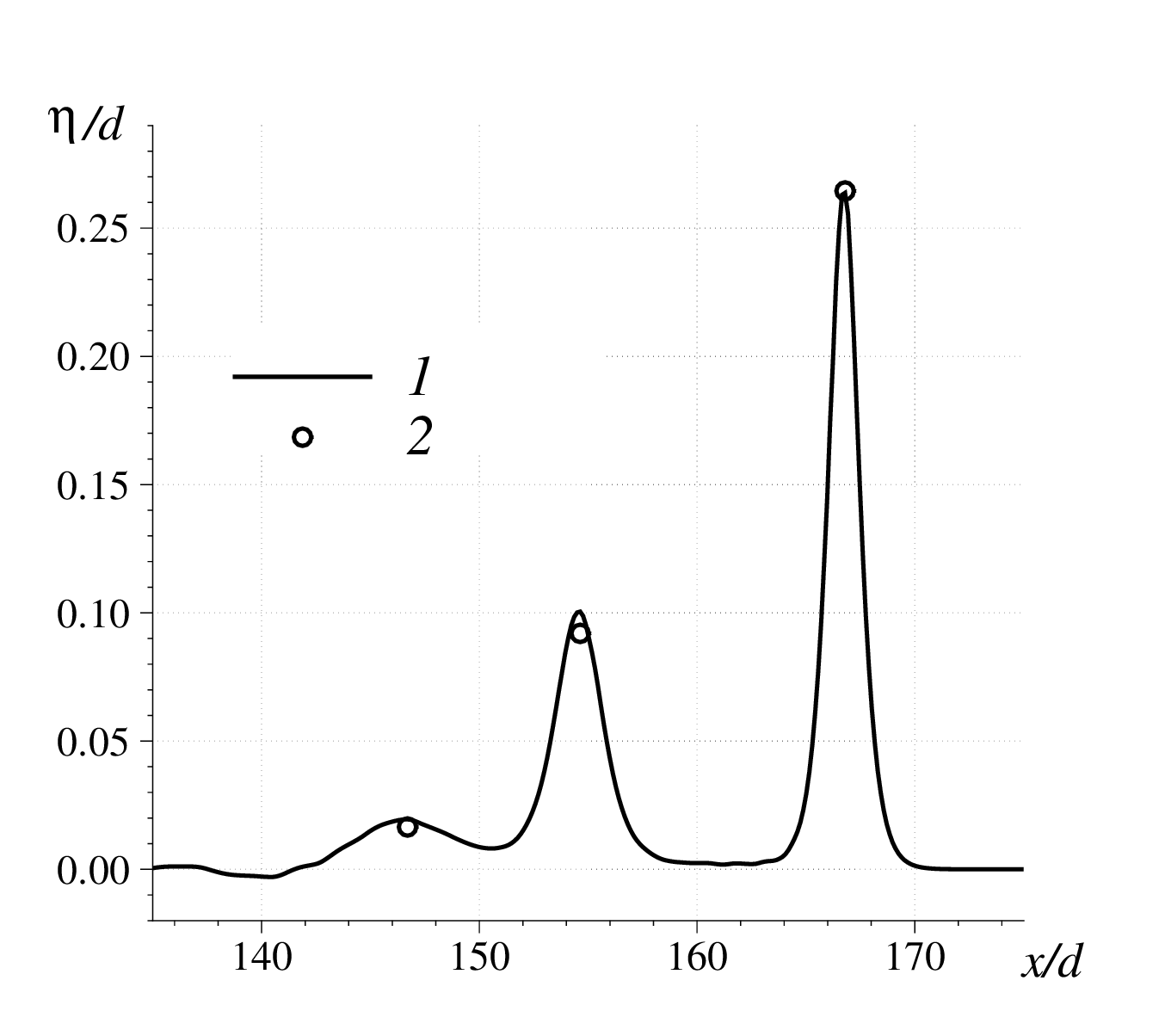}}
  \caption{\small\em Free surface profiles $y\ =\ \eta(x,\,t)$ during the interaction process of a solitary wave with an underwater step: (a) initial condition (1), $t\ =\ 1.5\;\s$ (2), $t\ =\ 2.0\;\s$ (3), $t\ =\ 2.5\;\s$ (4), $t\ =\ 3.0\;\s$ (5), smoothed bottom profile given by formula \eqref{eq:bsmooth} (6); (b) free surface profile $y\ =\ \eta(x,\,T)$ (1) at the final simulation time $t\ =\ T\,$. The experimental points (2) are taken from \cite{Seabra-Santos1987}, experiment \No 24. Numerical parameters are provided in Table~\ref{tab:params}.}
  \label{fig:profiS}
\end{figure}

Below we focus on one particular case of $\alpha\ =\ 3.65$ $\cm$. All other parameters are given in Table~\ref{tab:params}. It corresponds to the experiment \No 24 from \cite{Seabra-Santos1987}. The free surface dynamics is depicted in Figure~\ref{fig:ssurface}(\textit{a}) and the trajectories of every second grid node are shown in Figure~\ref{fig:ssurface}(\textit{b}). For the mesh adaptation we use the monitor function \eqref{eq:om1} with $\coef_1\ =\ \coef_2\ =\ 10\,$. In particular, one can see that three solitary waves are generated over the step. This fact agrees well with the theoretical predictions \cite{Kabbaj1985, Pelinovsky2010}. Moreover, one can see that the distribution of grid points follows perfectly all generated waves (over the step \emph{and} the reflected wave). Figure~\ref{fig:profiS}(\textit{a}) shows the free surface dynamics in the vicinity of the bottom step. In particular, one can see that the wave becomes notoriously steep by the time instance $t\ =\ 3$ $\s$ and during later times it splits into one reflected and three transmitted waves. The free surface profile at the final simulation time $y\ =\ \eta(x,\,T)$ is depicted in Figure~\ref{fig:profiS}(\textit{b}). On the same panel the experimental measurements are shown with empty circles $\circ\,$, which show a very good agreement with our numerical simulations.

\begin{figure}
  \centering
  \subfigure[]{\includegraphics[width=0.48\textwidth]{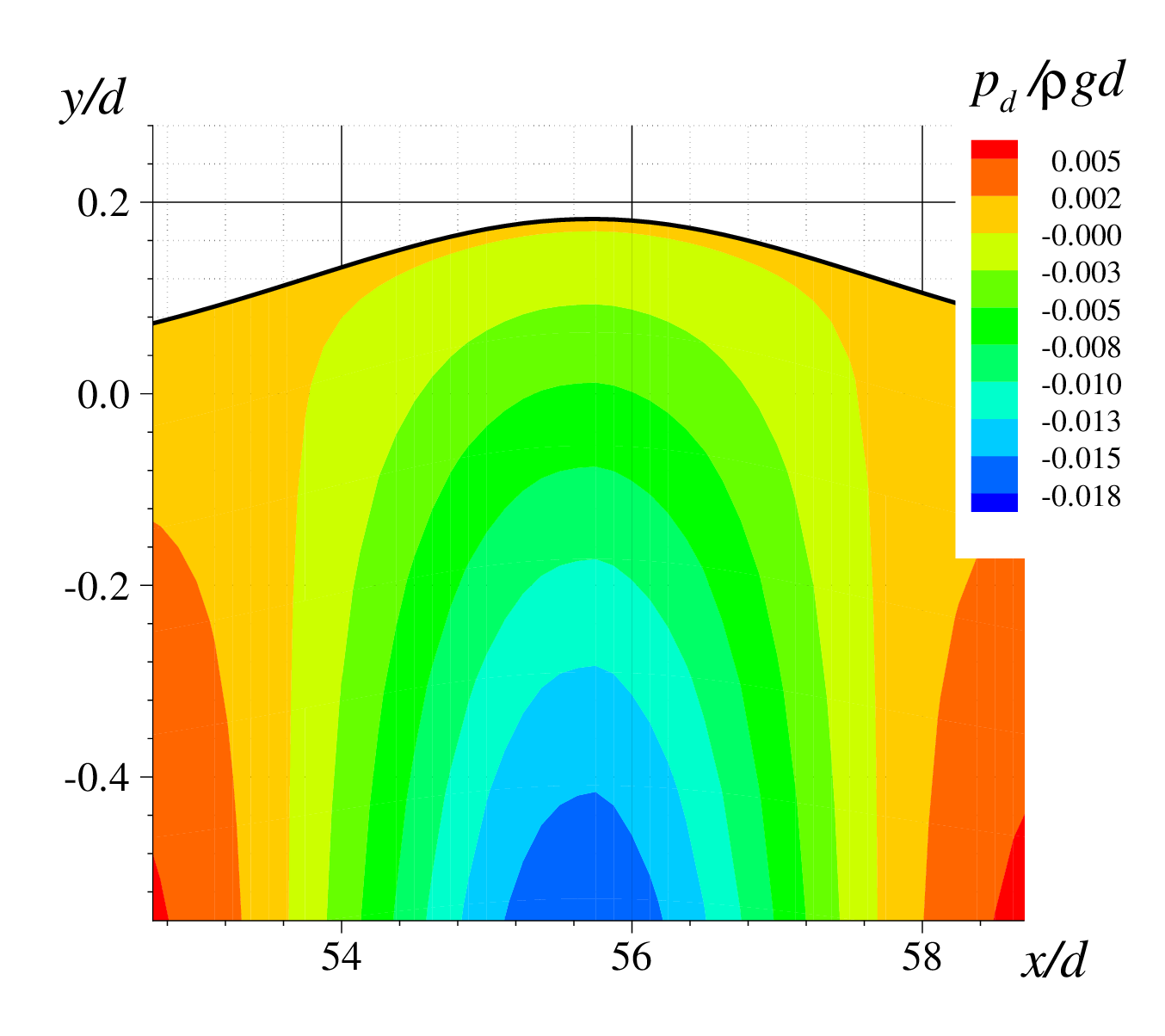}}
  \subfigure[]{\includegraphics[width=0.48\textwidth]{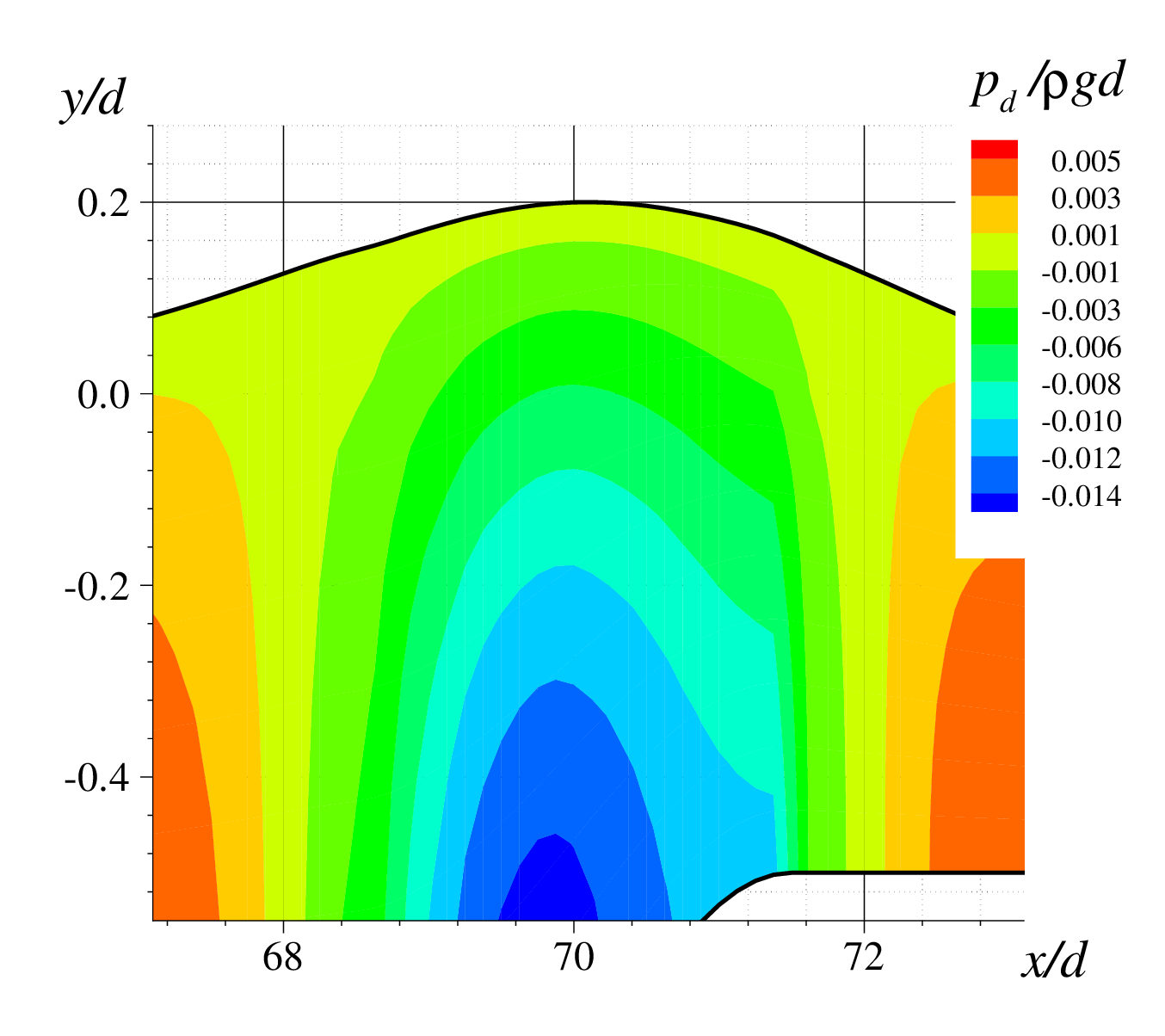}}
  \subfigure[]{\includegraphics[width=0.48\textwidth]{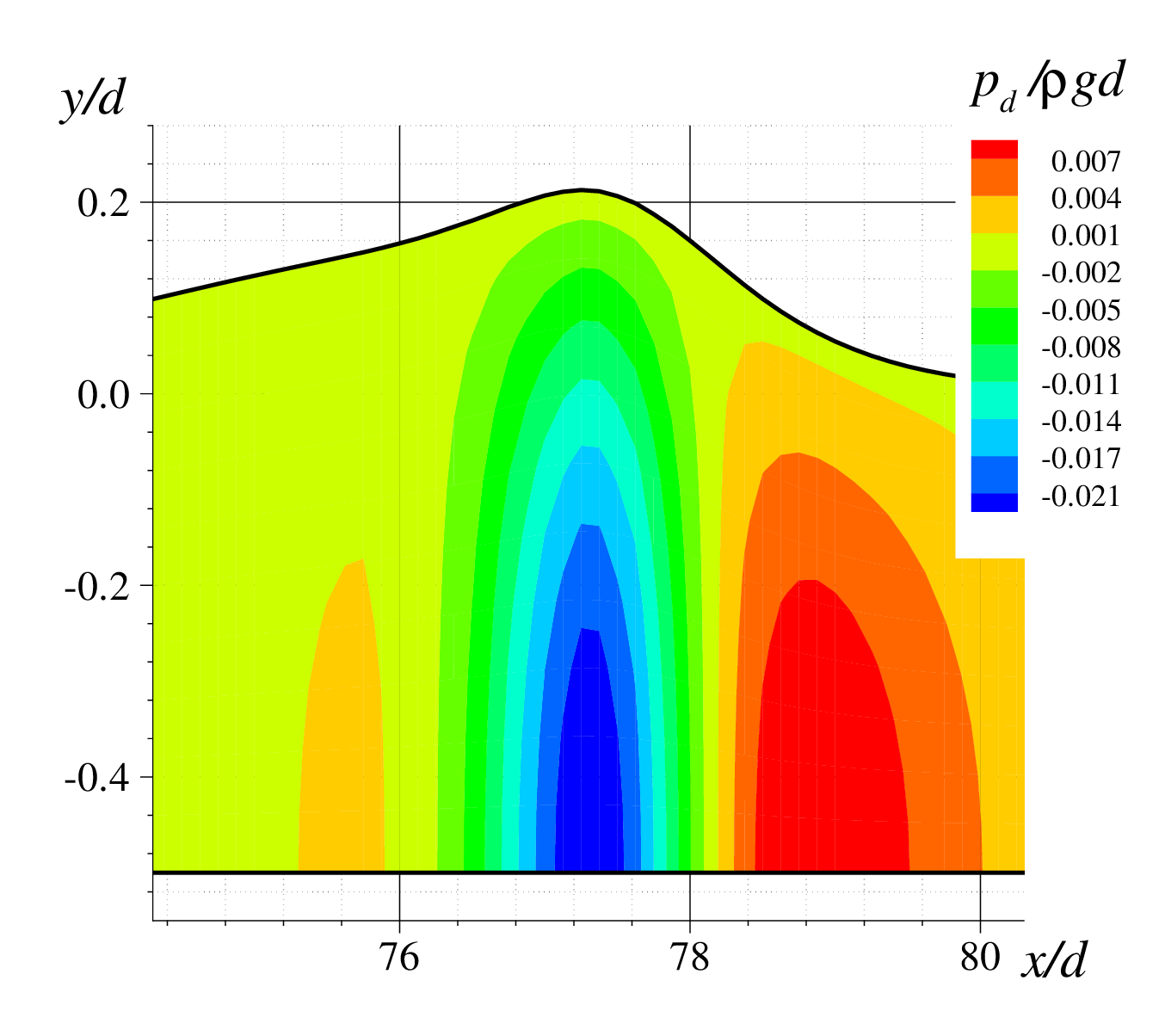}}
  \subfigure[]{\includegraphics[width=0.48\textwidth]{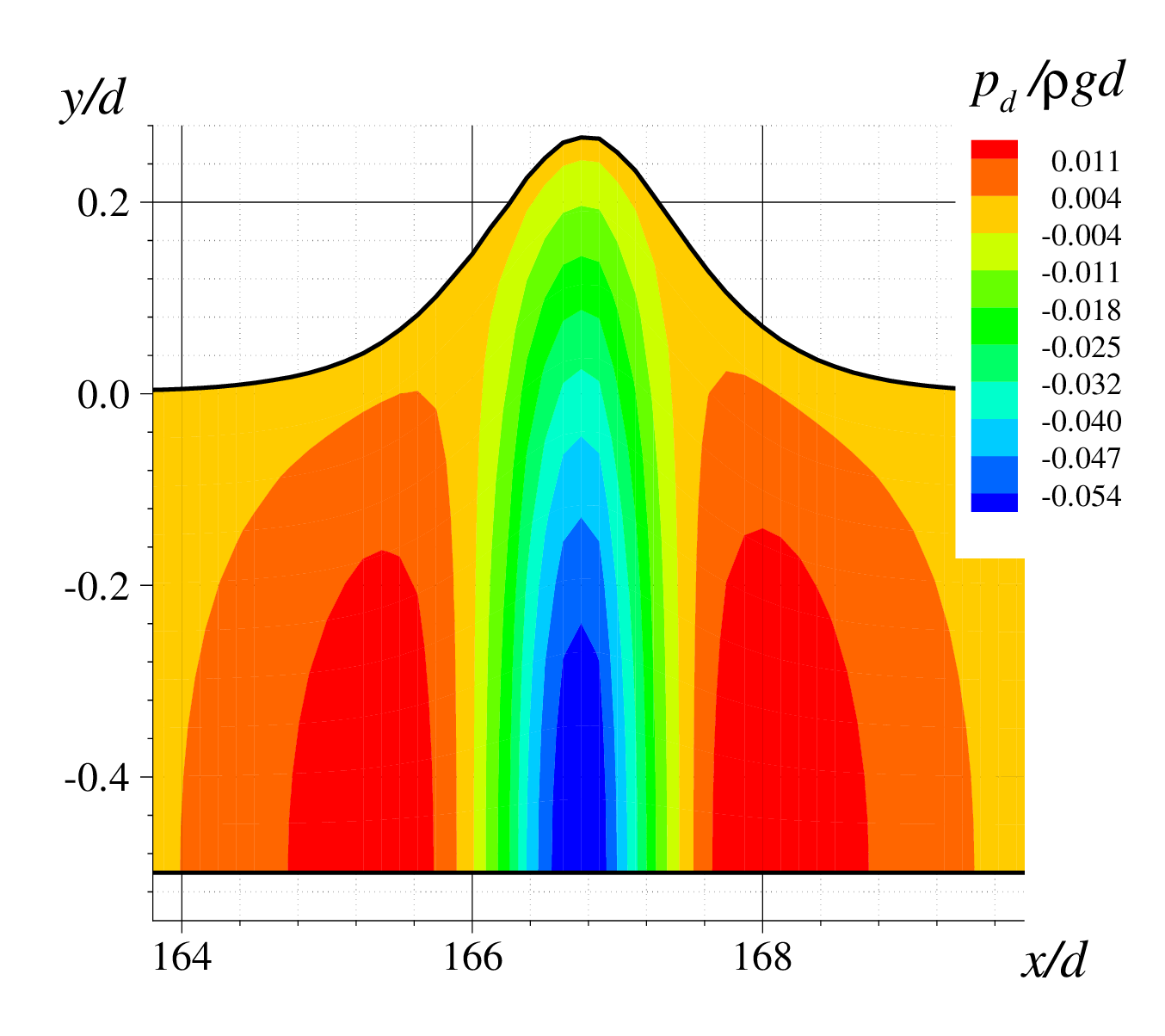}}
  \caption{\small\em Non-hydrostatic pressure distribution during a solitary wave/underwater step interaction process at different instances of time: (a) $t\ =\ 0.1\;\s$, (b) $t\ =\ 2.0\;\s$, (c) $t\ =\ 3.0\;\s$, (d) $t\ =\ 17.5\;\s$. Numerical parameters are provided in Table~\ref{tab:params}.}
  \label{fig:pressures}
\end{figure}

In our numerical experiments we go even further since a nonlinear dispersive wave model (such as the SGN equations employed in this study) can provide also information about the internal structure of the flow (\ie beneath the free surface). For instance, the non-hydrostatic component of the pressure field can be easily reconstructed\footnote{Please, notice that formula \eqref{eq:flatP} is not applicable here, since the bottom is not flat anymore.}:
\begin{equation}\label{eq:press}
  \frac{p_d(x,\,y,\,t)}{\rho}\ =\ -(\eta\ -\ y)\,\Bigl[\,\frac{\eta\ +\ y\ +\ 2\,h}{2}\cdot\Rr_1\ +\ \Rr_2\,\Bigr]\,, \qquad -h(x)\ \leq\ y\ \leq\ \eta(x,\,t)\,.
\end{equation}
where the quantities $\Rr_{1,2}$ are defined in \eqref{eq:pnh} as (see also the complete derivation in \cite{Khakimzyanov2016c}):
\begin{align*}
  \Rr_1\ &=\ u_{xt}\ +\ u\,u_{xx}\ -\ u_x^{\,2}\,, \\
  \Rr_2\ &=\ u_t\,h_x\ +\ u\,[\,u\,h_x\,]_x\,.
\end{align*}
We do not consider the hydrostatic pressure component since its variation is linear with water depth $y$:
\begin{equation*}
  p_{h}\ =\ \rho\,g\,(\eta\ -\ y)\,.
\end{equation*}
Even if the dispersive pressure component $p_d$ might be negligible comparing to the hydrostatic one $p_{h}$, its presence is crucial to balance the effects of nonlinearity, which results in the existence of solitary waves, as one of the most widely known effects in dispersive wave propagation \cite{John}. The dynamic pressure field and several other physical quantities under a solitary wave were computed and represented graphically in the framework of the full \textsc{Euler} equations in \cite{Dutykh2013b}. A good qualitative agreement with our results can be reported. The balance of dispersive and nonlinear effects results also in the symmetry of the non-hydrostatic pressure distribution with respect to the wave crest. It can be seen in Figure~\ref{fig:pressures}(\textit{a,d}) before and after the interaction process. On the other hand, during the interaction process the symmetry is momentaneously broken (see Figure~\ref{fig:pressures}(\textit{b,c})). However, with the time going on, the system relaxes again to a symmetric\footnote{The symmetry here is understood with respect to the vertical axis passing by the wave crest.} pressure distribution shown in Figure~\ref{fig:pressures}(\textit{d}).

\begin{figure}
  \centering
  \subfigure[]{\includegraphics[width=0.48\textwidth]{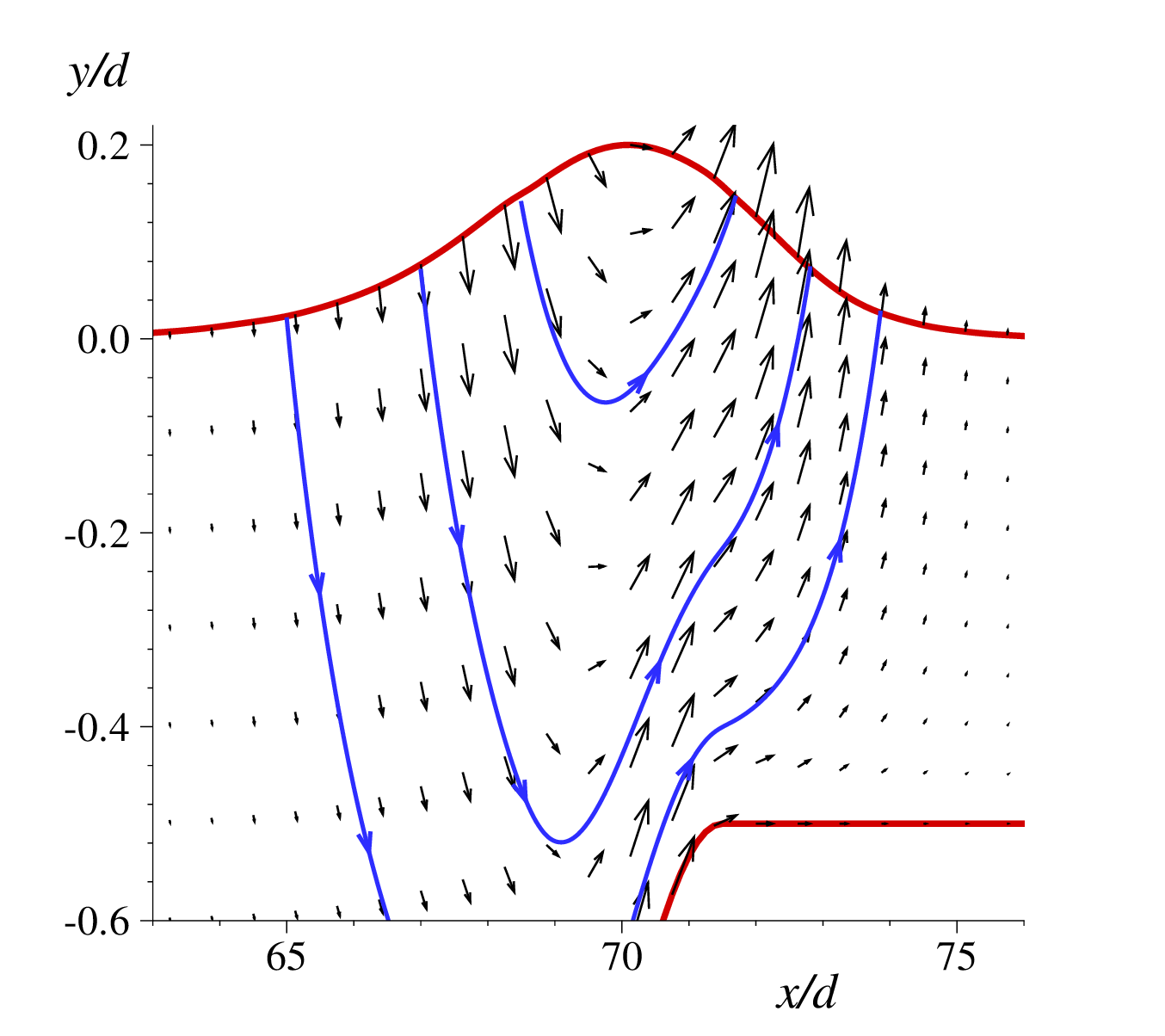}}
  \subfigure[]{\includegraphics[width=0.48\textwidth]{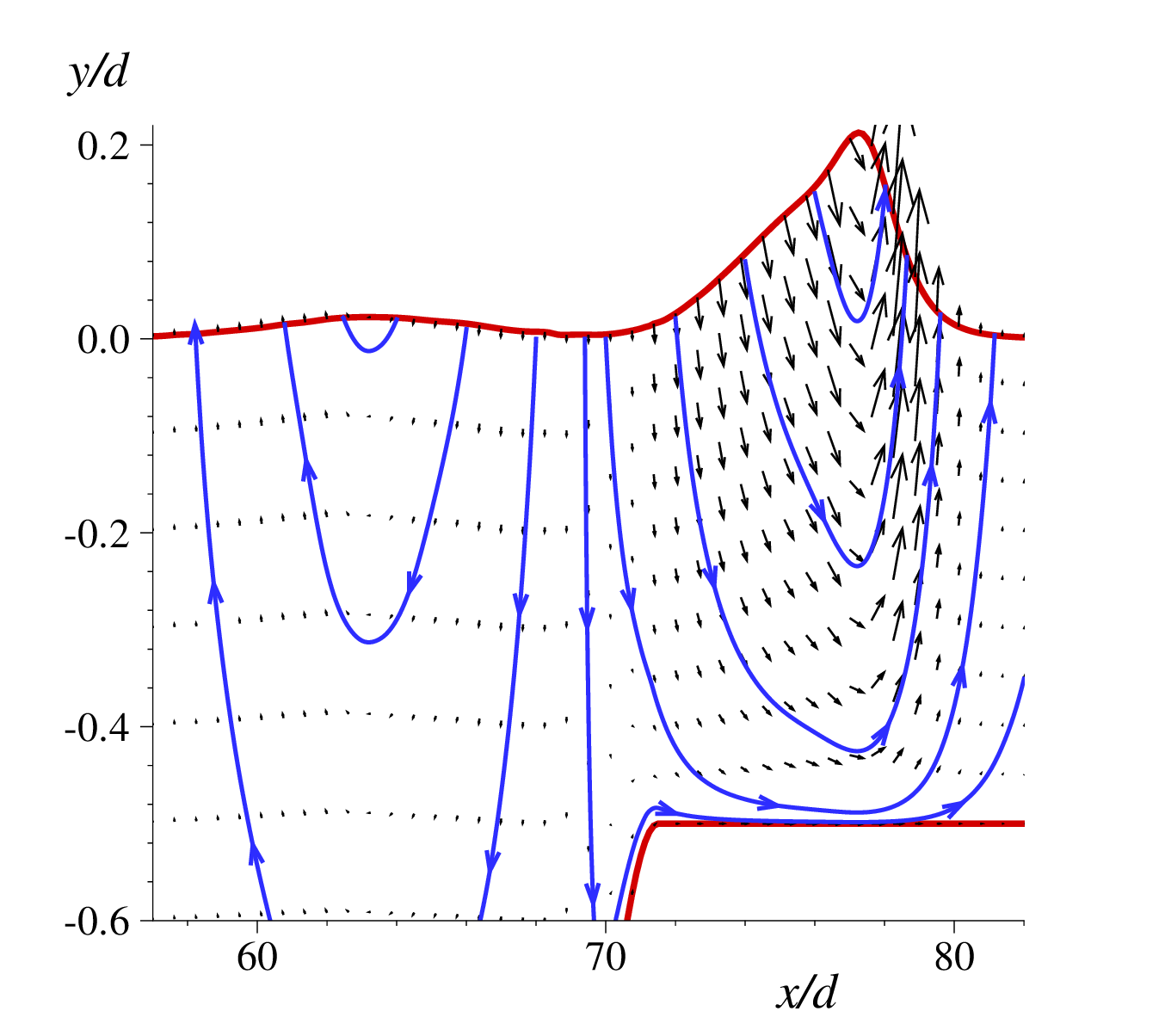}}
  \caption{\small\em Reconstructed velocity field in the fluid during the solitary wave interaction process with an underwater step: (a) $t\ =\ 2.0\;\s\,$, (b) $t\ =\ 3.0\;\s\,$. Solid blue lines show a few streamlines. Numerical parameters are provided in Table~\ref{tab:params}.}
  \label{fig:vel}  
\end{figure}

Knowledge of the solution to the SGN equations allows to reconstruct also the velocity field\footnote{This information can be used later to compute fluid particle trajectories \cite{Flierl1981}, for example.} $\bigl(\tilde{u}(x,\,y,\,t),\; \tilde{v}(x,\,y,\,t)\bigr)$ in the fluid bulk. Under the additional assumption that the flow is potential, one can derive the following asymptotic (neglecting the terms of the order $\O(\mu^4)\ \equiv\ \O\bigl(\frac{d^4}{\lambda^4}\bigr)$ in the horizontal velocity $\tilde{u}(x,\,y,\,t)$ and of the order $\O(\mu^2)\ \equiv\ \O\bigl(\frac{d^2}{\lambda^2}\bigr)$ for the vertical one $\tilde{v}(x,\,y,\,t)$) representation formula \cite{Fedotova2009} (see also the derivation in \cite{Khakimzyanov2016c} for the 3D case with moving bottom):
\begin{align}\label{eq:rec1}
  \tilde{u}(x,\,y,\,t)\ &=\ u\ +\ \Bigl(\,\frac{\H}{2}\ -\ y\ -\ h\,\Bigr)\cdot\bigl(\,[\,u\,h_x\,]_x\ +\ u_x\,h_x\bigr)\ +\ \Bigl(\,\frac{\H^{\,2}}{6}\ -\ \frac{(y + h)^2}{2}\,\Bigr)\,u_{xx}\,, \\
  \tilde{v}(x,\,y,\,t)\ &=\ -u\,h_x\ -\ (y\ +\ h)\,u_x\,. \label{eq:rec2}
\end{align}
The formulas above allow to compute the velocity vector field in the fluid domain at any time (when the solution $\bigl(\H(x,\,t),\;u(x,\,t)\bigr)$ is available) and in any point $(x,\,y)$ above the bottom $y\ =\ -h(x)$ and beneath the free surface $y\ =\ \eta(x,\,t)\,$. Figure~\ref{fig:vel} shows a numerical application of this reconstruction technique at two different moments of time $t\ =\ 2$ and $3$ $\s$ during the interaction process with the bathymetry change. In particular, in Figure~\ref{fig:vel}(\textit{a}) one can see that important vertical particle velocities emerge during the interaction with the bottom step. In subsequent time moments one can see the division of the flow in two structures (see Figure~\ref{fig:vel}(\textit{b})): the left one corresponds to the reflected wave, while the right structure corresponds to the transmitted wave motion. The reconstructed velocity fields via the SGN model compare fairly well with the 2D \textsc{Navier}--\textsc{Stokes} predictions \cite{Pelinovsky2010}. However, the computational complexity of our approach is significantly lower than the simulation of the full \textsc{Navier}--\textsc{Stokes} equations. This is probably the main advantage of the proposed modelling methodology.

\subsection{Wave generation by an underwater landslide}
\label{sec:slide}

As the last illustration of the proposed above numerical scheme, we model wave generation by the motion of an underwater landslide over uneven bottom. This test-case is very challenging since it involves rapid bottom motion (at least of its part). We recall that all previous tests were performed on a static bottom (\ie $h_t\ \equiv\ 0\,$). The numerical simulation of underwater landslides is an important application where the inclusion of non-hydrostatic effects is absolutely crucial \cite{Lynett2002}. Moreover, the accurate prediction of generated waves allows to assess more accurately the natural hazard induced by unstable sliding masses (rockfalls, debris, ground movements) \cite{Ward}.

Usually, the precise location of unstable underwater masses is unknown and the numerical simulation is a preferred tool to study these processes. The landslide can be modelled as a solid undeformable body moving down the slope \cite{Watts2000, Chubarov2005, Grilli2005, Enet2007}. Another possibility consists in representing the landslide as another fluid layer of higher density (possibly also viscosity) located near the bottom \cite{Fernandez-Nieto2007, Castro2013}. In some works the landslide motion was not simulated (\eg \cite{Ioualalen2010}) and the initial wave field generated by the landslide motion was determined using empirical formulas \cite{Vogel}. Then, this initial condition was propagated using an appropriate water wave model \cite{Ioualalen2010}. However, strictly speaking the employed empirical models are valid only for an absolutely rigid landslide sliding down a constant slope. Subsequent numerical simulations showed that the bottom shape influences quite substantially the generated wave field \cite{Beisel2011}. Consequently, for realistic modelling of real world cases one needs to take into account the actual bathymetry \cite{Liu2005a} and even possible deformations of the landslide during its motion \cite{Lindstrom2014}. In a recent experimental work \cite{Lindstrom2014} the deformability of the landslide was achieved by composing it with four solid parts interconnected by springs. The idea to represent a landslide as a finite number of blocks was used in numerical \cite{Ranguelov2008} and theoretical \cite{Tinti1997} investigations. In the present study we use the quasi-deformable\footnote{This model can be visualized if you imagine a landslide composed of infinitely many solid blocks.} landslide model \cite{Beisel2012, Dutykh2012, Dutykh2011d}. In this model the landslide deforms according to encountered bathymetry changes, however, at every time instance, all components of the velocity vector are the same in all particles which constitute the landslide (as in a solid rigid body). We shall use two long wave models:
\begin{itemize}
  \item The SGN equations (fully nonlinear non-hydrostatic weakly dispersive model)
  \item NSWE equations\footnote{The numerical algorithm to solve NSW equations on a moving grid was presented and validated in \cite{Khakimzyanov2015a}.} (standard hydrostatic dispersionless model)
\end{itemize}
The advantage of the SGN equations over other frequently used long wave models \cite{Lynett2002, Watts2003, Ioualalen2010} are:
\begin{itemize}
  \item The \textsc{Galilean} invariance
  \item The energy balance equation (consistent with the full \textsc{Euler} \cite{Fedotova2014})
\end{itemize}
NSWE were employed in \cite{Assier-Rzadkieaicz2000a} to model the real world 16\up{th} October 1979 Nice event. It looks like the consensus on the importance of dispersive effects in landslide modeling is far from being achieved. For example, in \cite{Ioualalen2010} the authors affirm that the inclusion of dispersion gives results very similar to NSWE. In other works \cite{Lovholt2008, Tappin2008, Glimsdal2013} the authors state that dispersive effects significantly influence the resulting wave field, especially during long time propagation. Consequently, in the present study we use both the SGN and NSWE equations to shed some light on the r\^ole of dispersive effects.

Consider a 1D fluid domain bounded from below by the solid (static) impermeable bottom given by the following function:
\begin{equation}\label{eq:static}
  h_0(x)\ =\ \frac{h_+\ +\ h_-}{2}\ +\ \frac{h_+\ -\ h_-}{2}\;\tanh\bigl[\,\digamma(x\ -\ \xi_\digamma)\,\bigr]\,,
\end{equation}
where $h_+$ and $h_-$ are water depths at $\pm\,\infty$ correspondingly (the domain we take is finite, of course). We assume for definiteness that
\begin{equation*}
  h_+\ <\ h_-\ <\ 0\,.
\end{equation*}
We have also by definition
\begin{equation*}
  \digamma\ \eqdef\ \frac{2\,\tan\theta_0}{h_-\ -\ h_+}\ >\ 0\,, \qquad
  \xi_\digamma\ \eqdef\ \frac{1}{2\,\digamma}\;\ln\Bigl[\,\frac{h_0\ -\ h_+}{h_-\ -\ h_+}\,\Bigr]\ >\ 0\,,
\end{equation*}
where $h_0\ \equiv\ h_0(0)$ is water depth in $x\ =\ 0$ and $\theta_0$ is the maximal slope angle, which is reached at the inflection point $\xi_\digamma\,$. It can be easily checked that
\begin{equation*}
  \frac{h_+\ +\ h_-}{2}\ <\ h_0\ <\ h_-\,.
\end{equation*}
Equation \eqref{eq:static} gives us the static part of the bottom shape. The following equation prescribes the shape of the bathymetry including the unsteady component:
\begin{equation*}
  y\ =\ -h(x,\,t)\ =\ h_0(x)\ +\ \zeta(x,\,t)\,,
\end{equation*}
where function $\zeta(x,\,t)$ prescribes the landslide shape. In the present study we assume that the landslide initial shape is given by the following analytical formula:
\begin{equation*}
  \zeta(x,\,0)\ =\ \begin{dcases}
    \dfrac{\hbar}{2}\;\biggl[\,1\ +\ \cos\,\Bigl[\,\frac{2\,\pi\,\bigl(x\ -\ x_c(0)\bigr)}{\upnu}\,\Bigr]\,\biggr]\,, & \abs{x\ -\ x_c(0)}\ \leq\ \dfrac{\upnu}{2}\,, \\
    0\,, & \abs{x\ -\ x_c(0)}\ >\ \dfrac{\upnu}{2}\,,
  \end{dcases}
\end{equation*}
where $x_c(0)$, $\hbar$ and $\upnu$ are initial landslide position, height and width (along the axis $Ox$) correspondingly. Initially we put the landslide at the unique\footnote{This point is \emph{unique} since the static bathymetry $h_0(x)$ is a monotonically increasing function of its argument $x\,$.} point where the water depth is equal to $h_0\ =\ 100$ $\m$, \ie
\begin{equation*}
  x_c(0)\ =\ \xi_\digamma\ -\ \frac{1}{2\,\digamma}\;\ln\Bigl[\,\frac{h_0\ -\ h_+}{h_-\ -\ h_+}\,\Bigr]\ \approx\ 8\;323.5\ \m\,.
\end{equation*}

For $t\ >\ 0$ the landslide position $x_c(t)$ and its velocity $v(t)$ are determined by solving a second order ordinary differential equation which describes the balance of all the forces acting on the sliding mass \cite{Beisel2012}. This model is well described in the literature \cite{Dutykh2012, Dutykh2011d} and we do not reproduce the details here.

\begin{table}
  \centering
  \begin{tabular}{l|c}
    \hline\hline
    \textit{Parameter} & \textit{Value} \\
    \hline\hline
    Fluid domain length, $\ell$ & $80\;000$ $\m$ \\
    Water depth, $h_0(0)$ & $-5.1$ $\m$ \\
    Rightmost water depth, $h_+$ & $-500$ $\m$ \\
    Leftmost water depth, $h_-$ & $-5$ $\m$ \\
    Maximal bottom slope, $\theta_0$ & $6^\circ$ \\
    Landslide height, $\hbar$ & $20$ $\m$ \\
    Landslide length, $\upnu$ & $5000$ $\m$ \\
    Initial landslide position, $x_c(0)$ & $8\;323.5$ $\m$ \\
    Added mass coefficient, $C_w$ & $1.0$ \\
    Hydrodynamic resistance coefficient, $C_d$ & $1.0$ \\
    Landslide density, $\rho_{\mathrm{sl}}/\rho_{\mathrm{w}}$ & $1.5$ \\
    Friction angle, $\theta^\ast$ & $1^\circ$ \\
    Final simulation time, $T$ & $1000$ $\s$ \\
    Number of grid points, $N$ & $400$ \\
    Monitor function parameter, $\coef_0$ & $200$ \\
    \hline\hline
  \end{tabular}
  \bigskip
  \caption{\small\em Numerical and physical parameters used in landslide simulation.}
  \label{tab:params2}
\end{table}

\begin{figure}
  \centering
  \subfigure[]{\includegraphics[width=0.48\textwidth]{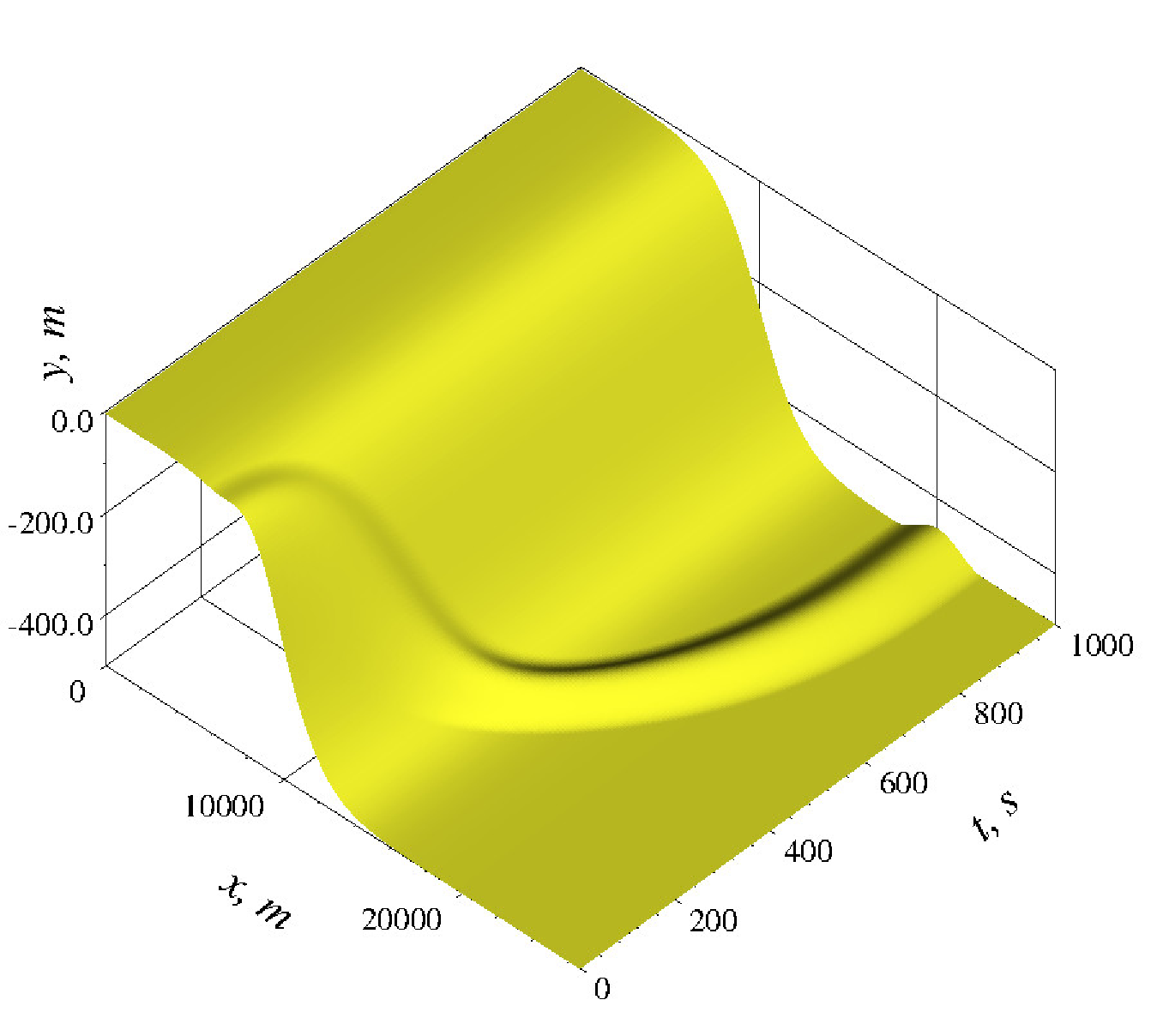}}
  \subfigure[]{\includegraphics[width=0.48\textwidth]{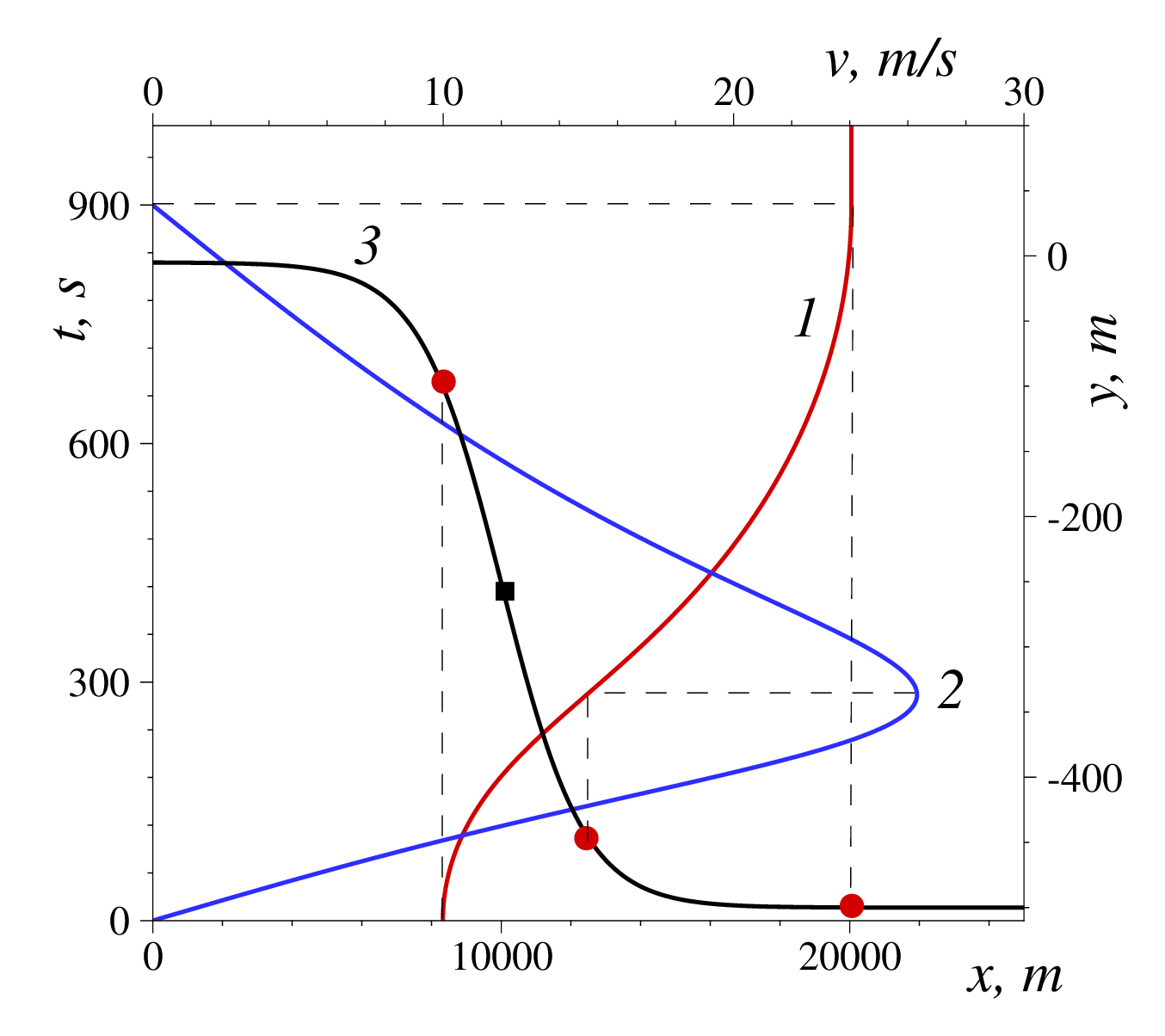}}
  \caption{\small\em Generation of surface waves by an underwater landslide motion: (a) dynamics of the moving bottom; (b) graphics of functions (1) $x\ =\ x_c(t)\,$, (2) $v\ =\ v(t)\,$, (3) $y\ =\ h_0(x)\,$. Two outer red circles denote landslide initial $t\ =\ 0$ and terminal $t\ =\ 897.6$ $\s$ positions. Middle red circle denotes landslide position at the moment of time $t\ =\ 284.2$ $\s$ where landslide's speed is maximal $v_{\max}\ \approx\ 26.3$ $\m/\s\,$. The black square shows the inflection point $\xi_\digamma$ position. The maximal speed is achieved well below the inflection point $\xi_\digamma\,$. Numerical parameters are given in Table~\ref{tab:params2}.}
  \label{fig:slide}
\end{figure}

In Figure~\ref{fig:slide}(\textit{a}) we show the dynamics of the moving bottom from the initial condition at $t\ =\ 0$ to the final simulation time $t\ =\ T$. All parameters are given in Table~\ref{tab:params2}. It can be clearly seen that landslide's motion significantly depends on the underlying static bottom shape. In Figure~\ref{fig:slide}(\textit{b}) we show landslide's barycenter trajectory $x\ =\ x_c(t)$ (line 1), its velocity $v\ =\ v(t)$ (line 2) and finally the static bottom profile $y\ =\ h_0(x)$ (line 3). From the landslide speed plot in Figure~\ref{fig:slide}(\textit{b}) (line 2), one can see that the mass is accelerating during the first $284.2$ $\s$ and slows down during $613.4$ $\s$. The distances traveled by the landslide during these periods have approximatively the same ratio $\approx 2\,$. It is also interesting to notice that the landslide stops abruptly its motion with a negative (\ie nonzero) acceleration.

In order to simulate water waves generated by the landslide, we take the fluid domain $\I\ =\ [0,\,\ell]$. For simplicity, we prescribe wall boundary conditions\footnote{It would be better to prescribe transparent boundary conditions here, but this question is totally open for the SGN equations.} at $x\ =\ 0$ and $x\ =\ \ell$. Undisturbed water depth at both ends is $h_0$ and $\approx\ h_+$ respectively. The computational domain length $\ell$ is chosen to be sufficiently large to avoid any kind of reflections from the right boundary. Initially the fluid is at rest with undisturbed free surface, \ie 
\begin{equation*}
  \eta(x,\,0)\ \equiv\ 0\,, \qquad u(x,\,0)\ \equiv\ 0\,.
\end{equation*}
Segment $\I$ is discretized using $N\ =\ 400$ points. In order to redistribute optimally mesh nodes, we employ the monitor function defined in equation \eqref{eq:om0}, which refines the grid where the waves are large (regardless whether they are of elevation or depression type).

\begin{figure}
  \centering
  \subfigure[]{\includegraphics[width=0.48\textwidth]{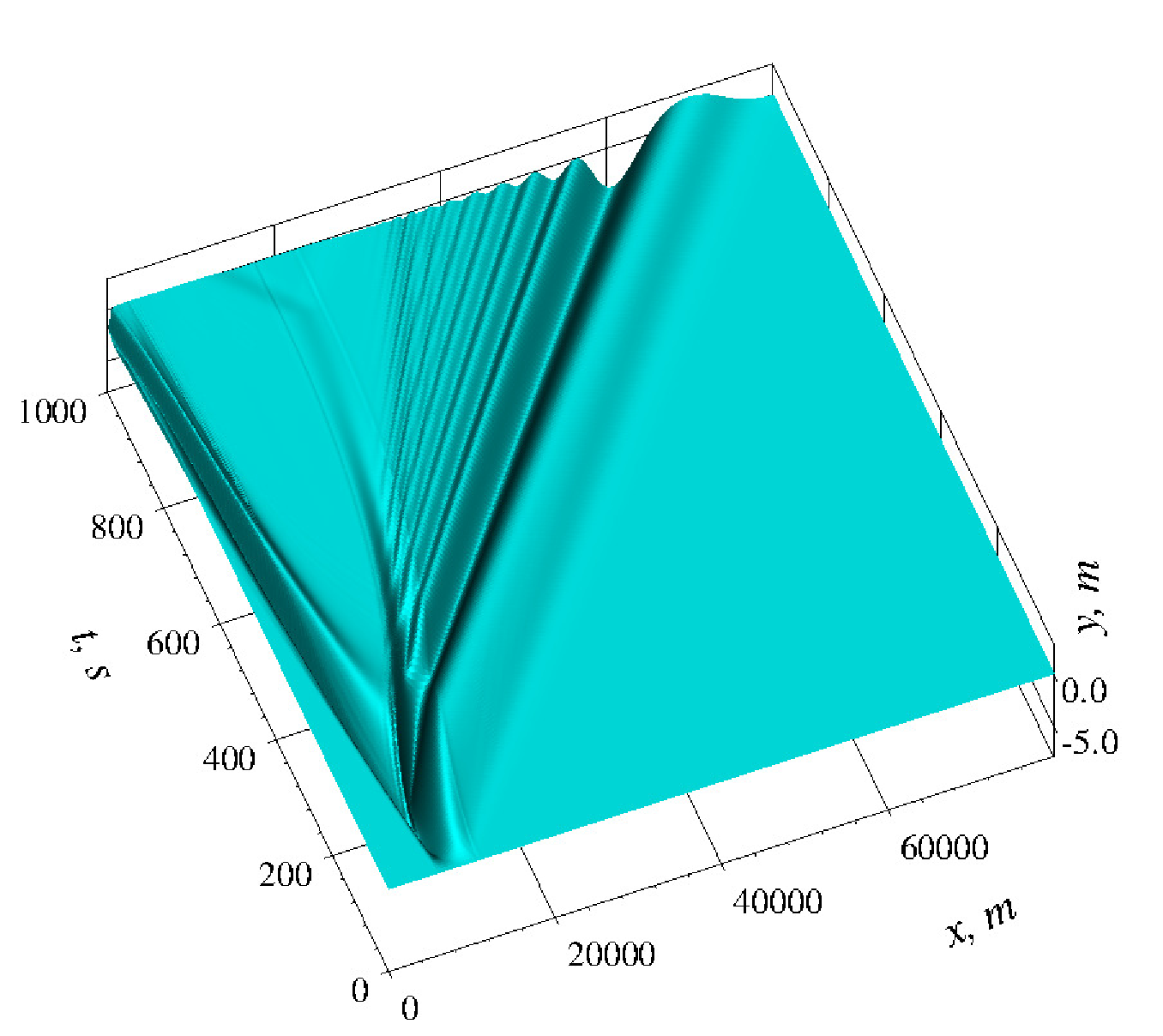}}
  \subfigure[]{\includegraphics[width=0.48\textwidth]{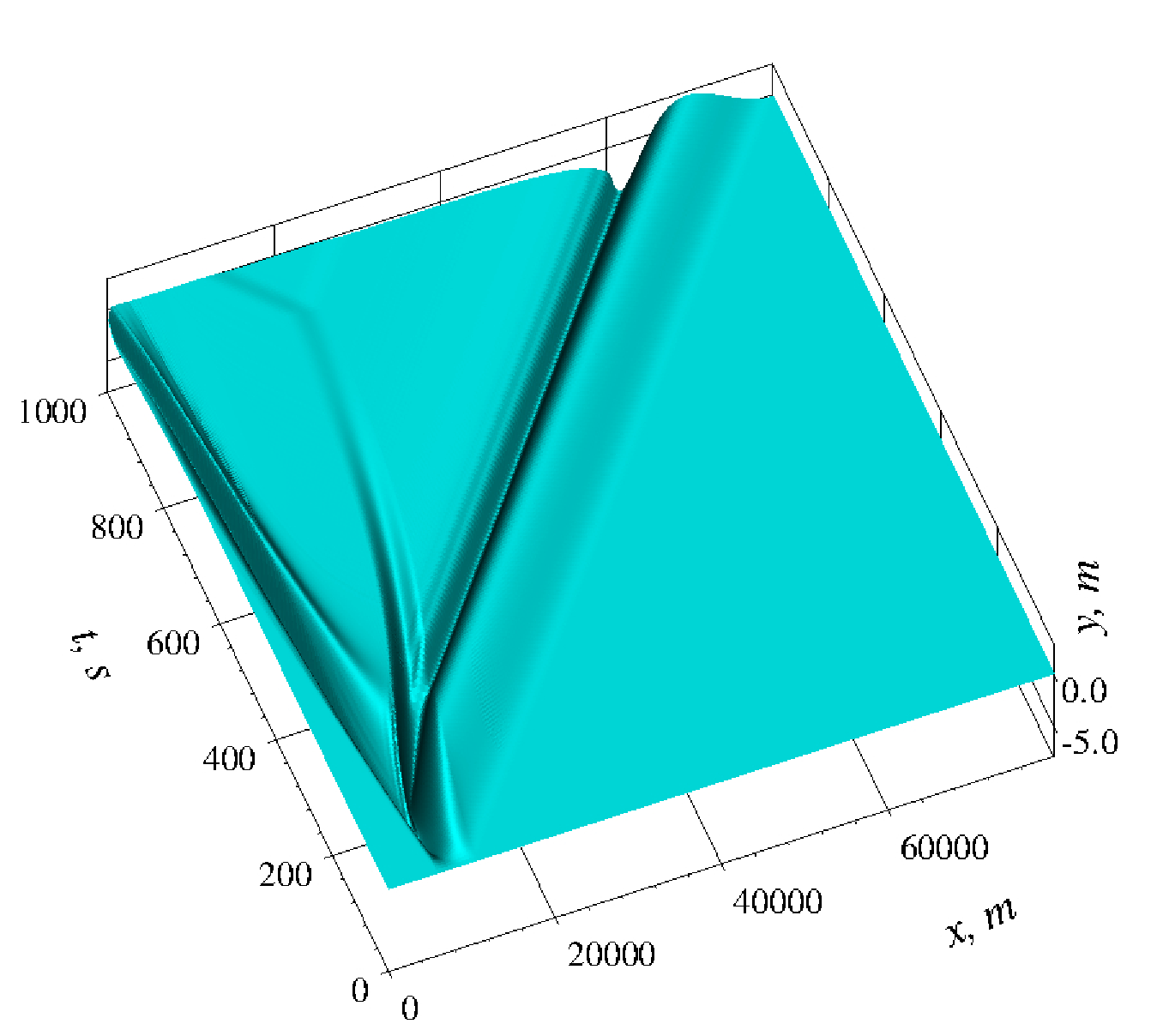}}
  \caption{\small\em Generation of surface waves $y\ =\ \eta(x,\,t)$ by an underwater landslide motion: (a) the SGN model (dispersive); (b) NSWE equations (dispersionless). Numerical parameters are given in Table~\ref{tab:params2}.}
  \label{fig:slides}
\end{figure}

\begin{figure}
  \centering
  \includegraphics[width=0.99\textwidth]{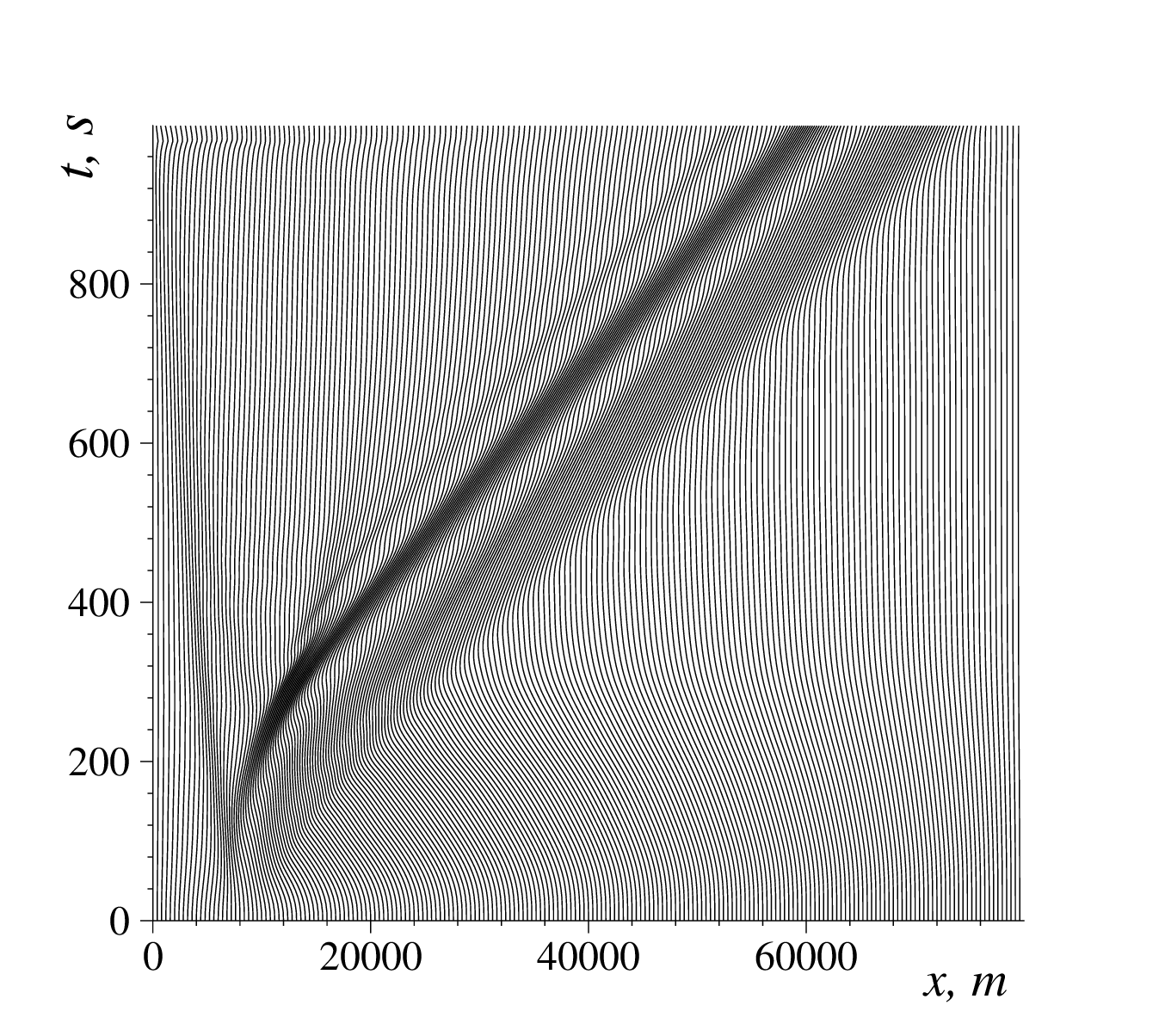}
  \caption{\small\em Trajectories of every second grid node during the underwater landslide simulation in the framework of the SGN equations. Numerical parameters are given in Table~\ref{tab:params2}.}
  \label{fig:grid}
\end{figure}

In Figure~\ref{fig:slides} we show the surface $y\ =\ \eta(x,\,t)$ in space-time, which shows the main features of the generated wave field. The left panel (\textit{a}) is the dispersive SGN prediction, while (\textit{b}) is the computation with NSWE that we include into this study for the sake of comparison. For instance, one can see that the dispersive wave system is much more complex even if NSWE seem to reproduce the principal wave components. The dispersive components follow the main wave travelling rightwards. There is also at least one depression wave moving towards the shore. The motion of grid points is shown in Figure~\ref{fig:grid}. The initial grid was chosen to be uniform, since the free surface was initially flat. However, during the wave generation process the grid adapts to the solution. The numerical method redistributes the nodes according to the chosen monitor function $\om_0[\,\eta\,]\,(x,\,t)\,$, \ie where the waves are large (regardless whether they are of elevation or depression type). We would like to underline the fact that in order to achieve a similar accuracy on a uniform grid, one would need about $4\,N$ points.

\begin{figure}
  \centering
  \subfigure[]{\includegraphics[width=0.48\textwidth]{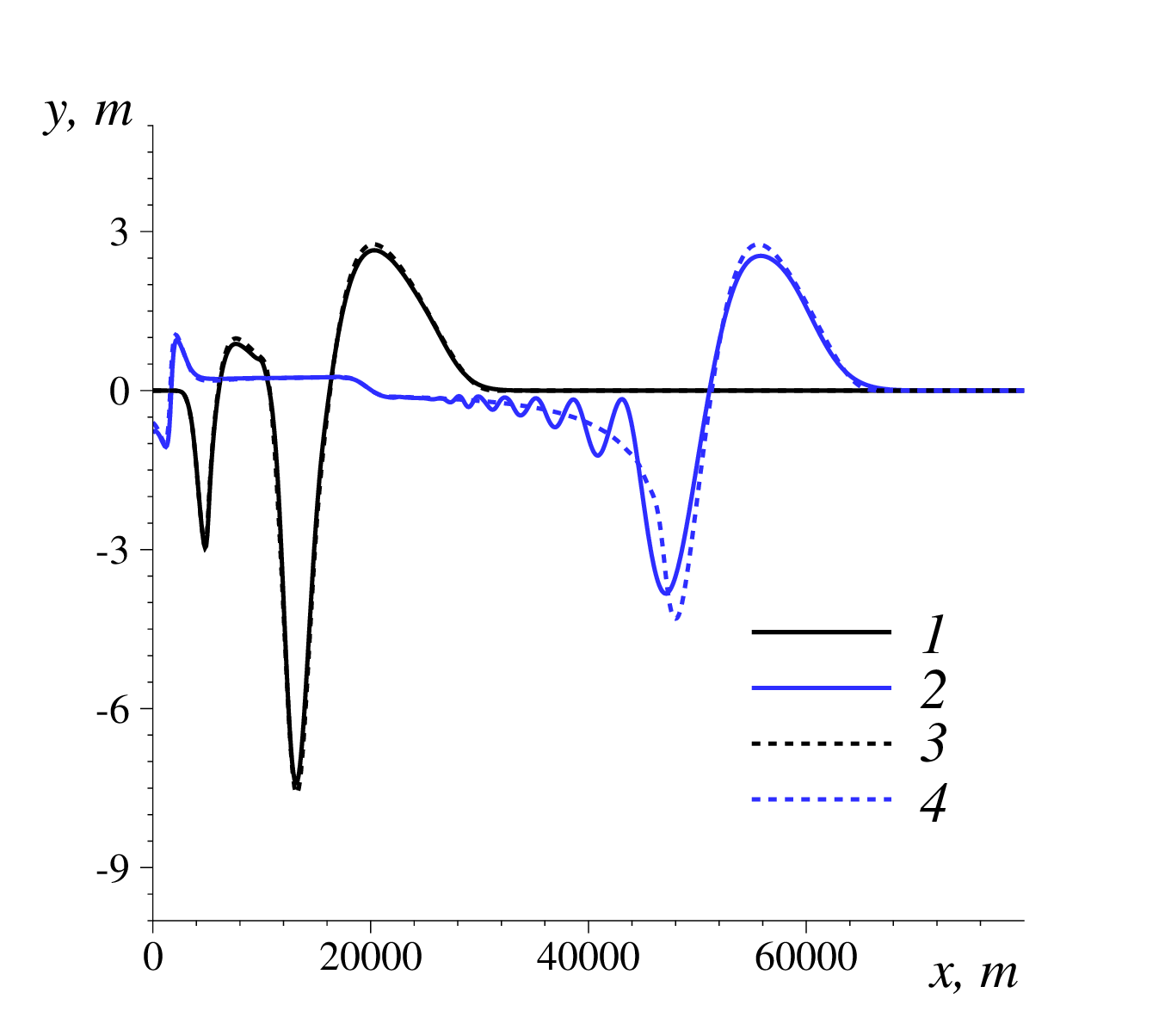}}
  \subfigure[]{\includegraphics[width=0.48\textwidth]{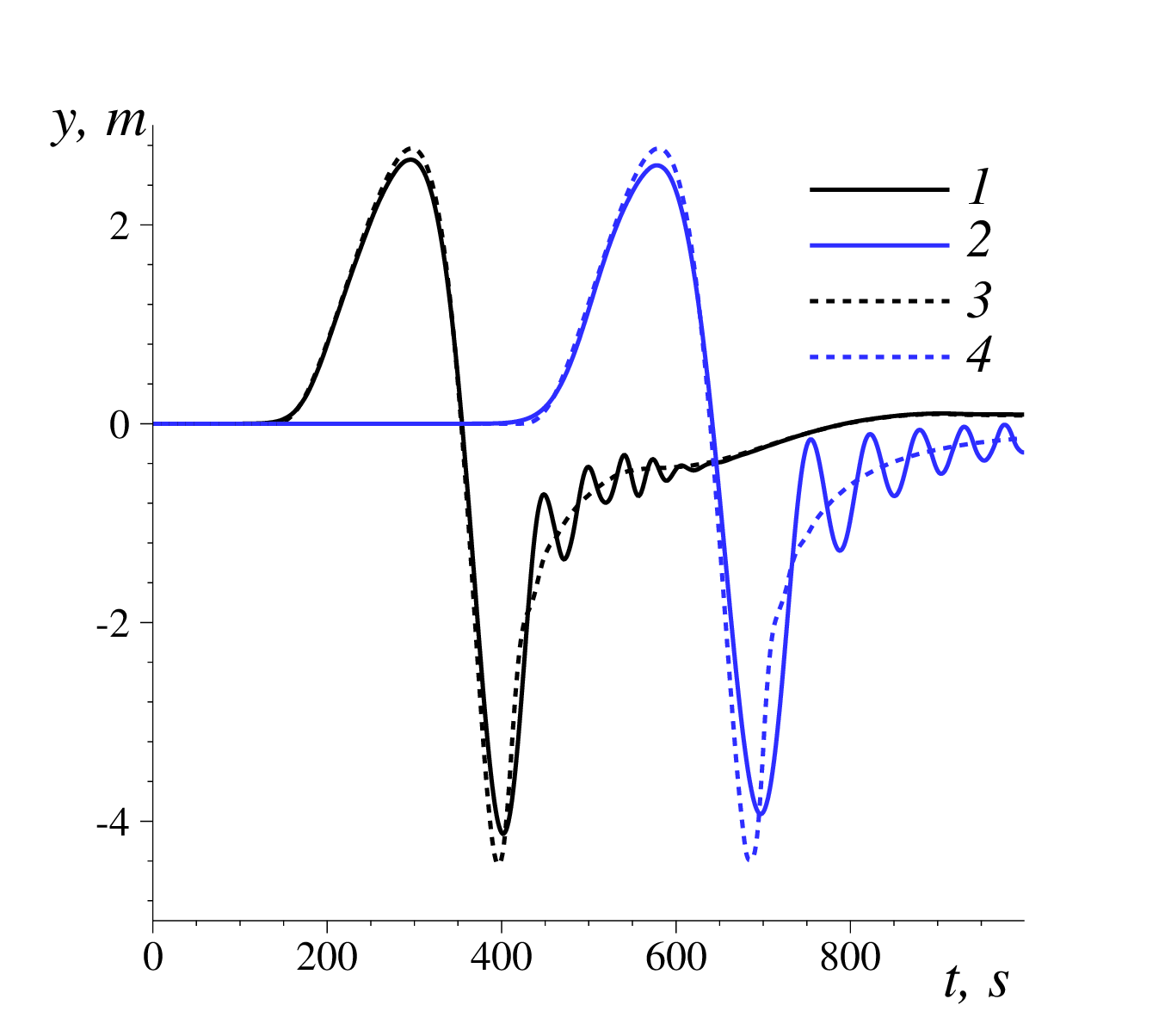}}
  \caption{\small\em Generation of surface waves by an underwater landslide: (a) free surface elevation profiles $y\ =\ \eta(x,\,t_{1,\,2})$ at $t_1\ =\ 300$ $\s$ (1,3) and $t_2\ =\ 800$ $\s$ (2,4); (b) free surface elevation $y\ =\ \eta(x_{1,\,2},\,t)$ as a function of time in two spatial locations $x_1\ =\ 20 000$ $\m$ (1,3) and $x_2\ =\ 40 000$ $\m$ (2,4). The SGN predictions are represented with solid lines (1,2) and NSWE with dashed lines (3,4). Numerical parameters are given in Table~\ref{tab:params2}.}
  \label{fig:profi}
\end{figure}

In Figure~\ref{fig:profi} we show two snapshots of the free surface elevation at two moments of time (\textit{a}) and wave gauge records collected at two different spatial locations (\textit{b}). In particular, we observe that there is a better agreement between NSWE and the SGN model in shallow regions (\ie towards $x\ =\ 0$), while a certain divergence between two models becomes more apparent in deeper regions (towards the right end $x\ =\ \ell$).

\begin{figure}
  \centering
  \subfigure[]{\includegraphics[width=0.48\textwidth]{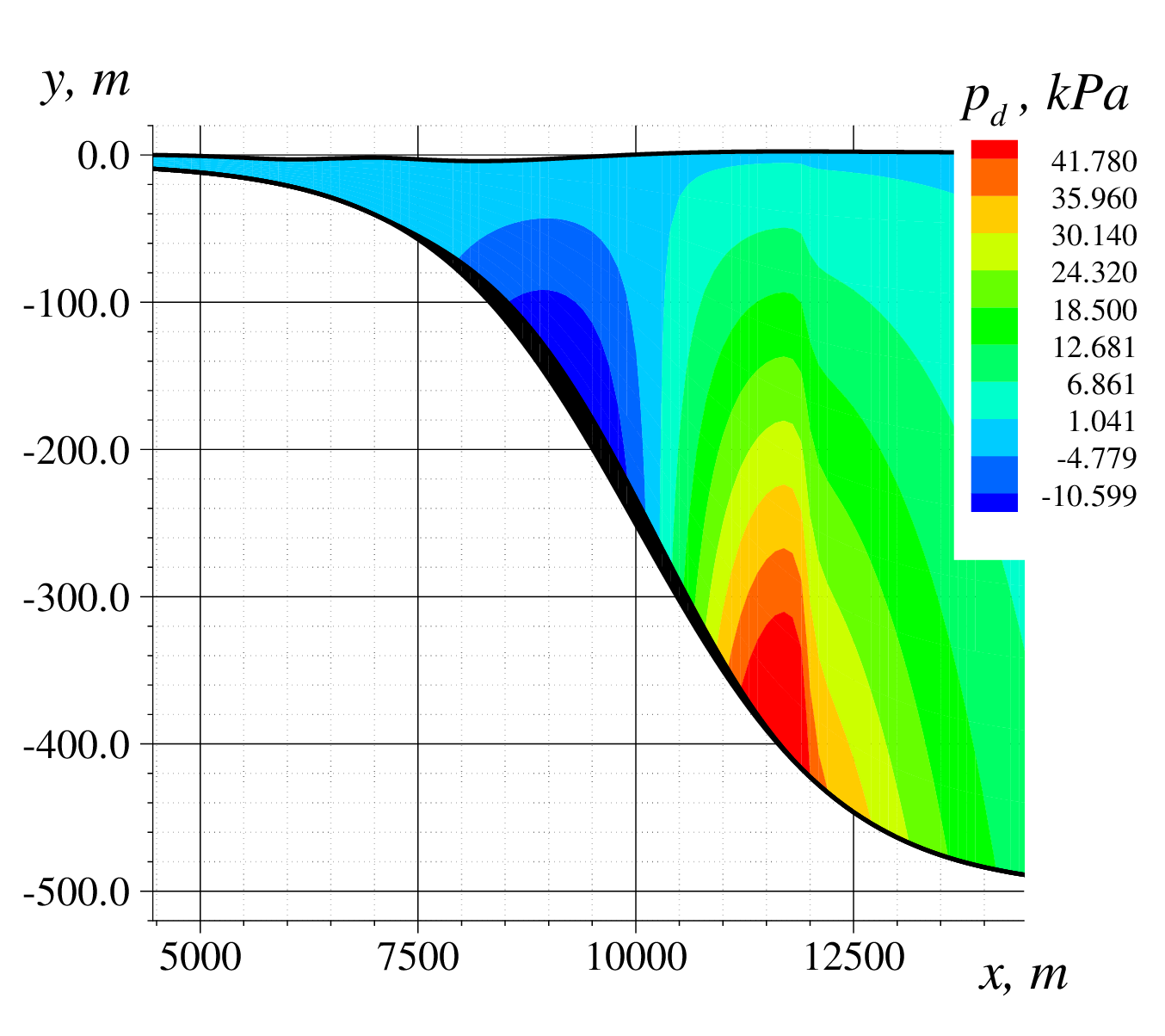}}
  \subfigure[]{\includegraphics[width=0.48\textwidth]{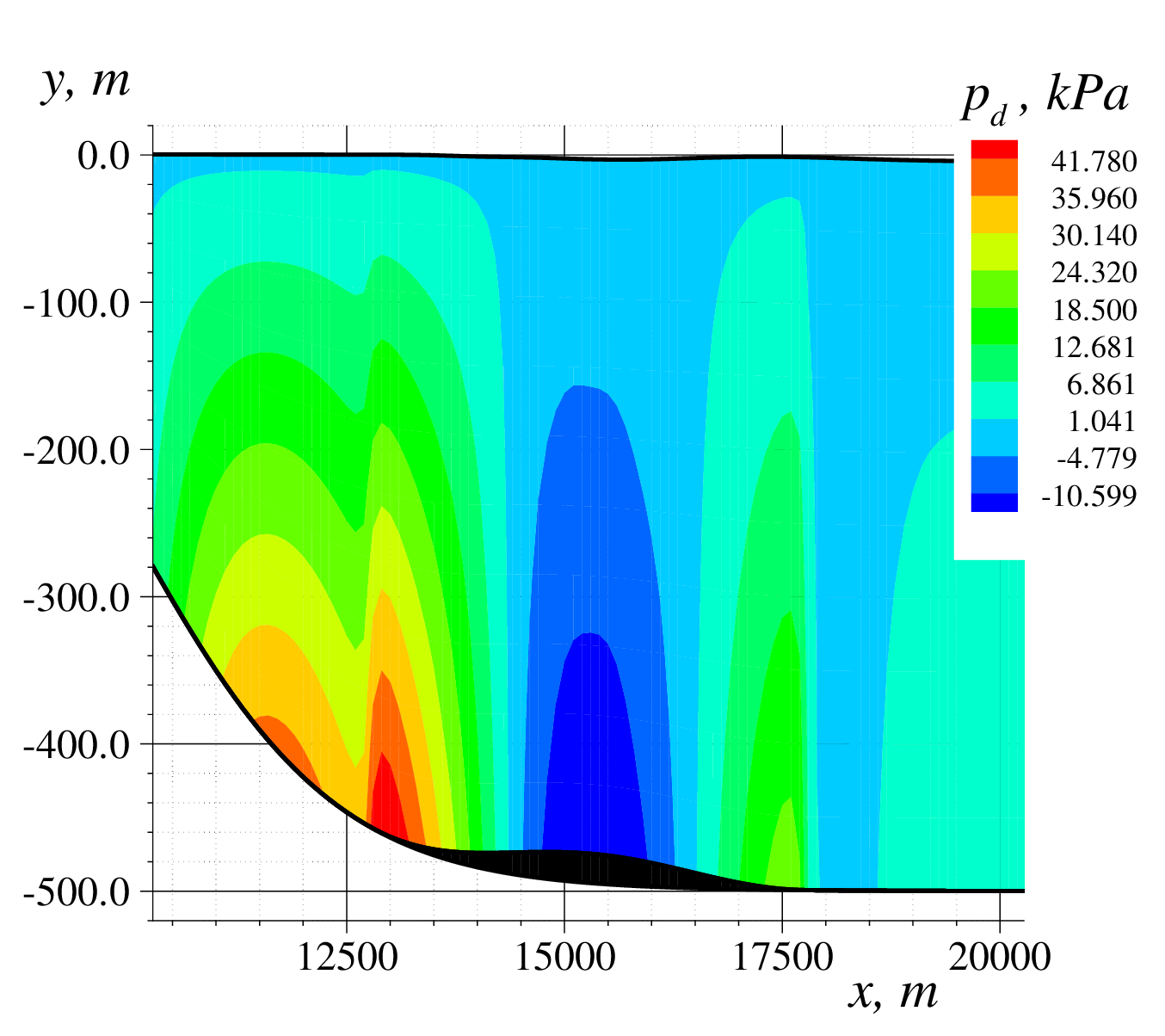}}
  \caption{\small\em Generation of surface waves by an underwater landslide. Isolines of the non-hydrostatic pressure at two moments of time: $t\ =\ 150\;\s$ (a); $t\ =\ 400\;\s$ (b). Numerical parameters are given in Table~\ref{tab:params2}.}
  \label{fig:slideP}
\end{figure}

\begin{figure}
  \centering
  \subfigure[]{\includegraphics[width=0.48\textwidth]{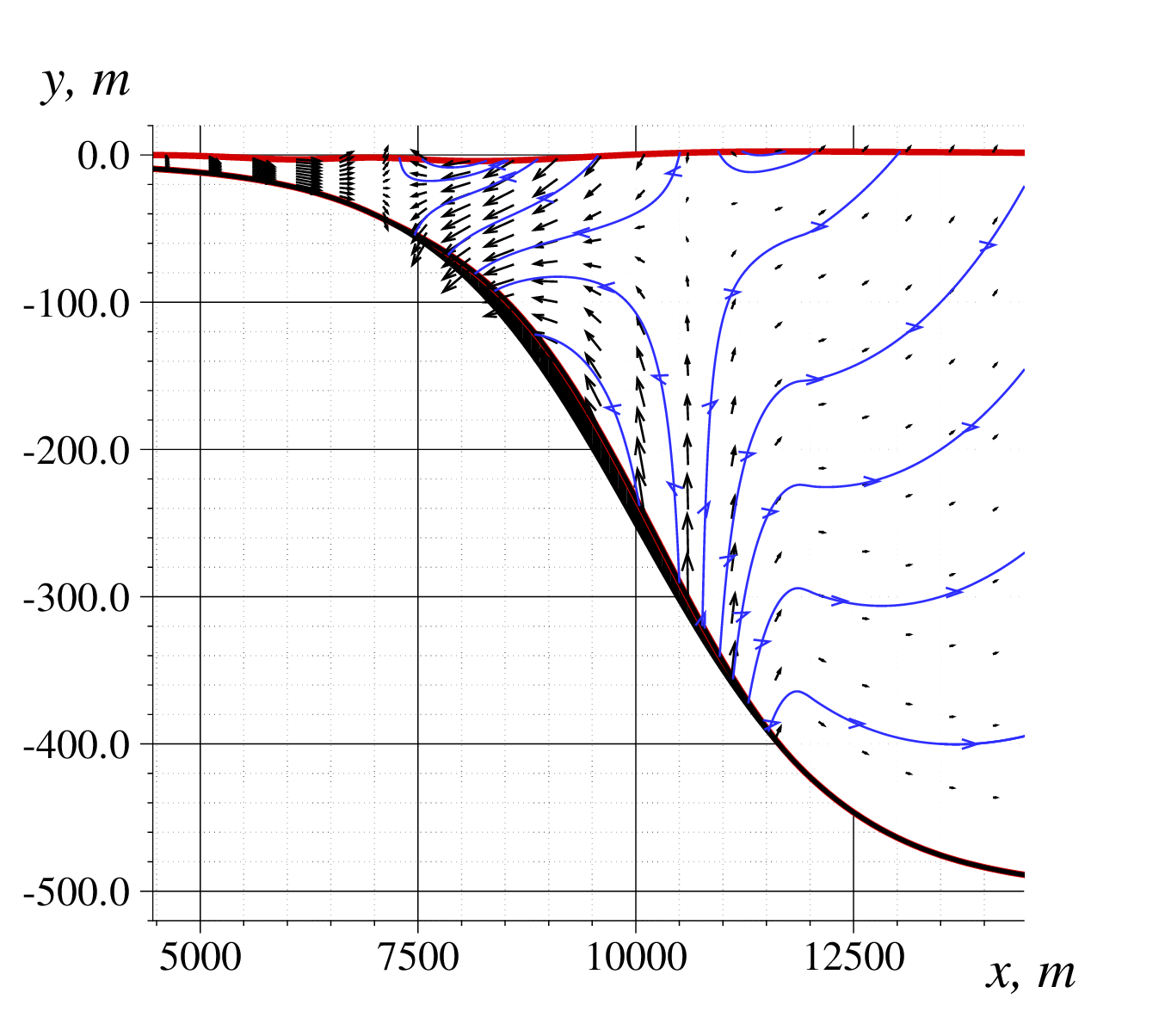}}
  \subfigure[]{\includegraphics[width=0.48\textwidth]{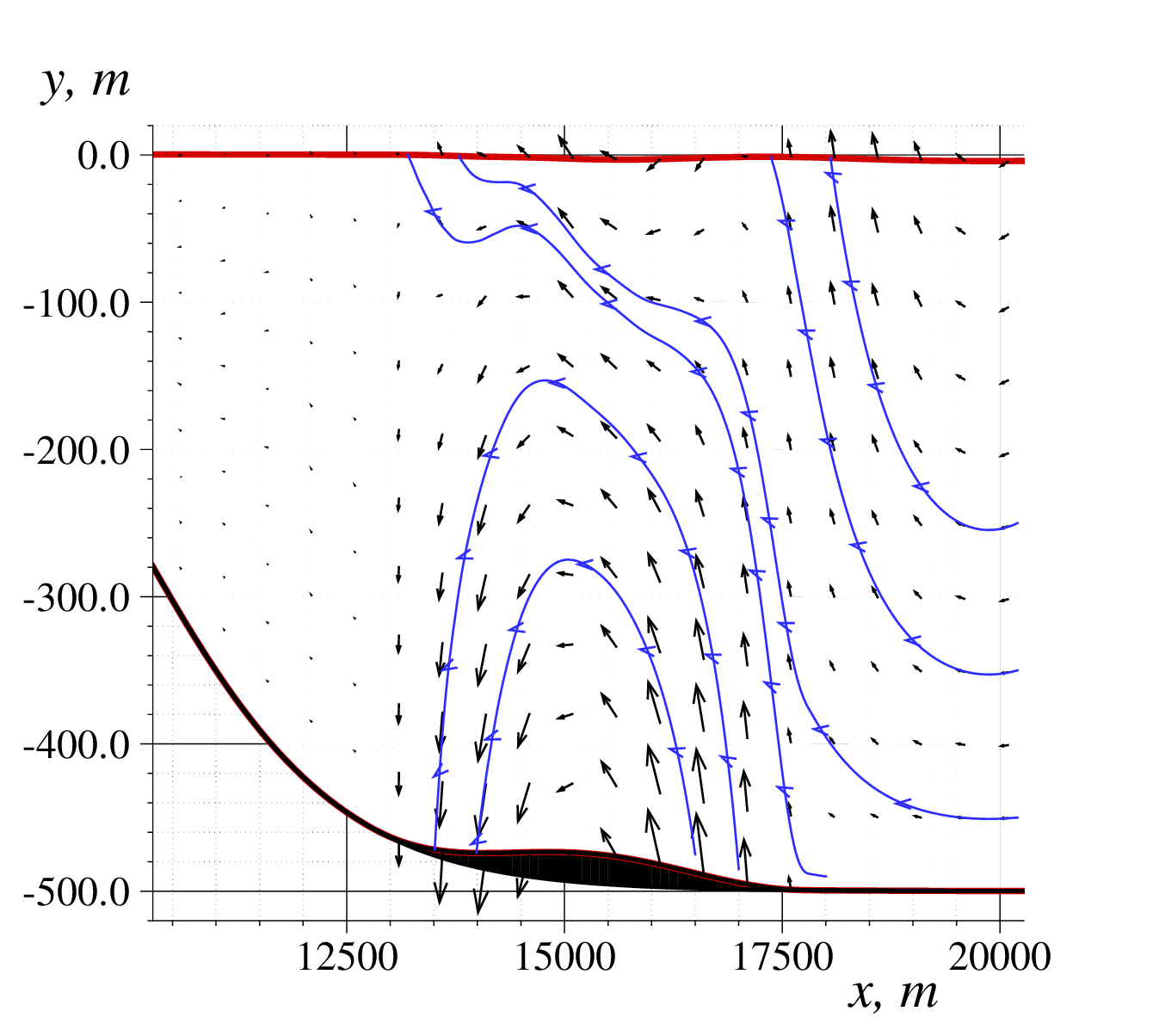}}
  \caption{\small\em Generation of surface waves by an underwater landslide. The reconstructed velocity field at two moments of time: $t\ =\ 150\;\s$ (a); $t\ =\ 400\;\s$ (b). Numerical parameters are given in Table~\ref{tab:params2}.}
  \label{fig:slideV}
\end{figure}

In the previous Section~\ref{sec:step} we showed the internal flow structure during nonlinear transformations of a solitary wave over a static step. In this Section we show that SGN equations can be used to reconstruct and to study the physical fields in situations where the bottom moves abruptly. In order to reconstruct the non-hydrostatic field between moving bottom and the free surface, one can use formula \eqref{eq:press}, but the quantity $\Rr_2$ has some extra terms due to the bottom motion:
\begin{equation*}
  \Rr_2\ =\ u_t\,h_x\ +\ u\,[\,u\,h_x\,]_x\ +\ h_{tt}\ +\ 2\,u\,h_{xt}\,.
\end{equation*}
In Figure~\ref{fig:slideP} we show the non-hydrostatic pressure field at two different moments of time. More precisely, we show a zoom on the area of interest around the landslide only. In panel (\textit{a}) $t\ =\ t_1\ =\ 150\;\s$ and the landslide barycenter is located at $x_c(t_1)\ =\ 9456\;\m\,$. Landslide moves downhill with the speed $v(t_1)\ =\ 15.72\;\m/\s$ and it continues to accelerate. In particular, one can see that there is a zone of positive pressure in front of the landslide and a zone of negative pressure just behind. This fact has implications on the fluid particle trajectories around the landslide. In right panel (\textit{b}) we show the moment of time $t\ =\ t_2\ =\ 400\;\s\,$. At this moment $x_c(t_2)\ =\ 15\,264\;\m$ and $v(t_2)\ =\ 21.4\;\m/\s\,$. The non-hydrostatic pressure distribution qualitatively changed. Zones of positive and negative pressure switched their respective positions. Moreover, in Figure~\ref{fig:profi} we showed that dispersive effects start to be noticeable at the free surface only after $t\ \geq\ 400\;\s\,$ and by $t\ =\ 800\;\s$ they are flagrant. In Figure~\ref{fig:slideV} we show the velocity fields in the fluid bulk at corresponding moments of time $t_1$ and $t_2$. We notice some similarities between the fluid flow around a landslide with an air flow around an airfoil. To our knowledge the internal hydrodynamics of landslide generated waves on a general non-constant sloping bottom and in the framework of SGN equations has not been shown before.

We remind that in the presence of moving bottom one should use the following reconstruction formulas for the velocity field (which are slightly different from \eqref{eq:rec1}, \eqref{eq:rec2}):
\begin{align*}
  \tilde{u}(x,\,y,\,t)\ &=\ u\ +\ \Bigl(\,\frac{\H}{2}\ -\ y\ -\ h\,\Bigr)\cdot\bigl(\,[\,h_t\ +\ u\,h_x\,]_x\ +\ u_x\,h_x\bigr)\ +\ \Bigl(\,\frac{\H^{\,2}}{6}\ -\ \frac{(y + h)^2}{2}\,\Bigr)\,u_{xx}\,, \\
  \tilde{v}(x,\,y,\,t)\ &=\ -h_t\ -\ u\,h_x\ -\ (y\ +\ h)\,u_x\,.
\end{align*}
These formulas naturally become \eqref{eq:rec1}, \eqref{eq:rec2} if the bottom is static, \ie $h_t\ \equiv\ 0\,$.


\section{Discussion}
\label{sec:concl}

Above we presented a detailed description of the numerical algorithm and a number of numerical tests which illustrate its performance. The main conclusions and perspectives of this study are outlined below.

\subsection{Conclusions}

In the second part of our series of papers we focused on the development of numerical algorithms for shallow water propagation over globally flat spaces (\ie we allow some variations of the bathymetry in the limits discussed in Section~\ref{sec:well}). The main distinction of our work is that the proposed algorithm allows for local mesh adaptivity by moving the grid points where they are needed. The performance of our method was illustrated on several test cases ranging from purely academic ones (\eg propagation of a solitary waves, which allowed us to estimate the overall accuracy of the scheme) to more realistic applications with landslide-generated waves \cite{Beisel2012}. The mathematical model chosen in this study allows us to have a look into the distribution of various physical fields in the fluid bulk. In particular, in some interesting cases we reconstructed the velocity field and the non-hydrostatic pressure distribution beneath the free surface.

We studied the linear stability of the proposed finite volume discretization. It was shown that the resulting scheme possesses an interesting and possible counter-intuitive property: smaller we take the spatial discretization step $\dx$, less restrictive is becoming the stability CFL-type condition on the time step $\tau$. This result was obtained using the classical \textsc{von Neumann} analysis \cite{Charney1950}. However, we show (and we compute it) that there exists the upper limit of allowed time steps. Numerical schemes with such properties seem to be new.

We considered also in great detail the question of wall boundary conditions\footnote{The wall boundary condition for the velocity component $u(x,\,t)$ is straightforward, \ie $\left.u(x,\,t)\right|_{x\,=\,0}^{x\,=\,\ell}\ =\ 0$. However, there was an open question of how to prescribe the boundary conditions for the elliptic part of the equations.} for the SGN system. It seems that this issue was not properly addressed before. The wall boundary condition for the elliptic part of equations follows naturally from the form of the momentum equation we chose in this study.

Finally, in numerical experiments we showed how depth-integrated SGN equations can be used to study nonlinear transformations of water waves over some bathymetric features (such as an underwater step or a sliding mass). Moreover, we illustrated clearly that SGN equations (and several other approximate dispersive wave models) can be successfully used to reconstruct the flow field under the wave. The accuracy of this reconstruction will be studied in future works by direct comparisons with the full \textsc{Euler} equations where these quantities are resolved.

\subsection{Perspectives}

The main focus of our study was set on the adaptive spatial discretization. The first natural continuation of our study is the generalization to 3D physical problems (\ie involving two horizontal dimensions). The main difficulty is to generalize the mesh motion algorithm to this case, even if some ideas have been proposed in the literature \cite{Azarenok2003}.

In the present computations the time step was chosen to ensure the linear CFL condition. In other words, it was chosen in order to satisfy the numerical solution stability. In future works we would like to incorporate an adaptive time stepping procedure along the lines of \eg \cite{Soderlind2006} aimed to meet the prescribed error tolerance. Of course, the extension of the numerical method presented in this study to three-dimensional flows (\ie two horizontal dimensions) represents the main important extension of our work. Further improvement of the numerical algorithm can be expected if we include also some bathymetric features (such as $\grad h$) into the monitor function $\om[\eta,\,h](x,\,t)\,$. Physically this improvement is fully justified since water waves undergo constant transformations over bottom irregularities (as illustrated in Sections~\ref{sec:step} \& \ref{sec:slide}). A priori, everything is ready to perform these further numerical experiments.

Ideally, we would like to generalize the algorithm presented in this study for the \textsc{Serre}--\textsc{Green}--\textsc{Naghdi} (SGN) equations to the base model in its most general form \eqref{eq:base3}, \eqref{eq:base4}. In this way we would be able to incorporate several fully nonlinear shallow water models (discussed in Part~\textsc{I} \cite{Khakimzyanov2016a}) in the same numerical framework. It would allow the great flexibility in applications to choose and to assess the performance of various approximate models.

Moreover, in the present study we raised the question of boundary conditions for SGN equations. However, non-reflecting (or transparent) boundary conditions would allow to take much smaller domains in many applications. Unfortunately, this question is totally open to our knowledge for the SGN equations (however, it is well understood for NSWE). In future works we plan to fill in this gap as well.

Finally, the SGN equations possess a number of variational structures. The \textsc{Hamiltonian} formulation can be found \eg in \cite{Johnson2002}. Various \textsc{Lagrangians} can be found in \cite{Miles1985, Fedotova1996, Kim2001, Clamond2009}. Recently, a multi-symplectic formulation for SGN equations has been proposed \cite{Chhay2016}. All these available variational structures raise an important question: after the discretization can we preserve them at the discrete level as well? It opens beautiful perspectives for the development of structure-preserving numerical methods as it was done for the classical \textsc{Korteweg}--\textsc{de Vries} \cite{Dutykh2013a} and nonlinear Schr\"odinger \cite{Chen2002} equations.

In the following parts of this series of papers we shall discuss the derivation of the SGN equations on a sphere \cite{Khakimzyanov2016a} and their numerical simulation using the finite volume method \cite{Khakimzyanov2016b}.


\subsection*{Acknowledgments}
\addcontentsline{toc}{subsection}{Acknowledgments}

This research was supported by RSCF project No 14--17--00219. The authors would like to thank Prof.~Emmanuel~\textsc{Audusse} (Universit\'e Paris 13, France) who brought our attention to the problem of boundary conditions for the SGN equations.


\appendix
\section{Derivation of the non-hydrostatic pressure equation}
\label{app:press}

In this Appendix we give some hints for the derivation of the non-hydrostatic pressure equation \eqref{eq:ell} and relation \eqref{eq:relpb}. Let us start with the latter. For this purpose we rewrite equation \eqref{eq:nswe} in a more compact form using the total derivative operator:
\begin{equation}\label{eq:nswe2}
  \D\u\ =\ -g\,\grad\eta\ +\ \frac{\grad\Pnh\ -\ \pb\grad h}{\H}\,,
\end{equation}
By definition of non-hydrostatic quantities $\Pnh$ and $\pb$ (see equations \eqref{eq:pnh} and \eqref{eq:pb} correspondingly) we obtain:
\begin{equation*}
  \pb\ =\ \frac{3\,\Pnh}{2\,\H}\ +\ \frac{\H}{4}\;\Rr_2\,.
\end{equation*}
We have to substitute into the last relation the expression for $\Rr_2$:
\begin{equation*}
  \Rr_2\ =\ (\D\u)\scal\grad h\ +\ \u\scal\bigl((\u\scal\grad)\grad h\bigr)\ +\ h_{tt}\ +\ 2\,\u\scal\grad h_t\,,
\end{equation*}
along with the expression \eqref{eq:nswe2} for the horizontal acceleration $\D\u$ of fluid particles. After simple algebraic computations one obtains \eqref{eq:relpb}.

The derivation of equation \eqref{eq:ell} is somehow similar. First, from definitions \eqref{eq:pnh}, \eqref{eq:pb} we obtain another relation between non-hydrostatic pressures:
\begin{equation}\label{eq:rel2}
  \Pnh\ =\ \frac{\H^{\,3}}{12}\;\Rr_1\ +\ \frac{\H}{2}\;\pb\,,
\end{equation}
with $\Rr_1$ rewritten in the following form:
\begin{equation*}
  \Rr_1\ =\ \div(\D\u)\ -\ 2\,(\div\u)^2\ +\ 2\,\begin{vmatrix}
    u_{1\,x_1} & u_{1\,x_2} \\
    u_{2\,x_1} & u_{2\,x_2}
  \end{vmatrix}\,.
\end{equation*}
Substituting into equation \eqref{eq:rel2} the just shown relation \eqref{eq:relpb} with the last expression for $\Rr_1$, yields the required equation \eqref{eq:ell}.


\section{Acronyms}

In the text above the reader could encounter the following acronyms:

\begin{description}
  \item[SW] Solitary Wave
  \item[AMR] Adaptive Mesh Refinement
  \item[BBM] \textsc{Benjamin--Bona--Mahony}
  \item[BVP] Boundary Value Problem
  \item[CFL] \textsc{Courant--Friedrichs--Lewy}
  \item[IVP] Initial Value Problem
  \item[MOL] Method Of Lines
  \item[ODE] Ordinary Differential Equation
  \item[PDE] Partial Differential Equation
  \item[SGN] \textsc{Serre--Green--Naghdi}
  \item[TVD] Total Variation Diminishing
  \item[NSWE] Nonlinear Shallow Water Equations
\end{description}


\bigskip\bigskip
\addcontentsline{toc}{section}{References}
\bibliographystyle{abbrv}

\begin{thebibliography}{}

\end{thebibliography}


\begin{thebibliography}{100}

\bibitem{Alalykin1970}
G.~B. Alalykin, S.~K. Godunov, L.~L. Kireyeva, and L.~A. Pliner.
\newblock {\em {Solution of One-Dimensional Problems in Gas Dynamics on Moving
  Grids}}.
\newblock Nauka, Moscow, 1970.

\bibitem{AntunesDoCarmo1993}
J.~S. {Antunes Do Carmo}, F.~J. {Seabra Santos}, and E.~Barth{\'{e}}lemy.
\newblock {Surface waves propagation in shallow water: A finite element model}.
\newblock {\em Int. J. Num. Meth. Fluids}, 16(6):447--459, mar 1993.

\bibitem{Arvanitis2006}
C.~Arvanitis and A.~I. Delis.
\newblock {Behavior of Finite Volume Schemes for Hyperbolic Conservation Laws
  on Adaptive Redistributed Spatial Grids}.
\newblock {\em SIAM J. Sci. Comput.}, 28(5):1927--1956, jan 2006.

\bibitem{Arvanitis2010}
C.~Arvanitis, T.~Katsaounis, and C.~Makridakis.
\newblock {Adaptive Finite Element Relaxation Schemes for Hyperbolic
  Conservation Laws}.
\newblock {\em ESAIM: Mathematical Modelling and Numerical Analysis},
  35(1):17--33, 2010.

\bibitem{Assier-Rzadkieaicz2000a}
S.~Assier-Rzadkieaicz, P.~Heinrich, P.~C. Sabatier, B.~Savoye, and J.~F.
  Bourillet.
\newblock {Numerical Modelling of a Landslide-generated Tsunami: The 1979 Nice
  Event}.
\newblock {\em Pure Appl. Geophys.}, 157(10):1707--1727, oct 2000.

\bibitem{Azarenok2003}
B.~N. Azarenok, S.~A. Ivanenko, and T.~Tang.
\newblock {Adaptive Mesh Redistibution Method Based on Godunov's Scheme}.
\newblock {\em Commun. Math. Sci.}, 1(1):152--179, 2003.

\bibitem{Bakhvalov1969}
N.~S. Bakhvalov.
\newblock {The optimization of methods of solving boundary value problems with
  a boundary layer}.
\newblock {\em USSR Computational Mathematics and Mathematical Physics},
  9(4):139--166, jan 1969.

\bibitem{Barakhnin1997}
V.~B. Barakhnin and G.~S. Khakimzyanov.
\newblock {On the algorithm for one nonlinear dispersive shallow-water model}.
\newblock {\em Russ. J. Numer. Anal. Math. Modelling}, 12(4):293--317, 1997.

\bibitem{Barakhnin1999}
V.~B. Barakhnin and G.~S. Khakimzyanov.
\newblock {The splitting technique as applied to the solution of the nonlinear
  dispersive shallow-water equations}.
\newblock {\em Doklady Mathematics}, 59(1):70--72, 1999.

\bibitem{Barth2004}
T.~J. Barth and M.~Ohlberger.
\newblock {Finite Volume Methods: Foundation and Analysis}.
\newblock In E.~Stein, R.~de~Borst, and T.~J.~R. Hughes, editors, {\em
  Encyclopedia of Computational Mechanics}. John Wiley {\&} Sons, Ltd,
  Chichester, UK, nov 2004.

\bibitem{Barthelemy2004}
E.~Barth{\'{e}}l{\'{e}}my.
\newblock {Nonlinear shallow water theories for coastal waves}.
\newblock {\em Surveys in Geophysics}, 25:315--337, 2004.

\bibitem{Beisel2012}
S.~A. Beisel, L.~B. Chubarov, D.~Dutykh, G.~S. Khakimzyanov, and N.~Y. Shokina.
\newblock {Simulation of surface waves generated by an underwater landslide in
  a bounded reservoir}.
\newblock {\em Russ. J. Numer. Anal. Math. Modelling}, 27(6):539--558, 2012.

\bibitem{Beisel2011}
S.~A. Beisel, L.~B. Chubarov, and G.~S. Khakimzyanov.
\newblock {Simulation of surface waves generated by an underwater landslide
  moving over an uneven slope}.
\newblock {\em Russ. J. Numer. Anal. Math. Modelling}, 26(1):17--38, 2011.

\bibitem{bona}
T.~B. Benjamin, J.~L. Bona, and J.~J. Mahony.
\newblock {Model equations for long waves in nonlinear dispersive systems}.
\newblock {\em Philos. Trans. Royal Soc. London Ser. A}, 272:47--78, 1972.

\bibitem{Berger2011}
M.~J. Berger, D.~L. George, R.~J. LeVeque, and K.~T. Mandli.
\newblock {The GeoClaw software for depth-averaged flows with adaptive
  refinement}.
\newblock {\em Advances in Water Resources}, 34(9):1195--1206, sep 2011.

\bibitem{Bohr1920}
N.~Bohr.
\newblock {{\"{U}}ber die Serienspektra der Element}.
\newblock {\em Zeitschrift f{\"{u}}r Physik}, 2(5):423--469, oct 1920.

\bibitem{Bona1986}
J.~L. Bona, V.~A. Dougalis, and O.~A. Karakashian.
\newblock {Fully discrete Galerkin methods for the Korteweg-de Vries equation}.
\newblock {\em Computers {\&} Mathematics with Applications}, 12(7):859--884,
  jul 1986.

\bibitem{Bonneton2011}
P.~Bonneton, F.~Chazel, D.~Lannes, F.~Marche, and M.~Tissier.
\newblock {A splitting approach for the fully nonlinear and weakly dispersive
  Green-Naghdi model}.
\newblock {\em J. Comput. Phys.}, 230:1479--1498, 2011.

\bibitem{Bristeau2011}
M.-O. Bristeau, N.~Goutal, and J.~Sainte-Marie.
\newblock {Numerical simulations of a non-hydrostatic shallow water model}.
\newblock {\em Comput. {\&} Fluids}, 47(1):51--64, aug 2011.

\bibitem{Brocchini2013}
M.~Brocchini.
\newblock {A reasoned overview on Boussinesq-type models: the interplay between
  physics, mathematics and numerics}.
\newblock {\em Proc. R. Soc. A}, 469(2160):20130496, oct 2013.

\bibitem{Byatt-Smith1988}
J.~G.~B. Byatt-Smith.
\newblock {The reflection of a solitary wave by a vertical wall}.
\newblock {\em J. Fluid Mech.}, 197:503--521, 1988.

\bibitem{Carbone2013}
F.~Carbone, D.~Dutykh, J.~M. Dudley, and F.~Dias.
\newblock {Extreme wave run-up on a vertical cliff}.
\newblock {\em Geophys. Res. Lett.}, 40(12):3138--3143, 2013.

\bibitem{Castro2013}
M.~J. Castro, M.~de~la Asuncion, J.~Macias, C.~Par{\'{e}}s, E.~D.
  Fernandez-Nieto, J.~M. Gonzalez-Vida, and T.~{Morales de Luna}.
\newblock {IFCP Riemann solver: Application to tsunami modelling using CPUs}.
\newblock In M.~E. Vazquez-Cendon, A.~Hidalgo, P.~Garcia-Navarro, and L.~Cea,
  editors, {\em Numerical Methods for Hyperbolic Equations: Theory and
  Applications}, pages 237--244. CRC Press, Boca Raton, London, New York,
  Leiden, 2013.

\bibitem{Casulli1999}
V.~Casulli.
\newblock {A semi-implicit finite difference method for non-hydrostatic,
  free-surface flows}.
\newblock {\em Int. J. Num. Meth. Fluids}, 30(4):425--440, jun 1999.

\bibitem{Chambarel2009}
J.~Chambarel, C.~Kharif, and J.~Touboul.
\newblock {Head-on collision of two solitary waves and residual falling jet
  formation}.
\newblock {\em Nonlin. Processes Geophys.}, 16:111--122, 2009.

\bibitem{Chan1970a}
R.~K.-C. Chan and R.~L. Street.
\newblock {A computer study of finite-amplitude water waves}.
\newblock {\em J. Comp. Phys.}, 6(1):68--94, aug 1970.

\bibitem{Chang2001}
K.-A. Chang, T.-J. Hsu, and P.~L.-F. Liu.
\newblock {Vortex generation and evolution in water waves propagating over a
  submerged rectangular obstacle}.
\newblock {\em Coastal Engineering}, 44(1):13--36, sep 2001.

\bibitem{Charney1950}
J.~G. Charney, R.~Fj{\"{o}}rtoft, and J.~Neumann.
\newblock {Numerical Integration of the Barotropic Vorticity Equation}.
\newblock {\em Tellus}, 2(4):237--254, nov 1950.

\bibitem{Chazel2007}
F.~Chazel.
\newblock {Influence of bottom topography on long water waves}.
\newblock {\em M2AN}, 41:771--799, 2007.

\bibitem{ChazelLannes2010}
F.~Chazel, D.~Lannes, and F.~Marche.
\newblock {Numerical simulation of strongly nonlinear and dispersive waves
  using a Green-Naghdi model}.
\newblock {\em J. Sci. Comput.}, 48:105--116, 2011.

\bibitem{Chen2002}
J.-B. Chen, M.-Z. Qin, and Y.-F. Tang.
\newblock {Symplectic and multi-symplectic methods for the nonlinear
  Schr{\"{o}}dinger equation}.
\newblock {\em Computers {\&} Mathematics with Applications},
  43(8-9):1095--1106, apr 2002.

\bibitem{Chhay2016}
M.~Chhay, D.~Dutykh, and D.~Clamond.
\newblock {On the multi-symplectic structure of the Serre-Green-Naghdi
  equations}.
\newblock {\em J. Phys. A: Math. Gen}, 49(3):03LT01, jan 2016.

\bibitem{Chorin1968}
A.~Chorin.
\newblock {Numerical solution of the Navier-Stokes equations}.
\newblock {\em Math. Comp.}, 22:745--762, 1968.

\bibitem{Chubarov2005}
L.~B. Chubarov, S.~V. Eletsky, Z.~I. Fedotova, and G.~S. Khakimzyanov.
\newblock {Simulation of surface waves by an underwater landslide}.
\newblock {\em Russ. J. Numer. Anal. Math. Modelling}, 20(5):425--437, 2005.

\bibitem{Chubarov2000}
L.~B. Chubarov, Z.~I. Fedotova, Y.~I. Shokin, and B.~G. Einarsson.
\newblock {Comparative Analysis of Nonlinear Dispersive Shallow Water Models}.
\newblock {\em Int. J. Comp. Fluid Dyn.}, 14(1):55--73, jan 2000.

\bibitem{Chubarov1987}
L.~B. Chubarov and Y.~I. Shokin.
\newblock {The numerical modelling of long wave propagation in the framework of
  non-linear dispersion models}.
\newblock {\em Comput. {\&} Fluids}, 15(3):229--249, jan 1987.

\bibitem{Cienfuegos2007}
R.~Cienfuegos, E.~Barth{\'{e}}lemy, and P.~Bonneton.
\newblock {A fourth-order compact finite volume scheme for fully nonlinear and
  weakly dispersive Boussinesq-type equations. Part II: boundary conditions and
  validation}.
\newblock {\em Int. J. Num. Meth. Fluids}, 53(9):1423--1455, mar 2007.

\bibitem{Clamond2009}
D.~Clamond and D.~Dutykh.
\newblock {Practical use of variational principles for modeling water waves}.
\newblock {\em Phys. D}, 241(1):25--36, 2012.

\bibitem{Cooker1997}
M.~J. Cooker, P.~D. Weidman, and D.~S. Bale.
\newblock {Reflection of a high-amplitude solitary wave at a vertical wall}.
\newblock {\em J. Fluid Mech.}, 342:141--158, 1997.

\bibitem{Courant1928}
R.~Courant, K.~Friedrichs, and H.~Lewy.
\newblock {{\"{U}}ber die partiellen Differenzengleichungen der mathematischen
  Physik}.
\newblock {\em Mathematische Annalen}, 100(1):32--74, 1928.

\bibitem{Tkalich2007}
M.~H. Dao and P.~Tkalich.
\newblock {Tsunami propagation modelling - a sensitivity study}.
\newblock {\em Nat. Hazards Earth Syst. Sci.}, 7:741--754, 2007.

\bibitem{Davletshin1984}
V.~H. Davletshin.
\newblock {Force action of solitary waves on vertical structures}.
\newblock In {\em Tsunami meeting}, pages 41--43, Gorky, 1984. Institute of
  Applied Physics.

\bibitem{SV1871}
A.~J.~C. de~Saint-Venant.
\newblock {Th{\'{e}}orie du mouvement non-permanent des eaux, avec application
  aux crues des rivi{\`{e}}res et {\`{a}} l'introduction des mar{\'{e}}es dans
  leur lit}.
\newblock {\em C. R. Acad. Sc. Paris}, 73:147--154, 1871.

\bibitem{Dingemans1997}
M.~W. Dingemans.
\newblock {\em {Water wave propagation over uneven bottom}}.
\newblock World Scientific, Singapore, 1997.

\bibitem{Dougalis1985}
V.~A. Dougalis and O.~A. Karakashian.
\newblock {On Some High-Order Accurate Fully Discrete Galerkin Methods for the
  Korteweg-de Vries Equation}.
\newblock {\em Mathematics of Computation}, 45(172):329, oct 1985.

\bibitem{DMII}
V.~A. Dougalis and D.~E. Mitsotakis.
\newblock {Theory and numerical analysis of Boussinesq systems: A review}.
\newblock In N.~A. Kampanis, V.~A. Dougalis, and J.~A. Ekaterinaris, editors,
  {\em Effective Computational Methods in Wave Propagation}, pages 63--110. CRC
  Press, 2008.

\bibitem{DMS1}
V.~A. Dougalis, D.~E. Mitsotakis, and J.-C. Saut.
\newblock {On some Boussinesq systems in two space dimensions: Theory and
  numerical analysis}.
\newblock {\em Math. Model. Num. Anal.}, 41(5):254--825, 2007.

\bibitem{John}
P.~G. Drazin and R.~S. Johnson.
\newblock {\em {Solitons: An introduction}}.
\newblock Cambridge University Press, Cambridge, 1989.

\bibitem{Duran2013}
A.~Duran, D.~Dutykh, and D.~Mitsotakis.
\newblock {On the Galilean Invariance of Some Nonlinear Dispersive Wave
  Equations}.
\newblock {\em Stud. Appl. Math.}, 131(4):359--388, nov 2013.

\bibitem{Dutykh2013a}
D.~Dutykh, M.~Chhay, and F.~Fedele.
\newblock {Geometric numerical schemes for the KdV equation}.
\newblock {\em Comp. Math. Math. Phys.}, 53(2):221--236, 2013.

\bibitem{Dutykh2013b}
D.~Dutykh and D.~Clamond.
\newblock {Efficient computation of steady solitary gravity waves}.
\newblock {\em Wave Motion}, 51(1):86--99, jan 2014.

\bibitem{Dutykh2011a}
D.~Dutykh, D.~Clamond, P.~Milewski, and D.~Mitsotakis.
\newblock {Finite volume and pseudo-spectral schemes for the fully nonlinear 1D
  Serre equations}.
\newblock {\em Eur. J. Appl. Math.}, 24(05):761--787, 2013.

\bibitem{Dutykh2007}
D.~Dutykh and F.~Dias.
\newblock {Dissipative Boussinesq equations}.
\newblock {\em C. R. Mecanique}, 335:559--583, 2007.

\bibitem{Dutykh2009b}
D.~Dutykh and F.~Dias.
\newblock {Energy of tsunami waves generated by bottom motion}.
\newblock {\em Proc. R. Soc. A}, 465:725--744, 2009.

\bibitem{Dutykh2016}
D.~Dutykh and D.~Ionescu-Kruse.
\newblock {Travelling wave solutions for some two-component shallow water
  models}.
\newblock {\em J. Diff. Eqns.}, 261(2):1099--1114, jul 2016.

\bibitem{Dutykh2011d}
D.~Dutykh and H.~Kalisch.
\newblock {Boussinesq modeling of surface waves due to underwater landslides}.
\newblock {\em Nonlin. Processes Geophys.}, 20(3):267--285, may 2013.

\bibitem{Dutykh2011e}
D.~Dutykh, T.~Katsaounis, and D.~Mitsotakis.
\newblock {Finite volume schemes for dispersive wave propagation and runup}.
\newblock {\em J. Comput. Phys.}, 230(8):3035--3061, apr 2011.

\bibitem{Dutykh2010e}
D.~Dutykh, T.~Katsaounis, and D.~Mitsotakis.
\newblock {Finite volume methods for unidirectional dispersive wave models}.
\newblock {\em Int. J. Num. Meth. Fluids}, 71:717--736, 2013.

\bibitem{Dutykh2012}
D.~Dutykh, D.~Mitsotakis, S.~A. Beisel, and N.~Y. Shokina.
\newblock {Dispersive waves generated by an underwater landslide}.
\newblock In E.~Vazquez-Cendon, A.~Hidalgo, P.~Garcia-Navarro, and L.~Cea,
  editors, {\em Numerical Methods for Hyperbolic Equations: Theory and
  Applications}, pages 245--250. CRC Press, Boca Raton, London, New York,
  Leiden, 2013.

\bibitem{Enet2007}
F.~Enet and S.~T. Grilli.
\newblock {Experimental study of tsunami generation by three-dimensional rigid
  underwater landslides}.
\newblock {\em J. Waterway, Port, Coastal and Ocean Engineering},
  133(6):442--454, 2007.

\bibitem{Ertekin1986}
R.~C. Ertekin, W.~C. Webster, and J.~V. Wehausen.
\newblock {Waves caused by a moving disturbance in a shallow channel of finite
  width}.
\newblock {\em J. Fluid Mech.}, 169:275--292, aug 1986.

\bibitem{Fabien2014}
M.~S. Fabien.
\newblock {\em {Spectral Methods for Partial Dierential Equations that Model
  Shallow Water Wave Phenomena}}.
\newblock Master, University of Washington, 2014.

\bibitem{Fedotova1996}
Z.~I. Fedotova and E.~D. Karepova.
\newblock {Variational principle for approximate models of wave hydrodynamics}.
\newblock {\em Russ. J. Numer. Anal. Math. Modelling}, 11(3):183--204, 1996.

\bibitem{Fedotova2009}
Z.~I. Fedotova and G.~S. Khakimzyanov.
\newblock {Shallow water equations on a movable bottom}.
\newblock {\em Russ. J. Numer. Anal. Math. Modelling}, 24(1):31--42, 2009.

\bibitem{Fedotova2014}
Z.~I. Fedotova, G.~S. Khakimzyanov, and D.~Dutykh.
\newblock {Energy equation for certain approximate models of long-wave
  hydrodynamics}.
\newblock {\em Russ. J. Numer. Anal. Math. Modelling}, 29(3):167--178, jan
  2014.

\bibitem{Fedotova1997}
Z.~I. Fedotova and V.~Y. Pashkova.
\newblock {Methods of construction and the analysis of difference schemes for
  nonlinear dispersive models of wave hydrodynamics}.
\newblock {\em Russ. J. Numer. Anal. Math. Modelling}, 12(2), 1997.

\bibitem{Fenton1982}
J.~D. Fenton and M.~M. Rienecker.
\newblock {A Fourier method for solving nonlinear water-wave problems:
  application to solitary-wave interactions}.
\newblock {\em J. Fluid Mech.}, 118:411--443, apr 1982.

\bibitem{Fernandez-Nieto2007}
E.~D. Fernandez-Nieto, F.~Bouchut, D.~Bresch, M.~J. Castro-Diaz, and
  A.~Mangeney.
\newblock {A new Savage-Hutter type models for submarine avalanches and
  generated tsunami}.
\newblock {\em J. Comput. Phys.}, 227(16):7720--7754, 2008.

\bibitem{Flierl1981}
G.~R. Flierl.
\newblock {Particle motions in large-amplitude wave fields}.
\newblock {\em Geophysical {\&} Astrophysical Fluid Dynamics}, 18(1-2):39--74,
  aug 1981.

\bibitem{Glimsdal2006}
S.~Glimsdal, G.~K. Pedersen, K.~Atakan, C.~B. Harbitz, H.~P. Langtangen, and
  F.~Lovholt.
\newblock {Propagation of the Dec. 26, 2004, Indian Ocean Tsunami: Effects of
  Dispersion and Source Characteristics}.
\newblock {\em Int. J. Fluid Mech. Res.}, 33(1):15--43, 2006.

\bibitem{Glimsdal2013}
S.~Glimsdal, G.~K. Pedersen, C.~B. Harbitz, and F.~L{\o}vholt.
\newblock {Dispersion of tsunamis: does it really matter?}
\newblock {\em Natural Hazards and Earth System Science}, 13(6):1507--1526, jun
  2013.

\bibitem{Green1974}
A.~E. Green, N.~Laws, and P.~M. Naghdi.
\newblock {On the theory of water waves}.
\newblock {\em Proc. R. Soc. Lond. A}, 338:43--55, 1974.

\bibitem{Vogel}
S.~Grilli, S.~Vogelmann, and P.~Watts.
\newblock {Development of a 3D numerical wave tank for modeling tsunami
  generation by underwater landslides}.
\newblock {\em Engng Anal. Bound. Elem.}, 26:301--313, 2002.

\bibitem{Grilli2005}
S.~T. Grilli and P.~Watts.
\newblock {Tsunami Generation by Submarine Mass Failure. I: Modeling,
  Experimental Validation, and Sensitivity Analyses}.
\newblock {\em Journal of Waterway Port Coastal and Ocean Engineering},
  131(6):283, 2005.

\bibitem{Grue2008}
J.~Grue, E.~N. Pelinovsky, D.~Fructus, T.~Talipova, and C.~Kharif.
\newblock {Formation of undular bores and solitary waves in the Strait of
  Malacca caused by the 26 December 2004 Indian Ocean tsunami}.
\newblock {\em J. Geophys. Res.}, 113(C5):C05008, may 2008.

\bibitem{Hairer2009}
E.~Hairer, S.~P. N{\o}rsett, and G.~Wanner.
\newblock {\em {Solving ordinary differential equations: Nonstiff problems}}.
\newblock Springer, 2009.

\bibitem{Hairer1996}
E.~Hairer and G.~Wanner.
\newblock {\em {Solving Ordinary Differential Equations II. Stiff and
  Differential-Algebraic Problems}}.
\newblock Springer Series in Computational Mathematics, Vol. 14, 1996.

\bibitem{Hammack2004}
J.~Hammack, D.~Henderson, P.~Guyenne, and M.~Yi.
\newblock {Solitary wave collisions}.
\newblock In {\em Proc. 23rd International Conference on Offshore Mechanics and
  Arctic Engineering}, 2004.

\bibitem{Harlow1965}
F.~H. Harlow and J.~E. Welch.
\newblock {Numerical Calculation of Time-Dependent Viscous Incompressible Flow
  of Fluid with Free Surface}.
\newblock {\em Phys. Fluids}, 8:2182, 1965.

\bibitem{Hermes1973}
H.~Hermes.
\newblock {\em {Introduction to Mathematical Logic}}.
\newblock Universitext. Springer Berlin Heidelberg, Berlin, Heidelberg, 1973.

\bibitem{Higham2002}
N.~J. Higham.
\newblock {\em {Accuracy and Stability of Numerical Algorithms}}.
\newblock SIAM Philadelphia, 2nd ed. edition, 2002.

\bibitem{Horrillo2006}
J.~Horrillo, Z.~Kowalik, and Y.~Shigihara.
\newblock {Wave Dispersion Study in the Indian Ocean-Tsunami of December 26,
  2004}.
\newblock {\em Marine Geodesy}, 29(3):149--166, dec 2006.

\bibitem{Huang2001a}
W.~Huang.
\newblock {Practical Aspects of Formulation and Solution of Moving Mesh Partial
  Differential Equations}.
\newblock {\em J. Comp. Phys.}, 171(2):753--775, aug 2001.

\bibitem{Huang2001}
W.~Huang and R.~D. Russell.
\newblock {Adaptive mesh movement - the MMPDE approach and its applications}.
\newblock {\em J. Comp. Appl. Math.}, 128(1-2):383--398, mar 2001.

\bibitem{Ilin1969}
A.~M. Il'in.
\newblock {Differencing scheme for a differential equation with a small
  parameter affecting the highest derivative}.
\newblock {\em Mathematical Notes of the Academy of Sciences of the USSR},
  6(2):596--602, aug 1969.

\bibitem{Ioualalen2010}
M.~Ioualalen, S.~Migeon, and O.~Sardoux.
\newblock {Landslide tsunami vulnerability in the Ligurian Sea: case study of
  the 1979 October 16 Nice international airport submarine landslide and of
  identified geological mass failures}.
\newblock {\em Geophys. J. Int.}, 181(2):724--740, mar 2010.

\bibitem{Johnson2002}
R.~S. Johnson.
\newblock {Camassa-Holm, Korteweg-de Vries and related models for water waves}.
\newblock {\em J. Fluid Mech.}, 455:63--82, 2002.

\bibitem{Kabbaj1985}
A.~Kabbaj.
\newblock {\em {Contribution {\`{a}} l'{\'{e}}tude du passage des ondes de
  gravit{\'{e}} et de la g{\'{e}}n{\'{e}}ration des ondes internes sur un
  talus, dans le cadre de la th{\'{e}}orie de l'eau peu profonde}}.
\newblock Th{\`{e}}se, Universit{\'{e}} Scientifique et M{\'{e}}dicale de
  Grenoble, 1985.

\bibitem{Kazolea2013}
M.~Kazolea and A.~I. Delis.
\newblock {A well-balanced shock-capturing hybrid finite volume-finite
  difference numerical scheme for extended 1D Boussinesq models}.
\newblock {\em Appl. Numer. Math.}, 67:167--186, 2013.

\bibitem{Khakimzyanov2015b}
G.~Khakimzyanov and D.~Dutykh.
\newblock {On supraconvergence phenomenon for second order centered finite
  differences on non-uniform grids}.
\newblock {\em J. Comp. Appl. Math.}, 326:1--14, dec 2017.

\bibitem{Khakimzyanov2016a}
G.~S. Khakimzyanov, D.~Dutykh, and Z.~I. Fedotova.
\newblock {Dispersive shallow water wave modelling. Part III: Model derivation
  on a globally spherical geometry}.
\newblock {\em Submitted}, pages 1--40, 2017.

\bibitem{Khakimzyanov2016c}
G.~S. Khakimzyanov, D.~Dutykh, Z.~I. Fedotova, and D.~E. Mitsotakis.
\newblock {Dispersive shallow water wave modelling. Part I: Model derivation on
  a globally flat space}.
\newblock {\em Submitted}, pages 1--40, 2017.

\bibitem{Khakimzyanov2016b}
G.~S. Khakimzyanov, D.~Dutykh, and O.~Gusev.
\newblock {Dispersive shallow water wave modelling. Part IV: Numerical
  simulation on a globally spherical geometry}.
\newblock {\em Submitted}, pages 1--40, 2017.

\bibitem{Khakimzyanov2015a}
G.~S. Khakimzyanov, D.~Dutykh, D.~E. Mitsotakis, and N.~Y. Shokina.
\newblock {Numerical solution of conservation laws on moving grids}.
\newblock {\em Submitted}, pages 1--28, 2017.

\bibitem{Kim2001}
J.~W. Kim, K.~J. Bai, R.~C. Ertekin, and W.~C. Webster.
\newblock {A derivation of the Green-Naghdi equations for irrotational flows}.
\newblock {\em J. Eng. Math.}, 40(1):17--42, 2001.

\bibitem{Kim2000}
J.~W. Kim and R.~C. Ertekin.
\newblock {A numerical study of nonlinear wave interaction in regular and
  irregular seas: irrotational Green-Naghdi model}.
\newblock {\em Marine Structures}, 13(4-5):331--347, jul 2000.

\bibitem{Kreiss1992}
H.~O. Kreiss and G.~Scherer.
\newblock {Method of lines for hyperbolic equations}.
\newblock {\em SIAM Journal on Numerical Analysis}, 29:640--646, 1992.

\bibitem{Kurkin2015}
A.~A. Kurkin, S.~V. Semin, and Y.~A. Stepanyants.
\newblock {Transformation of surface waves over a bottom step}.
\newblock {\em Izvestiya, Atmospheric and Oceanic Physics}, 51(2):214--223, mar
  2015.

\bibitem{Lai2010}
Z.~Lai, C.~Chen, G.~W. Cowles, and R.~C. Beardsley.
\newblock {A nonhydrostatic version of FVCOM: 1. Validation experiments}.
\newblock {\em J. Geophys. Res.}, 115(C11):C11010, nov 2010.

\bibitem{LeMetayer2010}
O.~{Le M{\'{e}}tayer}, S.~Gavrilyuk, and S.~Hank.
\newblock {A numerical scheme for the Green-Naghdi model}.
\newblock {\em J. Comp. Phys.}, 229(6):2034--2045, 2010.

\bibitem{LeRoux2012}
D.~Y. {Le Roux}.
\newblock {Spurious inertial oscillations in shallow-water models}.
\newblock {\em J. Comp. Phys.}, 231(24):7959--7987, oct 2012.

\bibitem{Lindstrom2014}
E.~K. Lindstr{\o}m, G.~K. Pedersen, A.~Jensen, and S.~Glimsdal.
\newblock {Experiments on slide generated waves in a 1:500 scale fjord model}.
\newblock {\em Coastal Engineering}, 92:12--23, oct 2014.

\bibitem{Liu2005a}
P.~L.-F. Liu, T.-R. Wu, F.~Raichlen, C.~E. Synolakis, and J.~C. Borrero.
\newblock {Runup and rundown generated by three-dimensional sliding masses}.
\newblock {\em J. Fluid Mech.}, 536(1):107--144, jul 2005.

\bibitem{Lovholt2008}
F.~L{\o}vholt, G.~Pedersen, and G.~Gisler.
\newblock {Oceanic propagation of a potential tsunami from the La Palma
  Island}.
\newblock {\em J. Geophys. Res.}, 113(C9):C09026, sep 2008.

\bibitem{Lynett2002}
P.~Lynett and P.~L.~F. Liu.
\newblock {A numerical study of submarine-landslide-generated waves and
  run-up}.
\newblock {\em Proc. R. Soc. A}, 458(2028):2885--2910, dec 2002.

\bibitem{Madsen1969}
O.~S. Madsen and C.~C. Mei.
\newblock {The transformation of a solitary wave over an uneven bottom}.
\newblock {\em J. Fluid Mech.}, 39(04):781--791, dec 1969.

\bibitem{Madsen2002}
P.~A. Madsen, H.~B. Bingham, and H.~Liu.
\newblock {A new Boussinesq method for fully nonlinear waves from shallow to
  deep water}.
\newblock {\em J. Fluid Mech.}, 462:1--30, 2002.

\bibitem{Madsen1991}
P.~A. Madsen, R.~Murray, and O.~R. Sorensen.
\newblock {A new form of the Boussinesq equations with improved linear
  dispersion characteristics}.
\newblock {\em Coastal Engineering}, 15:371--388, 1991.

\bibitem{Madsen1992}
P.~A. Madsen and O.~R. Sorensen.
\newblock {A new form of the Boussinesq equations with improved linear
  dispersion characteristics. Part 2. A slowly-varying bathymetry}.
\newblock {\em Coastal Engineering}, 18:183--204, 1992.

\bibitem{Manoylin1989}
S.~V. Manoylin.
\newblock {Some experimental and theoretical methods of estimation of tsunami
  wave action on hydro-technical structures and seaports}.
\newblock Technical report, Siberian Branch of Computing Center, Krasnoyarsk,
  1989.

\bibitem{Maxworthy1976}
T.~Maxworthy.
\newblock {Experiments on collisions between solitary waves}.
\newblock {\em J. Fluid Mech}, 76:177--185, 1976.

\bibitem{Miles1985}
J.~W. Miles and R.~Salmon.
\newblock {Weakly dispersive nonlinear gravity waves}.
\newblock {\em J. Fluid Mech.}, 157:519--531, 1985.

\bibitem{Mirie1982}
S.~M. Mirie and C.~H. Su.
\newblock {Collision between two solitary waves. Part 2. A numerical study}.
\newblock {\em J. Fluid Mech.}, 115:475--492, 1982.

\bibitem{Mitsotakis2014}
D.~Mitsotakis, B.~Ilan, and D.~Dutykh.
\newblock {On the Galerkin/Finite-Element Method for the Serre Equations}.
\newblock {\em J. Sci. Comput.}, 61(1):166--195, feb 2014.

\bibitem{Pelinovsky2010}
E.~Pelinovsky, B.~H. Choi, T.~Talipova, S.~B. Wood, and D.~C. Kim.
\newblock {Solitary wave transformation on the underwater step: Asymptotic
  theory and numerical experiments}.
\newblock {\em Applied Mathematics and Computation}, 217(4):1704--1718, oct
  2010.

\bibitem{Peregrine1966}
D.~H. Peregrine.
\newblock {Calculations of the development of an undular bore}.
\newblock {\em J. Fluid Mech.}, 25(02):321--330, mar 1966.

\bibitem{Peregrine1967}
D.~H. Peregrine.
\newblock {Long waves on a beach}.
\newblock {\em J. Fluid Mech.}, 27:815--827, 1967.

\bibitem{Peregrine2003}
D.~H. Peregrine.
\newblock {Water-Wave Impact on Walls}.
\newblock {\em Annu. Rev. Fluid Mech.}, 35:23--43, 2003.

\bibitem{Ranguelov2008}
B.~Ranguelov, S.~Tinti, G.~Pagnoni, R.~Tonini, F.~Zaniboni, and A.~Armigliato.
\newblock {The nonseismic tsunami observed in the Bulgarian Black Sea on 7 May
  2007: Was it due to a submarine landslide?}
\newblock {\em Geophys. Res. Lett.}, 35(18):L18613, sep 2008.

\bibitem{Reddy1992}
S.~C. Reddy and L.~N. Trefethen.
\newblock {Stability of the method of lines}.
\newblock {\em Numerische Mathematik}, 62(1):235--267, 1992.

\bibitem{Sadaka2012}
G.~Sadaka.
\newblock {Solution of 2D Boussinesq systems with FreeFem++: the flat bottom
  case}.
\newblock {\em Journal of Numerical Mathematics}, 20(3-4):303--324, jan 2012.

\bibitem{Samarskii2001}
A.~A. Samarskii.
\newblock {\em {The Theory of Difference Schemes}}.
\newblock CRC Press, New York, 2001.

\bibitem{Schiesser1994}
W.~E. Schiesser.
\newblock {Method of lines solution of the Korteweg-de vries equation}.
\newblock {\em Computers Mathematics with Applications}, 28(10-12):147--154,
  1994.

\bibitem{Seabra-Santos1987}
F.~J. Seabra-Santos, D.~P. Renouard, and A.~M. Temperville.
\newblock {Numerical and Experimental study of the transformation of a Solitary
  Wave over a Shelf or Isolated Obstacle}.
\newblock {\em J. Fluid Mech}, 176:117--134, 1987.

\bibitem{Seabra-Santos1989}
F.~J. Seabra-Santos, A.~M. Temperville, and D.~P. Renouard.
\newblock {On the weak interaction of two solitary waves}.
\newblock {\em Eur. J. Mech. B/Fluids}, 8(2):103--115, 1989.

\bibitem{Serre1953a}
F.~Serre.
\newblock {Contribution {\`{a}} l'{\'{e}}tude des {\'{e}}coulements permanents
  et variables dans les canaux}.
\newblock {\em La Houille blanche}, 8:830--872, 1953.

\bibitem{Serre1956}
F.~Serre.
\newblock {Contribution to the study of long irrotational waves}.
\newblock {\em La Houille blanche}, 3:374--388, 1956.

\bibitem{Shampine1994}
L.~F. Shampine.
\newblock {ODE solvers and the method of lines}.
\newblock {\em Numerical Methods for Partial Differential Equations},
  10(6):739--755, 1994.

\bibitem{Shokin2006}
Y.~I. Shokin, Y.~V. Sergeeva, and G.~S. Khakimzyanov.
\newblock {Predictor-corrector scheme for the solution of shallow water
  equations}.
\newblock {\em Russ. J. Numer. Anal. Math. Modelling}, 21(5):459--479, jan
  2006.

\bibitem{Soderlind2006}
G.~S{\"{o}}derlind and L.~Wang.
\newblock {Adaptive time-stepping and computational stability}.
\newblock {\em J. Comp. Appl. Math.}, 185(2):225--243, 2006.

\bibitem{Sorensen2004}
O.~R. S{\o}rensen, H.~A. Sch{\"{a}}ffer, and L.~S. S{\o}rensen.
\newblock {Boussinesq-type modelling using an unstructured finite element
  technique}.
\newblock {\em Coastal Engineering}, 50(4):181--198, feb 2004.

\bibitem{Su1980}
C.~H. Su and R.~M. Mirie.
\newblock {On head-on collisions between two solitary waves}.
\newblock {\em J. Fluid Mech.}, 98:509--525, 1980.

\bibitem{Tappin2008}
D.~R. Tappin, P.~Watts, and S.~T. Grilli.
\newblock {The Papua New Guinea tsunami of 17 July 1998: anatomy of a
  catastrophic event}.
\newblock {\em Nat. Hazards Earth Syst. Sci.}, 8:243--266, 2008.

\bibitem{Thomas1979}
P.~D. Thomas and C.~K. Lombart.
\newblock {Geometric conservation law and its application to flow computations
  on moving grid}.
\newblock {\em AIAA Journal}, 17(10):1030--1037, 1979.

\bibitem{Tikhonov1961}
A.~N. Tikhonov and A.~A. Samarskii.
\newblock {Homogeneous difference schemes}.
\newblock {\em Zh. vych. mat.}, 1(1):5--63, 1961.

\bibitem{Tikhonov1962}
A.~N. Tikhonov and A.~A. Samarskii.
\newblock {Homogeneous difference schemes on non-uniform nets}.
\newblock {\em Zh. vych. mat.}, 2(5):812--832, 1962.

\bibitem{Tinti1997}
S.~Tinti, E.~Bortolucci, and C.~Vannini.
\newblock {A Block-Based Theoretical Model Suited to Gravitational Sliding}.
\newblock {\em Natural Hazards}, 16(1):1--28, 1997.

\bibitem{Touboul2014}
J.~Touboul and E.~Pelinovsky.
\newblock {Bottom pressure distribution under a solitonic wave reflecting on a
  vertical wall}.
\newblock {\em Eur. J. Mech. B/Fluids}, 48:13--18, nov 2014.

\bibitem{Walkley2002}
M.~Walkley and M.~Berzins.
\newblock {A finite element method for the two-dimensional extended Boussinesq
  equations}.
\newblock {\em Int. J. Num. Meth. Fluids}, 39(10):865--885, aug 2002.

\bibitem{Ward}
S.~N. Ward.
\newblock {Landslide tsunami}.
\newblock {\em J. Geophysical Res.}, 106:11201--11215, 2001.

\bibitem{Watts2003}
P.~Watts, S.~T. Grilli, J.~T. Kirby, G.~J. Fryer, and D.~R. Tappin.
\newblock {Landslide tsunami case studies using a Boussinesq model and a fully
  nonlinear tsunami generation model}.
\newblock {\em Natural Hazards And Earth System Science}, 3(5):391--402, 2003.

\bibitem{Watts2000}
P.~Watts, F.~Imamura, and S.~T. Grilli.
\newblock {Comparing model simulations of three benchmark tsunami generation
  cases}.
\newblock {\em Science of Tsunami Hazards}, 18(2):107--123, 2000.

\bibitem{Wei1995a}
G.~Wei and J.~T. Kirby.
\newblock {Time-Dependent Numerical Code for Extended Boussinesq Equations}.
\newblock {\em J. Waterway, Port, Coastal and Ocean Engineering},
  121(5):251--261, sep 1995.

\bibitem{Zagryadskaya1980}
N.~N. Zagryadskaya, S.~V. Ivanova, L.~S. Nudner, and A.~I. Shoshin.
\newblock {Action of long waves on a vertical obstacle}.
\newblock {\em Bulletin of VNIIG}, 138:94--101, 1980.

\bibitem{Zakharov1991}
V.~E. Zakharov.
\newblock {\em {What Is Integrability?}}
\newblock Springer Series in Nonlinear Dynamics, 1991.

\bibitem{Zhang2013}
Y.~Zhang, A.~B. Kennedy, N.~Panda, C.~Dawson, and J.~J. Westerink.
\newblock {Boussinesq-Green-Naghdi rotational water wave theory}.
\newblock {\em Coastal Engineering}, 73:13--27, mar 2013.

\bibitem{Zhao2015}
B.~B. Zhao, R.~C. Ertekin, and W.~Y. Duan.
\newblock {A comparative study of diffraction of shallow-water waves by
  high-level IGN and GN equations}.
\newblock {\em J. Comput. Phys.}, 283:129--147, feb 2015.

\bibitem{Zheleznyak1985a}
M.~I. Zheleznyak.
\newblock {Influence of long waves on vertical obstacles}.
\newblock In E.~N. Pelinovsky, editor, {\em Tsunami Climbing a Beach}, pages
  122--139. Applied Physics Institute Press, Gorky, 1985.

\bibitem{Zheleznyak1985}
M.~I. Zheleznyak and E.~N. Pelinovsky.
\newblock {Physical and mathematical models of the tsunami climbing a beach}.
\newblock In E.~N. Pelinovsky, editor, {\em Tsunami Climbing a Beach}, pages
  8--34. Applied Physics Institute Press, Gorky, 1985.

\end{thebibliography}

\bigskip\bigskip

\end{document}